\documentclass[onecolumn]{IEEEtran}
 
\usepackage{amsmath, amsfonts, amssymb, amsthm}
\usepackage{bbm}
\usepackage{bbold}
\usepackage{graphicx,dblfloatfix}
\usepackage[space]{cite}
\usepackage{hyperref}
\usepackage{xcolor}
\usepackage{enumerate}
 
\newtheorem{theorem}{Theorem}
\newtheorem{corollary}[theorem]{Corollary}
\newtheorem{lemma}[theorem]{Lemma}
\newtheorem{proposition}[theorem]{Proposition}
\newtheorem{definition}{Definition}

\newtheorem{remark}{Remark}

\renewcommand{\b}[1]{\boldsymbol{#1}} 
\newcommand{\E}{{\mathbb{E} }}
\newcommand{\e}{{\sf{e}}}
\newcommand{\h}[1]{{\hat{#1}}} 
\renewcommand{\t}[1]{{\tilde{#1}}}							
\newcommand{\indicator}[1]{\mathbbm{1}{\left\{ {#1} \right\} }} 
\newcommand{\markov}{{\ \leftrightarrow\ }} 
\newcommand{\N}{{\sf{N}}}
\renewcommand{\Pr}{{\mathbb{P}}} 

\newcommand{\reals}{{\mathbb{R}}}
\newcommand{\s}[1]{\mathcal{#1}} 
\newcommand{\step}[2]{\stackrel{\textnormal{#1}}{#2}}
\renewcommand{\sf}[1]{{\mathsf{#1}}} 
\newcommand{\supp}[1]{{\text{supp}(#1)}}

\newcommand{\ellc}{{\L\backslash\ell}}

\newcommand{\n}{^{\hspace{-0.2mm}\scriptscriptstyle (\hspace{-0.2mm}n\hspace{-0.2mm})}}

\newcommand{\f}{{\sf{f}}}
\newcommand{\enc}{{\phi\n_\ell}}
\newcommand{\encc}{{\phi_\text{c}\n}}
\newcommand{\dec}{{\varphi\n_\ell}}

\renewcommand{\L}{{\s{L}}}

\newcommand{\sM}{{\s{M}\n}}
\newcommand{\Mc}{{M_\text{c}\n}}
\newcommand{\sMc}{{\s{M}_\text{c}\n}}
\newcommand{\Ml}{{M\n_\ell}}

\renewcommand{\S}{{\s{S}}}

\renewcommand{\a}{{\b{a}}}

\newcommand{\X}{{\b{X}}}
\newcommand{\x}{{\b{x}}}
\newcommand{\hX}{{\hat{X}}}
\newcommand{\tX}{{\t{X}}}
\newcommand{\hx}{{\hat{x}}}
\newcommand{\tx}{{\t{x}}}
\newcommand{\sX}{{\s{X}}}
\newcommand{\sXL}{{\b{\s{X}}}}
\newcommand{\hsX}{{\h{\sX}}}
\newcommand{\tsX}{{\t{\sX}}}
\newcommand{\hXL}{{\h{\X}}}
\newcommand{\hsXL}{{\b{\hsX}}}
\newcommand{\XS}{{X_\S}}

\newcommand{\Y}{{\b{Y}}}
\newcommand{\hY}{{\h{\Y}}}
\newcommand{\y}{{\b{y}}}

\newcommand{\sU}{{\s{U}}}

\newcommand{\mw}[1]{{\color{black}#1}}

\newcommand{\ssb}[1]{{\color{blue}#1}}

\newcommand{\ve}{{\varepsilon}}
\newcommand{\eps}{{\epsilon}}

\newcommand{\R}{{\sf{R}}} 
\newcommand{\RR}{{\b{R}}}

\newcommand{\Rexp}{{\R^\dag_{\bf}}}
\newcommand{\Rexc}{{\R^\ddag_{\bf}}}
\newcommand{\Rmaxexc}{\mw{\tilde{\R}^\dag_{\bf,\textnormal{max-exc}}}}

\newcommand{\RexpG}{{\R^\dag_\text{G}}} 
\newcommand{\RG}{{\R_\text{G}}}
\newcommand{\RGjoint}{{\R_{\text{G},X_1X_2}}}

\newcommand{\Rgenie}{{\sf{g}}}
\newcommand{\Rsuper}{{\sf{s}}}

\newcommand{\RGW}{{\s{R}_\text{GW}}}

\newcommand{\fl}{{\f_\ell}}
\renewcommand{\bf}{{\b{\f}}}
\newcommand{\bd}{{\b{\d}}}

\newcommand{\bfd}{{\overline{\f\d}}}
\newcommand{\bfdl}{{\bfd_\ell}}
\newcommand{\bfdlp}{\mw{\bfd_{\ell'}}}

\newcommand{\D}{{\boldsymbol{D}}}
\newcommand{\Dmax}{{D_\text{max}}}
\newcommand{\0}{{\b{0}}}
\renewcommand{\d}{{\sf{d}}}
\newcommand{\dl}{{\sf{d}_\ell}}
\newcommand{\db}{{\hspace{2pt} \overline{\d}}}
\newcommand{\dbl}{{\bar{\d}_\ell}}

\newcommand{\DS}{{D_\S}}

\newcommand{\K}{{\b{K}}}

\newcommand{\KXS}{{\K_{\XS}}}
\newcommand{\Ktwo}{{\K_{X_1 X_2}}}

\newcommand{\KW}{{\sf{K}_\text{W}}}
\newcommand{\KGK}{{\sf{K}_\text{GK}}} 
 
\newcommand{\Ncc}{{N_\text{cc}}}

\newcommand{\bbar}[1]{{\bar{\bar{#1}}}} 
\newcommand{\tN}{{\tilde{N}}} 

\newcommand{\Kgenie}{{\sf{C}_\sf{g}}}
\newcommand{\Ksgenie}{{\sf{C}_\sf{g}^\ast}}
\newcommand{\Ksuper}{{\sf{C}_\sf{s}}}
\newcommand{\Kssuper}{{\sf{C}^*_\sf{s}}}

\newcounter{tempEquationCounter} 
\newcounter{thisEquationNumber}
\newenvironment{floatEq}
{\setcounter{thisEquationNumber}{\value{equation}}\addtocounter{equation}{1}
\begin{figure*}[!t]
\normalsize\setcounter{tempEquationCounter}{\value{equation}}
\setcounter{equation}{\value{thisEquationNumber}}
}
{\setcounter{equation}{\value{tempEquationCounter}}
\hrulefill\vspace*{4pt}
\end{figure*}
}

\allowdisplaybreaks[4] 
\graphicspath{{.}{./Figures/}} 

\begin{document}

\title{A Rate-Distortion Approach to Caching}

\author{Roy Timo, Shirin Saeedi Bidokhti, Mich\`{e}le Wigger  and Bernhard C. Geiger
\thanks{R.~Timo is with Ericsson Research, Stockholm, roy.timo@ericsson.com. S.~Saeedi Bidokhti is with the Department of Electrical Engineering, Stanford University, saeedi@stanford.edu. M.~Wigger is with the Communications and Electronics Dsepartment, Telecom ParisTech, michele.wigger@telecom-paristech.fr. B.~Geiger is with the Institute for Communications Engineering, Technical University of Munich, bernhard.geiger@tum.de.}
\thanks{Some of the material in this paper was completed by R. Timo at the Technical University of Munich and presented at the International Zurich Seminar on Communications (IZS), March, 2016.}
\thanks{S.~Saeedi Bidokhti was supported by the Swiss National Science Foundation Fellowship no. 158487. Bernhard C. Geiger was supported by the Erwin Schr\"odinger Fellowship J 3765 of the Austrian Science Fund.}
}

\maketitle

\begin{abstract}
This paper takes a rate-distortion approach to understanding the information-theoretic laws governing cache-aided communications systems. Specifically, we characterise the optimal tradeoffs between the \emph{delivery rate}, \emph{cache capacity} and \emph{reconstruction distortions} for a single-user problem and some special cases of a two-user problem. Our analysis considers discrete memoryless sources, expected- and excess-distortion constraints, and separable and $\bf$-separable distortion functions. We also establish a strong converse for separable-distortion functions, and we show that lossy versions of common information (G\'{a}cs-K\"{o}rner and Wyner) play an important role in caching.  Finally, we illustrate and explicitly evaluate these laws for multivariate Gaussian sources and binary symmetric sources. 
\end{abstract}


\section{Introduction}

\IEEEPARstart{T}{his} paper takes a rate-distortion approach to understanding the information-theoretic laws governing cache-aided communications systems. To fix ideas, let us start by outlining some of the applications that motivated our study.

\emph{On-demand media streaming:} Imagine an on-demand internet media provider, and consider the problem of streaming media to millions of users. A common problem  is that the users will most likely request and stream media during periods of high network congestion. For example, most users would prefer to watch a movie during the evening, rather than during the early hours of the morning. Downloading bandwidth hungry media files during such periods leads to further congestion, high latency, and poor user experience.  

To help overcome this problem, content providers often cache useful information about the media database in small storage systems at the network edge (with fast user connections) during periods of low network congestion. The basic idea is that information placed in these caches will not have to be transported later over a congested network. Naturally these small storage systems cannot host the entire media library, so the provider must carefully cache information that will be most useful to the users' future requests.

\emph{Distributed databases:} Now imagine a large database that is distributed over a vast global disk-storage network. Such a database might contain measurements taken by weather or traffic sensors spread across several countries; the time-series prices of companies' stock (or, FX prices) at different exchanges; the shopping history of customers; the browsing history of users; or the mobility patterns (or, channel-state measurements) of mobile devices in cellular networks. 

Now suppose that a user queries the database and requests an approximate copy of one file (or, perhaps, a function of several files). Since the database is large and distributed, we can expect that it will need to make several network calls to load relevant data in memory before it can communicate the file to the user. Such network calls are performance bottlenecks, potentially leading to high latency and network traffic costs.  

Modern database systems handle such problems by smartly caching the most common queries in fast memory. If, for example, it is known in advance that the user will request the weather forecast of a particular city, then we can simply cache part or all of this forecast in memory. Obviously, however, we cannot always know in advance what data will be requested, so we should carefully cache information that is useful to many different requests. 

The main purpose of this paper is to help develop a better understanding of such cache-aided communications systems. We will focus on single-user systems, and we will try to determine the ``most useful'' information to place in the cache.

In the spirit of Maddah-Ali and Niesen~\cite{Maddah-May-2014-A,Niesen-Jul-2015-A}, we will break the problem into two distinct phases: A \emph{caching phase} concerning the pre-placement of information in the \emph{cache}, and a \emph{delivery phase} concerning the reliable communication of the particular source (or, file) requested by the user. 

Since the caching phase occurs before the user makes its request, it seems reasonable that the information placed in the cache should be common to many different sources in the library. Moreover, to minimise overhead and latency during peak-congestion times, it seems reasonable that the delivery-phase message should not duplicate any information already stored in the cache. With this in mind, we will focus on the following problems. For a given library of sources and a given cache capacity:
\begin{itemize}
\item What ``common information'' should be put in the cache? 
\item What is the minimum delivery-phase rate needed to achieve a given fidelity requirement at the user?  
\end{itemize}
Our study will make the following specific assumptions.
\begin{itemize}

\item The library consists of $L$ different sources, and each source consists of $n$ symbols. Here $L$ is any fixed positive integer, and we consider the information-theoretic limits of cache-aided communications in the limit $n \to \infty$.

\item The cache can reliably store up to $nC$ bits, and it is said to have capacity $C$.

\item The fidelity of the user's reconstruction of the requested source can be meaningfully measured by a separable distortion function or, more generally, by an $\bf$-separable distortion function.\footnote{\mw{Roughly speaking, separable distortion functions can be expressed as an average of a given per-letter distortion function over the sequence of pairs of source and reconstruction symbols. In $\bf$-separable distortion functions this average is replaced by a more general function. Precise definitions are given in Definitions~\ref{def:sep} and \ref{def:fsep} in Section~\ref{sec:sep}. The class of $\bf$-separable distortion functions was recently introduced by Shkel and Verd\'u~\cite{Shkel-Feb-2016-C}, and it has a rather appealing axiomatic motivation that we review later in Section~\ref{Sec:OperationalRDC}}} . 
 
\item The $L$ sources are generated by an arbitrary $L$-component discrete memoryless source (DMS). This assumption is quite common in the multi-terminal information theory literature, admits rigorous proofs and nevertheless gives a great deal of insight about more complicated models. Although it seems restrictive, some important transformations (e.g.~Burrows-Wheeler) are known to emit almost memoryless processes~\cite{Visweswariah-Jun-2000-C,Effros-May-2002-A}.
\end{itemize}

Our paper is most related to Wang, Lim and Gastpar~\cite{Wang-Apr-2015-A}. A key difference to~\cite{Wang-Apr-2015-A}, however, is the source request model: Wang~\emph{et al.} assumed that each user randomly selects a symbol from each source at each time in an independent and identically distributed (iid) manner. They then leveraged connections to several classic multi-terminal problems to establishe some interesting tradeoffs between the optimal compression rate and cache capacity under a lossless\footnote{Specifically, Wang \emph{et al.} required that a function of the source is reliably reconstructed (otherwise known as a deterministic distortion function).} reconstruction constraint. In contrast to~\cite{Wang-Apr-2015-A}, we will require that the user requests one source in its entirety, we do not place prior probabilities on the user's selection, and we allow for lossy reconstructions. We thus consider a lossy worst-demand (i.e., compound source) scenario, while~\cite{Wang-Apr-2015-A} considered an ergodic iid-demand scenario. 

Hassanzadeh, Erkip, Llorca and Tulino~\cite{Hassanzadeh-Nov-2015} recently studied cache-aided communications systems for transmitting independent memoryless Gaussian sources under mean-squared error distortion constraints. Their caching schemes exploited successive-refinement techniques to minimise the mean-squared error of the users' reconstructions, and they presented a useful ``reverse filling-type solution'' to the minimum distortion problem. Yang and G\"und\"uz~\cite{Yang-Apr-2016-A} consider the same cache-aided Gaussian problem, but instead focussed on the minimum delivery-phase rate for a given distortion requirement. They presented a numerical method to determine the minimum delivery rate, and proposed two efficient caching algorithms. 

This paper will try to broaden the above work and improve our understanding of cache-aided communications. We will do this by, for example, considering arbitrarily correlated sources and lossy reconstructions with respect the general class of $\bf$-separable distortion functions.  

\emph{Organisation:} The problem setup is formally described in Section~\ref{Sec:OperationalRDC}. Our main results for separable distortion functions are presented in Sections~\ref{Sec:InformationalRDC} to~\ref{Sec:StrongConverse}. Specifically, Section~\ref{Sec:InformationalRDC} presents the optimal rate-distortion-cache tradeoffs; Section~\ref{Sec:CommonInformation} shows that two information-theoretic notions of common information (G\'acs-K\"orner and Wyner) play an important role in caching; Section~\ref{Sec:GrayWyner} relates caching to Gray and Wyner's seminal paper \emph{``Source coding for a simple network''}~\cite{Gray-Nov-1974-A}; Section~\ref{Sec:StrongConverse} presents a strong converse (in the sense of Kieffer~\cite{Kieffer-Mar-1991-A}); and Section~\ref{Sec:Examples} considers Gaussian and binary sources. Finally, Section~\ref{Sec:fSeparableDistortions} presents the optimal rate-distortion-cache tradeoffs for $\bf$-separable distortion functions, and Section~\ref{Sec:TwoUsers} considers a two-user version of the problem. 


\section{Problem Setup}\label{Sec:OperationalRDC}

\begin{figure}[t]
\begin{center}
\includegraphics[width=0.5\columnwidth]{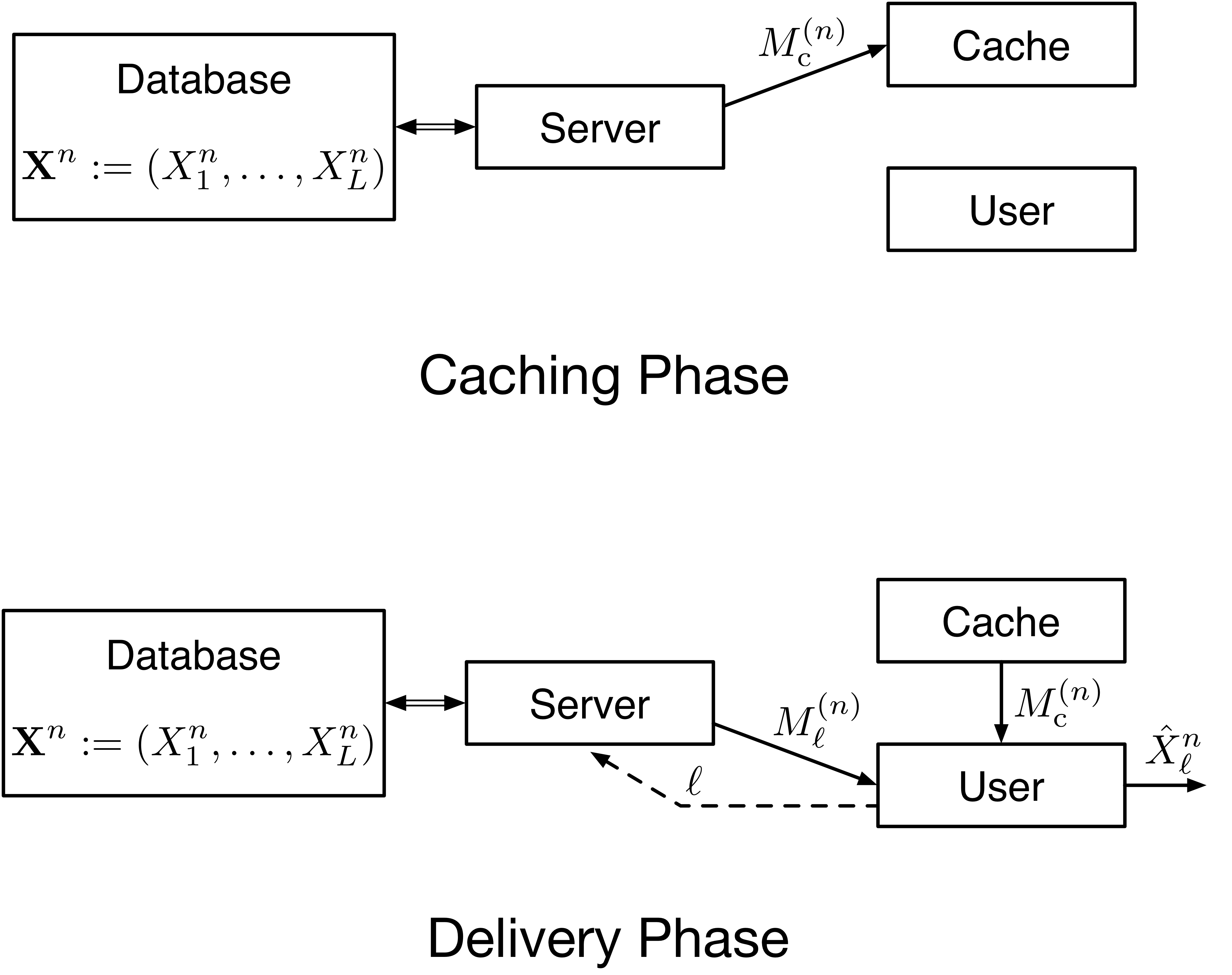}
\caption{A cache-aided communications system with a single user.}
\label{Fig:RD-Cache}
\end{center}
\end{figure}


\subsection{Source Model and RDC Codes}

Let $\L = \{1,\ldots,L\}$ denote the set of indices corresponding to the $L$ sources in the library. We assume throughout that the $\ell$-th source consists of $n$ symbols,
\begin{equation*}
X^n_\ell = (X_{\ell,1},X_{\ell,2},\ldots,X_{\ell,n}),
\end{equation*}
with each drawn from a finite alphabet $\sX_\ell$. The $L$ sources 
\begin{equation*}
\X^n = (X_{1}^n, X_{2}^n, \dots, X_{L}^n)
\end{equation*}
are randomly generated by a $L$-component \emph{discrete memoryless source (DMS)}; that is, $\X^n$ is a sequence of $n$ independent and identically distributed (iid) tuples $\X = (X_1,\ldots,X_L)$ characterized by an arbitrary joint pmf $p_\X(\x)$ defined on the Cartesian product space $\sXL = \sX_1 \times \cdots \times \sX_L$. Let $\hsX_\ell$ be an appropriate finite reconstruction alphabet for the $\ell$-th source at the receiver and let $\hsXL = \hsX_1 \times \cdots \times \hsX_L$.

A joint \emph{rate-distortion-cache (RDC) code} for a given blocklength $n$ is a collection of $(2L+1)$ mappings:
\begin{enumerate}[(i)]
\item A \emph{cache-phase encoder} at the server 
\begin{equation*}
\encc : \sXL^n \to \sMc.
\end{equation*}
Here $\sMc$ is a finite (index) set with an appropriate cardinality for the cache capacity. 
\item A \emph{delivery-phase encoder} at the server
\begin{equation*}
\enc : \sXL^n \to \sM
\end{equation*}
for each user request $\ell \in \L$. Here $\sM$ is a finite (index) set with an appropriate cardinality for the delivery phase.
\item A \emph{delivery-phase decoder} at the user
\begin{equation*}
\dec : \sM \times \sMc \to \hsX_\ell^n
\end{equation*}
for each possible user request $\ell \in \L$.
\end{enumerate}
We call the above collection of encoders and decoders an $(n,$ $\sM,\sMc)$-\emph{code}.  

During the \emph{caching phase} (before the user requests a particular source), the server places the message 
$\Mc = \encc(\X^n)$ in the cache. At some later time (the \emph{delivery phase}), the user picks $\ell \in \L$ arbitrarily and requests the corresponding source $X^n_\ell$ from the server. The server responds to the user's request with the message $\Ml = \enc(\X^n)$, and the user attempts to reconstruct $X^n_\ell$ by computing $\hX^n_\ell = \dec(\Ml,\Mc)$. This \emph{cache-aided} encoding and decoding process is illustrated in Figure~\ref{Fig:RD-Cache}. 

A key point to note here is that the cache message $\Mc$ is encoded and stored in the cache before the user selects $\ell \in \L$ and requests the corresponding source $X^n_\ell$ from the server. Hence, the cached message $\Mc$ should (in a sense to be made precise later) efficiently represent useful \emph{common information} shared between sources in the library.


\subsection{Separable and $\bf$-Separable Distortion Functions}\label{sec:sep}

As per the usual RD paradigm, let us first assume that the fidelity of the user's reconstruction $\hX_\ell^n$ of the $\ell$-th source $X^n_\ell$ can be meaningfully quantified by the arithmetic mean of single-symbol distortions. Specifically, for each $\ell \in \L$ let 
\begin{equation}\label{Eqn:SymbolDistortion}
\dl : \hsX_\ell \times \s{X}_\ell \to [0,\infty)
\end{equation}
be a \emph{single-symbol distortion function}. For example, we will often take $\dl$ to be the \emph{Hamming distortion function} where $\hsX_\ell = \sX_\ell$ and 
\begin{equation*}
\d_\ell(\hx_\ell,x_\ell) = 
\left\{
\begin{array}{rl}
1 & \text{ if } \hx_\ell \neq x_\ell\\
0 & \text{ if } \hx_\ell = x_\ell.
\end{array}
\right.
\end{equation*}

We assume throughout that each $\d_\ell$ satisfies the following two conditions:
\begin{itemize}
\item For each source symbol $x_\ell \in \s{X}_\ell$ there exists a reconstruction symbol $\h{x}_\ell \in \hsX_\ell$ such that $\dl(\h{x}_\ell,x_\ell) = 0$.
\item There exists a finite $\Dmax > 0$ such that $\dl(\hx_\ell,x_\ell) \leq \Dmax$ for all $x_\ell \in \sX_\ell$ and $\hx_\ell \in \hsX_\ell$.
\end{itemize}

\begin{definition}[Separable distortion function]\label{def:sep}
The $n$-symbol distortion between a particular source realisation $x^n_\ell \in \sX^n_\ell$ and reconstruction $\hx^n_\ell \in \hsX^n_\ell$ is
\begin{equation}\label{Eqn:SeparableDistortionFunctionDefinition}
\dbl(\hx_\ell^n, x_\ell^n) := \frac{1}{n} \sum_{i=1}^n \dl(\hx_{\ell,i},x_{\ell,i}).
\end{equation}
\end{definition}
Let $\bd = (\d_1,\ldots,\d_L)$ and $\bar{\bd} = (\db_1,\ldots,\db_L)$.

Although separable distortion functions are almost ubiquitous in the literature, it is also useful to consider the broader class of $\bf$-separable distortion functions recently introduced by Shkel and Verd\'u~\cite{Shkel-Feb-2016-C}. Specifically, let us define the following for each request $\ell \in \L$: Let 
\begin{equation}\label{Eqn:f-Functions}
\fl : [0,\infty) \to [0,\infty)
\end{equation}
be continuous and strictly increasing, and let $\dl$ be a single-symbol distortion function~\eqref{Eqn:SymbolDistortion}. 

\begin{definition}[$\bf$-Separable distortion function]\label{def:fsep}
The $n$ symbol distortion between $x^n_\ell \in \sX^n_\ell$ and $\hx^n_\ell \in \hsX^n_\ell$ is 
\begin{equation}\label{Eqn:fSeparableDistortionFunctions}
\bfdl(\hx^n, x^n) := \fl^{-1} \left(\frac{1}{n} \sum_{i = 1}^n \fl\big(\dl(\hx_{i},x_{i})\big) \right).
\end{equation}
\end{definition}
Let $\b{\bfd} = (\bfd_1,\ldots,\bfd_L)$.

The basic idea in~\eqref{Eqn:fSeparableDistortionFunctions} is to choose the function $\fl$ to assign appropriate (possibly non-linear) ``frequency costs'' to different quantisation error events. If $\fl$ is the identity mapping, then $\bfdl$ reduces to the usual separable distortion function $\db_\ell$ generated by $\dl$. Several interesting connections between $\f$-separable distortions and R\'enyi entropy, compression with linear costs, and sub-additive distortion functions are discussed in~\cite{Shkel-Feb-2016-C}. Moreover, $\f$-separable distortions have a rather pleasing axiomatic motivation based on the following observation by Kolmogorov~\cite{Tikhomirov-1991-Mean-A}.

\begin{proposition}
Let $\{a_1,\ldots,a_n\}$ be a set of $n$ real numbers and $\bar{\sf{M}}_n: \reals^n \to \reals$ satisfy the following four \emph{axioms of mean}:
\begin{enumerate}[(i)]
\item $\bar{\sf{M}}_n(a_1,\ldots,a_n)$ is a continuous and strictly increasing function of each argument $a_i$.  
\item $\bar{\sf{M}}_n(a_1,\ldots,a_n)$ is a symmetric function of its arguments. 
\item $\bar{\sf{M}}_n(a,\ldots,a) = a$.
\item For any integer $m \leq n$,
\begin{equation*}
\bar{\sf{M}}_n(a_1,\ldots,a_m,\ldots,a_n) \\
= \bar{\sf{M}}_n(a,\ldots,a,a_{m+1},\ldots,a_n), 
\end{equation*}
where $a = \bar{\sf{M}}_m(a_1,\ldots,a_m)$. 
\end{enumerate}
Then $\bar{\sf{M}}_n$ must take the form~\cite[p.~144]{Tikhomirov-1991-Mean-A}
\begin{equation*}
\bar{\sf{M}}_n(a_1,\ldots,a_n)  = \f^{-1} \left(\frac{1}{n} \sum_{i=1}^n \f(a_i) \right)
\end{equation*} 
for some continuous and strictly increasing $\f$. 
\end{proposition}


\subsection{Operational RDC Functions}

We will consider two different problem formulations: Optimal caching subject to an \emph{expected distortions} criteria, and optimal caching subject to an \emph{excess distortions} criteria. Throughout, let $\D=(D_1,\dots,D_L)$.

\begin{definition}\label{Def:RDCFunc:ExpDist}
We say that a rate-distortion-cache tuple $(R,$ $\D,C)$ is $\b{\bfd}$-\emph{achievable with respect to (w.r.t.) expected distortions} if there exists a sequence of $(n,\sM,\sMc)$-codes such that
\begin{subequations}\label{Eqn:Def:Ach:ExpDist}
\begin{align}
\label{Eqn:Def:Ach:ExpDist:CacheCapacity}
\limsup_{n \to \infty} \frac{1}{n} \log |\sMc| &\leq C,\\
\label{Eqn:Def:Ach:ExpDist:DeliveryRate}
\limsup_{n \to \infty} \frac{1}{n} \log |\sM| &\leq R, \text{ and }\\
\label{Eqn:Def:Ach:ExpDist:Distortions}
\limsup_{n\to\infty}
\E \Big[ \bfdl(\h{X}_\ell^n,X_\ell^n) \Big] &\leq D_\ell, 
\quad 
\forall\ \ell \in \L.
\end{align}
\end{subequations}
The \emph{RDC function w.r.t.~expected distortions} $\Rexp(\D,C)$ is the infimum of all rates $R \geq 0$ such that the rate-distortion-cache tuple $(R,\D,C)$ is $\b{\bfd}$-achievable.
\end{definition}

\begin{definition}\label{Def:RDCFunc:ExcDist}
We say that a rate-distortion-cache tuple $(R,$ $\D,C)$ is $\b{\bfd}$-\emph{achievable w.r.t.~excess distortions} if there exists a sequence of $(n,\sM,\sMc)$-codes such that~\eqref{Eqn:Def:Ach:ExpDist:CacheCapacity} and~\eqref{Eqn:Def:Ach:ExpDist:DeliveryRate} hold and
\begin{equation}\label{Eqn:Def:Ach:ExcDist:Distortions}
\lim_{n\to\infty}
\Pr \left[\ \bigcup_{\ell \in \L} \Big\{ \bfdl(\h{X}_\ell^n,X_\ell^n) \geq D_\ell \Big\} \right] = 0.
\end{equation}
The \emph{RDC function w.r.t.~excess distortions} $\Rexc(\D,C)$ is the infimum of all rates $R \geq 0$ such that the rate-distortion-cache tuple $(R,\D,C)$ is $\b{\bfd}$-achievable.
\end{definition}

For separable distortion functions, we will omit the subscript $\bf$ from the above definitions (because it is an identity mapping) and simply write $\R^\dag(\D,C)$ and $\R^\ddag(\D,C)$.


\section{\mw{Result:  RDC Function for} Separable Distortion Functions}\label{Sec:InformationalRDC}

This section presents a single-letter expression (the informational RDC function) for the expected and excess operational RDC functions defined in Section~\ref{Sec:OperationalRDC}.

\subsection{\mw{Preliminaries}}
We will need the following basic RD functions. The standard \emph{informational RD function} of the $\ell$-th source~$X_\ell$ w.r.t.~$\dl$ is
\begin{equation*}
\R_{X_\ell}(D_\ell): = 
\min_{p_{\h{X}_\ell|X_\ell} :\ \E[\dl(\hX_\ell,X_\ell)] \leq D_\ell} I(X_\ell;\h{X}_\ell),
\end{equation*} 
where the minimisation is over all test channels $p_{\h{X}_\ell|X_\ell}$ from~$\sX_\ell$ to $\hsX_\ell$ satisfying the indicated distortion constraint. 

The \emph{informational joint RD function of $\X$} w.r.t.~$\bd$ is~\cite{Gray-Oct-1972-A}
\begin{equation*}
\R_\X(\D) := \min_{p_{\hXL|\X}:\ \E[\dl(\hX_\ell,X_\ell)] \leq D_\ell,\ \forall \ell \in \L} I(\X;\hXL),
\end{equation*} 
where the minimisation is over all test channels $p_{\hXL | \X}$ from $\sXL$ to $\hsXL$ satisfying all $L$ of the indicated distortion constraints. 

The \emph{informational conditional RD function}~\cite{Gray-Oct-1972-A} of $X_\ell$ with side information $U$ is 
\begin{equation*}
\R_{X_\ell|U}(D_\ell) := 
\min_{p_{\hX_\ell|X_\ell U} :\ \E[ \dl(\hX_\ell,X_\ell)] \leq D_\ell}\ 
I(X_\ell ; \hX_\ell | U),
\end{equation*}
where the minimisation is over all test channels $p_{\hX_\ell|X_\ell U}$ from $\sX_\ell \times \sU$ to $\hsX_\ell$ satisfying the indicated distortion constraint. The above minima exist by the continuity of Shannon's information measures, the bounded single-symbol distortion functions $\bd$, and the fact that each (conditional) mutual information is minimised over a compact set.
 
 \subsection{\mw{Result}}
The next lemma summarises some basic properties of the operational RDC functions that we will use frequently. We omit the proof.

\begin{lemma}\label{Lem:BasicProperties}
The following statements are true for separable distortion functions.
\begin{enumerate}[(i)]

\setlength\itemsep{2pt}

\item 
$\R^\dag(\D,C)$ and $\R^\ddag(\D,C)$ are convex, non-increasing and continuous in $(\D,C) \in [0,\infty)^{L+1}$.

\item 
If $C > \R_\X(\D)$, then 
\begin{equation*}
\R^\dag(\D,C) = \R^\ddag(\D,C) = 0.
\end{equation*}

\item 
If $C = 0$, then 
\begin{equation*}
\R^\dag(\D,0) = \R^\ddag(\D,0) = \max_{\ell \in \L} \R_{X_\ell}(D_\ell).
\end{equation*} 

\item 
If an $(n,\sM,\sMc)$-code satisfies  
\begin{equation*}
\Pr \left[\ \bigcup_{\ell \in \L} \Big\{ \dbl(\h{X}_\ell^n,X_\ell^n) \geq D_\ell \Big\} \right] \leq \epsilon
\end{equation*}
for some $\epsilon > 0$, then the same code also satisfies 
\begin{equation*}
\E\Big[ \dbl(\hX^n_\ell, X^n_\ell) \Big] \leq D_\ell + \epsilon \Dmax,
\quad 
\forall\ \ell \in \L. 
\end{equation*}
\end{enumerate}
\end{lemma}

We now define  the informational RDC function of interest to our caching problem.
Let $\sU$ be a finite alphabet of cardinality $|\s{U}| \leq |\sXL| + 2L.$ The \emph{informational RDC function} of interest is 
\begin{equation}\label{Eqn:RDCFunctionSingleLetter}
\R(\D,C) := \min_{U :\ I(\X;U) \leq C}\ \max_{\ell \in \L}\ \R_{X_\ell | U}(D_\ell),
\end{equation}
where the minimisation is taken over the set of all auxiliary random variables $U$ on $\sU$ jointly distributed with $\X$ satisfying the indicated mutual information constraint.

\begin{theorem}\label{Thm:RDC} 
For separable distortion functions, \mw{the RDC functions for both expected and excess distortions coincide with the informational RDC function in \eqref{Eqn:RDCFunctionSingleLetter}:}
\begin{equation*}
\R^\dag(\D,C) = \R^\ddag(\D,C) = \R(\D,C).
\end{equation*}
\end{theorem}

\begin{IEEEproof}
Theorem~\ref{Thm:RDC} is proved in Appendix~\ref{Thm:RDC:Sec}.
\end{IEEEproof}

The next corollary particularises Theorem~\ref{Thm:RDC} to the (almost) lossless reconstruction setting. We omit the proof. 

\begin{corollary}\label{Thm:RDC:Cor}
For Hamming distortion functions, 
\begin{align*}
\R^\dag(\b{0},C) 
= \R^\ddag(\b{0},C) 
= \R(\b{0},C) 
= \min_{U:\ I(\X;U) \leq C}\ \max_{\ell \in \L}\ H(X_\ell|U),\\
\end{align*}
where $\sU$ can be restricted to $|\s{U}| \leq |\sXL| + L$.
\end{corollary}



\section{\mw{Caching Interpretations of the  G\'acs-K\"orner and Wyner's Common Information}}\label{Sec:CommonInformation}

We now consider two special `enhanced' caching setups and derive their RDC functions. We then show that the conditions under which the RDC function $\R(\D,C)$ of our original caching problem coincides with the RDC functions of these enhanced setups relates to the different common-information definitions of G\'acs, K\"orner, and Wyner. 


\subsection{Genie-Aided Caching and G\'acs-K\"orner Common Information}

Imagine that, before the caching phase, a genie tells the server which $\ell \in \L$ the user will choose in the future. The optimal caching strategy for this hypothetical \emph{genie-aided} system is obvious: We should compress the $\ell$-th source $X^n_\ell$ using an optimal RD code, cache $nC$ bits of the code's output, and then send the remaining bits during the delivery phase. The RDC function of the genie-aided problem is therefore
\begin{equation*}
\Rgenie(\D,C) = \Big[ \max_{\ell \in \L}\ \R_{X_\ell}(D_\ell) - C \Big]^+,
\end{equation*}
where $[a]^+ = \max\{0, a\}$.

In the main problem at hand, however, the server does not know in advance which $\ell \in \L$ the user will select, and this uncertainty may cost additional rate in either the caching or delivery phases. Since the optimal performance of the genie-aided system cannot be worse than that of our system, we immediately have the following lemma. 

\begin{lemma}\label{Lem:RDGenieLowerBound} 
$\R(\D,C) \geq \Rgenie(\D,C).$
\end{lemma}

\begin{figure}
\begin{center}
\includegraphics[width=.5\columnwidth]{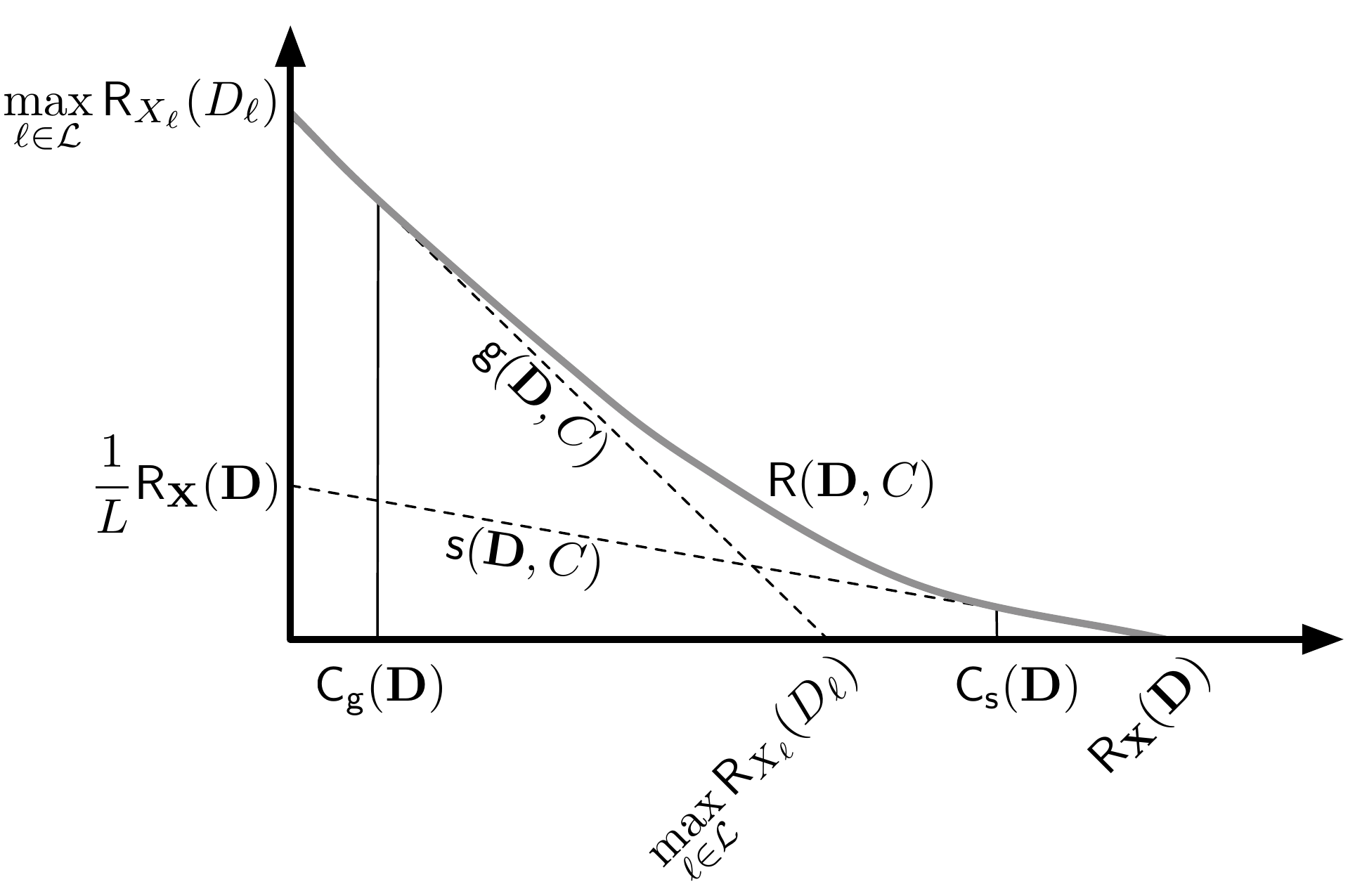}
\caption{An illustration of some typical characteristics of the RDC function $\R(\D,C)$ for a fixed distortion tuple $\D = (D_1,\ldots,D_L)$. The function describes the optimal (minimum) delivery-phase rate (vertical axis) for a given cache capacity (horizontal axis). The bounds in Lemmas~\ref{Lem:RDGenieLowerBound} and~\ref{Lem:SuperuserLowerBound} are plotted with dashed lines.}
\label{Fig:TypicalRDCFunction}
\end{center}
\end{figure}

Clearly, in the caching system, we can always achieve the genie bound at $C = 0$. It is therefore natural to consider the critical cache capacity
\begin{equation}\label{Eqn:Kgenie}
\Kgenie(\D): = \max\Big\{ C \geq 0 : \R(\D,C) = \Rgenie(\D,C) \Big\}.
\end{equation}   
That is, $\Kgenie(\D)$ is the largest \mw{cache} capacity so that there is no loss in performance because the transmitter has to fill the cache memory before learning the demand $\ell$. The maximum indicated in~\eqref{Eqn:Kgenie} exists because, for $0 \leq C \leq \R_\X(\D)$, $\R(\D,C)$ is convex and $\Rgenie(\D,C)$ is linear. Figure~\ref{Fig:TypicalRDCFunction} illustrates some typical characteristics of $\R(\D,C)$ and $\Rgenie(\D,C)$. 

We now consider the critical cache capacity $\Kgenie(\D)$. Let 
\begin{equation*}
\L^*(\D): = \Big\{ \ell^* \in \L : \R_{X_{\ell^*}}(D_{\ell^*}) = \max_{\ell \in \L} \R_{X_{\ell}}(D_\ell) \Big\}.
\end{equation*}
Define 
\begin{equation*}
\Ksgenie(\D) := \max_{U} I(\X;U),
\end{equation*}
where the maximisation is taken over the set of all auxiliary random variables $U$ on $\sU$ jointly distributed with $\X$ such that for all $\ell^* \in \L^*$ we have 
\begin{subequations}\label{Eqn:GenieCondition}
\begin{equation}\label{Eqn:GenieCondition1}
I(\X;U)
= 
\R_{X_{\ell^*}}(D_{\ell^*}) - \R_{X_{\ell^*}|U}(D_{\ell^*})
\end{equation}
and
\begin{equation}\label{Eqn:GenieCondition2}
\R_{X_{\ell^*}|U}(D_{\ell^*}) = \max_{\ell \in \L} \R_{X_{\ell}|U}(D_{\ell}).
\end{equation}
\end{subequations}

\begin{theorem}\label{Thm:Kgenie}
For separable distortion functions, 
\begin{equation*}
\Kgenie(\D) = \Ksgenie(\D).
\end{equation*}
\end{theorem}

\begin{corollary}\label{Thm:Kgenie:Cor}
For Hamming distortions
\begin{equation*}
\Kgenie(\b{0})
= \Ksgenie(\b{0}) 
= \max_U I(\X;U),
\end{equation*}
where the maximisation is taken over the set of all $U$ satisfying 
\begin{equation*}
U \markov X_{\ell^*} \markov X_{\L \backslash \ell^*}
\quad 
\forall\ \ell^* \in \L^*
\end{equation*}
and
\begin{equation*}
H(X_{\ell^*} | U) = \max_{\ell \in \L} H(X_\ell | U)
\quad 
\forall\ \ell^* \in \L^*.
\end{equation*}
\end{corollary}
\begin{IEEEproof}
Theorem~\ref{Thm:Kgenie} and Corollary~\ref{Thm:Kgenie:Cor} are proved in Appendices~\ref{Thm:Kgenie:Sec} and~\ref{Thm:Kgenie:Cor:Sec} respectively.
\end{IEEEproof}

We now explain Corollary~\ref{Thm:Kgenie:Cor} at hand of an easy example. Suppose that we only have two sources $X_1 = (A,B_1)$ and $X_2 = (A,B_2)$ where $A$, $B_1$ and $B_2$ are mutually independent and $H(B_1)=H(B_2)$. In this case, choose $U=A$. This way, $\Ksgenie(\b{0}) =H(A)$ and
\begin{equation*}
\R(\b{0},C) 
= \Rgenie(\b{0},C)
= H(A) + H(B_1)- C, \qquad C  \leq H(A).
\end{equation*}
In this example the optimal $U$ in Corollary~\ref{Thm:Kgenie:Cor} is ``common'' to both sources. In fact,  if we had placed additional private information about, say, $X_1$ (i.e., information about $B_1$) in the cache, then this information would be wasted whenever the user downloads $X_2$. 

The above idea naturally leads to a multivariate version of Viswanatham, Akyol and Rose's~\cite{Viswanatha-Jun-2014-A} definition of \emph{lossy G\'acs-K\"orner common information}. (See~\cite{Gacs-1973-A} for G\'acs and K\"orner's original treatment of common information.) 

\begin{definition}\label{Def:GacsKorner:Lossy}
Define the \emph{lossy G\'acs-K\"orner common information} of $\X$ w.r.t.~$\bd$ by
\begin{equation}\label{eq:GKlossy}
\KGK(\X;\D)
:= \max_{(U,\hXL)} I(\X;U),
\end{equation}
where the maximum is taken over all tuples $(U,\hXL)$ on $\sU \times \hsXL$ jointly distributed with $\X$ and satisfying 
\begin{enumerate}[(i)]
\setlength\itemsep{2pt}
\item $\forall\ \ell \in \L:\ $ $U \markov X_\ell \markov X_\ellc$
\item $\forall\ \ell \in \L:\ $ $U \markov \hX_\ell \markov X_\ell$
\item $\forall\ \ell \in \L:\ $ $\E[\dl(\hX_\ell,X_\ell)] \leq D_\ell$
\item $\forall\ \ell \in \L:\ $ $I(X_\ell;\hX_\ell) = \R_{X_\ell}(D_\ell)$.
\end{enumerate}
\end{definition}

The indicated maximum in Definition~\ref{Def:GacsKorner:Lossy} exists because the set of all tuples $(U,\hXL)$ satisfying (i)--(iv) can be viewed as a compact subset of the corresponding probability simplex. 

The next theorem relates the critical cache capacity $\Ksgenie(\D)$ to $\KGK(\X;\D)$, and it provides a new operational meaning for lossy G\'acs-K\"orner common information in caching systems.

\begin{theorem}\label{Thm:KgenieKGK} 
For separable distortion functions, 
\begin{equation*} 
\Ksgenie(\D) \geq \KGK(\X;\D),
\end{equation*}
with equality if $\R_{X_1}(D_1) = \cdots = \R_{X_L}(D_L)$.
\end{theorem}

\begin{IEEEproof}
Theorem~\ref{Thm:KgenieKGK} is proved in Appendix~\ref{Thm:KgenieKGK:Sec}.
\end{IEEEproof}

Now recall the multivariate extension of G\'acs and K\"orner's original (lossless) definition of common information in~\cite{Gacs-1973-A}:
\begin{definition}
G\'acs and K\"orner's  common information of the tuple $\X$ is defined as
\begin{equation}\label{eq:GK}
\KGK(\X): = \max_{U:\ U \markov X_\ell  \markov X_\ellc,\ \forall \ell \in \L} I(\X;U),
\end{equation}
where the maximisation is taken over all auxiliary random variables $U$ on $\sU$ jointly distributed with $\X$ satisfying the $L$ indicated Markov chains. 
\end{definition}

The definition of the lossy  G\'acs-K\"orner common information $\KGK(\X;\D)$ in \eqref{eq:GKlossy} and  the definition of the original   G\'acs-K\"orner common information $\KGK(\X)$ in \eqref{eq:GK} can apply both to discrete random vectors $\X$ as well as continuous random vectors $\X$. In this latter case, however, the lossy  G\'acs-K\"orner common information  is only defined when all rate-distortion functions in (iv) are finite, 
\begin{equation}\label{eq:finite}
\R_{X_\ell}(D_\ell) <\infty, \ \forall\ \ell \in \L.
\end{equation}
This finiteness can be guaranteed by, for example, our assumption that the single symbol distortion function $d_\ell$ satisfies $d_\ell(\hat{x}_\ell,x_\ell) < D_\text{max}$ for all $\hat{x}_\ell$ and $x_\ell$.
 
The optimization variable $U$ is subject to less constraints in \eqref{eq:GK} than \eqref{eq:GKlossy}, so 
\begin{equation*}
\KGK(\X;\D) \leq \KGK(\X)
\end{equation*}
whenever \eqref{eq:finite} holds. Moreover, for discrete random variables $\X$ and Hamming distortion functions $\bd$:
\begin{equation*}
\KGK(\X;\D=\0) = \KGK(\X).
\end{equation*}

\begin{lemma}\label{Lem:GacsKornerMarkovDefinition}
\mw{Let $\bd$ be Hamming distortion functions and consider zero distortions, $\D=\0$. Then,}
\begin{equation*}
\KGK(\X)
= \max_{U:\ H(U|X_\ell) = 0,\ \forall \ell \in \L}\ H(U)
\end{equation*}
\end{lemma}
\begin{IEEEproof}
Lemma~\ref{Lem:GacsKornerMarkovDefinition} is proved in Appendix~\ref{Lem:GacsKornerMarkovDefinition:Sec}.
\end{IEEEproof}

\mw{This leads the following corollary:

\begin{corollary}
For Hamming distortion functions we have 
\begin{equation*}
\Ksgenie(\b{0}) \geq \mw{\KGK(\X)}
\end{equation*}
with equality if $H(X_1)=H(X_2)=\ldots=H(X_L)$.
\end{corollary}}


\subsection{The Superuser \mw{Setup} and Wyner's Common Information}

Now imagine that a \emph{superuser} is connected to the server by $L$ independent rate $R$ noiseless links, and suppose that the superuser requests every source. The optimal caching strategy for this superuser problem is again clear: Take an optimal code for the joint RD function of $\X$, cache $C$ bits of the code's output, and distribute the remaining bits equally over the $L$ links in the delivery phase. The RDC function of this superuser problem is  
\begin{equation}
\Rsuper(\D,C) = \left[\frac{\R_\X(\D) - C}{L}\right]^+.
\end{equation}

Returning to our main problem: The server only receives one source request and, therefore, it cannot distribute and average the compressed bits of the joint RD code over $L$ noiseless channels. This intuition leads to the following bound. 

\begin{lemma}\label{Lem:SuperuserLowerBound}
\begin{equation*}
\R(\D,C) \geq \Rsuper(\D,C).
\end{equation*}
\end{lemma}

Clearly the superuser bound is achievable by the caching system at $C = \R_\X(\D)$ and, similarly to the previous subsection, it is natural to consider the smallest cache capacity for which there is \emph{no rate loss} with respect to the optimal superuser system:
\begin{equation}\label{Eqn:Ksuper}
\Ksuper(\D) := \min\big\{ C \geq 0 :  \R(\D,C) = \Rsuper(\D,C) \big\}.
\end{equation}
The minimum in~\eqref{Eqn:Ksuper} exists because, for $0 \leq C \leq \R_\X(\D)$, $\R_\X(\D)$ is convex and $\Rsuper(\D,C)$ is linear. Figure~\ref{Fig:TypicalRDCFunction} depicts the superuser bound and the critical cache capacity $\Ksuper(\D)$. 

We now characterise $\Ksuper(\D)$. For a given $\D$, let 
\begin{equation*}
\Kssuper(\D) := \min_{(U,\hXL)} I(\X;U)
\end{equation*}
where the minimum is taken over all tuples $(U,\hXL)$ on $\sU \times \hsXL$ such that the following five properties hold
\begin{enumerate}[(i)]
\setlength\itemsep{2pt}
\item $\X \markov \hXL \markov U$
\item $I(X_1 ; \hX_1 | U)  = \cdots = I(X_L ; \hX_L | U)$
\item $\forall\ \ell \in \L:\ $ $\hX_\ell \markov U \markov \hX_\ellc$
\item $\forall\ \ell \in \L:\ $ $\E[\dl(\hX_\ell,X_\ell)] \leq D_\ell$
\item $I(\X;\hXL) = \R_{\X}(\D)$.
\end{enumerate}

\begin{theorem}\label{Thm:Ksuper}
For separable distortion functions, 
\begin{equation*}
\Ksuper(\D) = \Kssuper(\D).
\end{equation*}
\end{theorem}

\begin{IEEEproof}
Theorem~\ref{Thm:Ksuper} is proved in Appendix~\ref{Thm:Ksuper:Sec}.
\end{IEEEproof}

The above quantity $\Kssuper(\D)$ is closely related to the natural multivariate extension of Viswanatha, Akyol and Rose's~\cite{Viswanatha-Jun-2014-A} informational definition of \emph{lossy Wyner common information}. 

\begin{definition}
For a given distortion tuple $\D$ and single-symbol distortion functions $\bd$, define the \emph{lossy Wyner common information} of $\X$ by 
\begin{equation*}
\KW(\X;\D) := \min_{(U,\hXL)} I(\X;U)
\end{equation*}
where the minimum is taken over all tuples $(U,\hXL)$ on $\sU \times \hsXL$ such that the following four properties hold
\begin{enumerate}[(i)]
\setlength\itemsep{2pt}
\item $\X \markov \hXL \markov U$
\item $\forall\ \ell \in \L:\ $ $\hX_\ell \markov U \markov \hX_\ellc$
\item $\forall\ \ell \in \L:\ $ $\E[\dl(\hX_\ell,X_\ell)] \leq D_\ell$
\item $I(\X;\hXL) = \R_{\X}(\D)$.
\end{enumerate}
\end{definition} 
The next theorem follows trivially from the above definitions, and it gives an operational meaning for lossy Wyner common information for caching. 
\begin{theorem}
For separable distortion functions, 
\begin{equation*} 
\Kssuper(\D) \geq \KW(\X;\D).
\end{equation*}
\end{theorem}

Now recall the natural multivariate extension of Wyner's original definition of common information \cite{Wyner-Mar-1975-A}:
\begin{definition} 
Wyner's common information of the tuple $\X$ is defined as
\begin{equation*}
\KW(\X): = \min_{U:\ X_\ell \markov U \markov X_\ellc,\ \forall \ell \in \L} I(\X;U),
\end{equation*}
where the minimum is taken over all auxiliary random variables satisfying all $L$ indicated Markov chains. 
\end{definition}
The lossy Wyner common information $\KW(\X;\D)$ as well as Wyner's original common information $\KW(\X)$ are both defined for discrete  and  continuous random vectors $\X$. In the latter case, the lossy Wyner common information is only defined when the rate-distortion function in (iv) is finite, $\R_\X(\D)<\infty$.
\begin{remark}
At this point it is worth noting that, in general, the lossy Wyner common information $\KW(\X;\D)$ is neither convex/concave nor monotonic in $\D$. Moreover, it is generally the case that $\KW(\X;\D)$ can be larger/smaller than the Wyner common information $\KW(\X)$. A nice treatment of this issue for $L = 2$ variables is given by Viswanatha \emph{et al.}~in~\cite[Sec.~III.B]{Viswanatha-Jun-2014-A}.
\end{remark}

Let now $\X$ be a discrete random vector and $\bd$ be Hamming distortion functions. Consider zero distortions, $\D=\0$. Then,
\begin{equation}
\KW(\X) = \KW(\X;\D=\0),
\end{equation}
which implies the following corollary.

\begin{corollary}
For Hamming distortions we have that 
\begin{equation*}
\Kssuper(\b{0}) \geq \mw{\KW(\X)}
\end{equation*}
with equality if $\mw{\KW(\X)} = I(\X;U^*)$ for some $U^*$ satisfying 
\begin{itemize}
\item $H(X_1|U^*) = \cdots = H(X_L|U^*)$ 
\item $\forall\ \ell \in \L :\ $ $X_\ell \markov U^* \markov U^*_\ellc$.
\end{itemize} 
\end{corollary}


\subsection{The Super-Genie Lower Bound}

We conclude this section by combining and generalising the genie and superuser lower bounds. For each subset $\S \subseteq \L$, let 
\begin{equation*}
\DS = (D_\ell;\ \ell \in \S)
\quad 
\text{and}
\quad 
\XS = (X_\ell;\ \ell \in \S),
\end{equation*}
and let
\begin{equation*}
\R_{\XS}(\DS) = \min_{p_{\hX_\S|X_S}}\ I(X_\S;\hX_\S)
\end{equation*}
denote the \emph{joint RD function} of $X_\s{S}$, where the minimum is taken over all test channels $p_{\hX_\S|X_S}$ from $\sX_\S$ to $\hsX_\S$ such that $\E[\dl(\hX_\ell,X_\ell)] \leq D_\ell$ for all $\ell \in \S$.

\begin{lemma}\label{Lem:SuperuserSLowerBound}
\begin{equation*}
\R(\D,C) \geq \max_{\s{S} \subseteq \L}\ \left[ \frac{\R_{X_\s{S}}(D_\s{S}) - C}{|\s{S}|} \right]^+
\end{equation*}
\end{lemma}

\begin{IEEEproof}
Lemma~\ref{Lem:SuperuserSLowerBound} is proved in Appendix~\ref{Lem:SuperuserSLowerBound:Sec}. 
\end{IEEEproof}


\section{Connections to Gray and Wyner's ``Source coding for a simple network''}\label{Sec:GrayWyner}

\begin{figure}[t]
\begin{center}
\includegraphics[width=0.5\columnwidth]{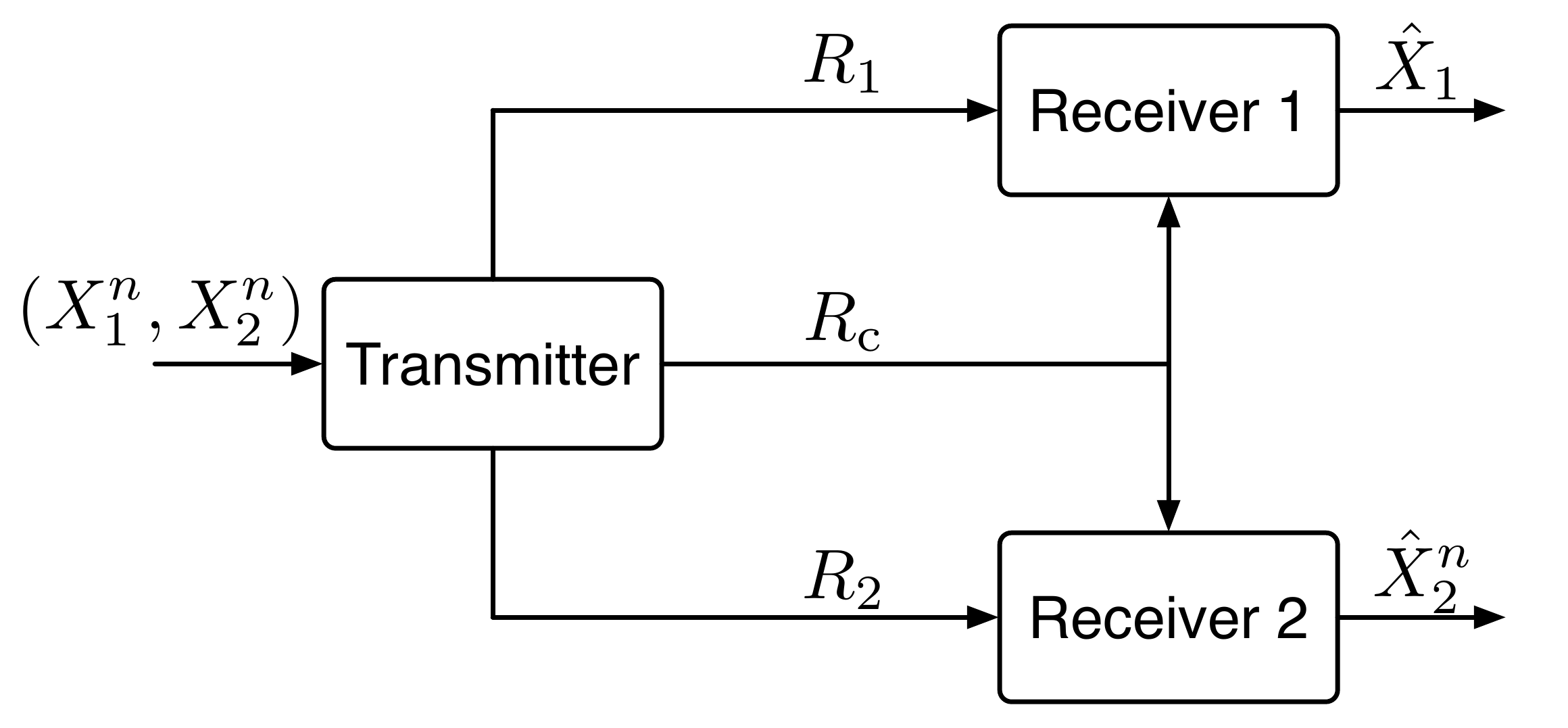}
\caption{Lossy source coding for a simple network with two receivers. }
\label{Fig:GrayWynerSimpleNetwork}
\end{center}
\end{figure}

The RDC function in~\eqref{Eqn:RDCFunctionSingleLetter} is closely related to Gray and Wyner's classic \emph{``source coding for a simple network''} problem~\cite{Gray-Nov-1974-A} illustrated in Figure~\ref{Fig:GrayWynerSimpleNetwork}. A transmitter is connected to two different  receivers via a common link of rate $R_\text{c}$ and two private links of rates $R_1$ and $R_2$. The set of all achievable rate tuples $(R_\text{c},R_1,R_2)$ for which receivers~1 and 2 can respectively reconstruct $X^n_1$ and $X^n_2$ to within distortions $D_1$ and $D_2$ is given by~\cite[Thm.~8]{Gray-Nov-1974-A}
\begin{align}
\notag
\RGW(D_1,D_2) := 
&
\bigcup_{p_{U|X_1,X_2}}
\Bigg\{
(R_\text{c},R_1,R_2) : 
&
\qquad
\left.
\begin{array}{rcl}
R_\text{c} &\geq& I(X_1,X_2;U)\\
R_1 &\geq& \R_{X_1|U}(D_1)\\ 
R_2 &\geq& \R_{X_2|U}(D_2)
\end{array}
\right\},
\end{align}
where the union is over all test channels $p_{U|X_1,X_2}$ from $\sX_1 \times \sX_2$ to an auxiliary alphabet $\sU$ of cardinality $|\s{U}| \leq |\sX_1| |\sX_2| + 4$. 

It is possible to extend Gray and Wyner's ``simple network'' from $L = 2$ receivers to an arbitrary number of receivers.  More specifically suppose that the transmitter is connected to $L$ different receivers via a \emph{single} common link with rate $R_\text{c}$ and $L$ private links of rates $\RR = (R_1,R_2,\ldots,R_L)$ respectively, where $R_\ell$ denotes the rate to the $\ell$-th receiver. It is not too hard to show that the set of all achievable rate tuples $(R_\text{c},\RR)$ for which every receiver $\ell$ can reconstruct~$X^n_\ell$ to within a distortion $D_\ell$ is given by
\begin{align*}
\RGW(\D) := 
\bigcup_{p_{U|\X}}
\Big\{
(R_\text{c},\RR) : 
\left.
\begin{array}{rcl}
R_\text{c} &\geq& I(\X;U)\\
R_\ell &\geq& \R_{X_\ell|U}(D_\ell)\quad \forall\ \ell \in \L.
\end{array}
\right\},
\end{align*}
where the union is over all test channels $p_{U|X_1,X_2}$ from $\sXL$ to $\sU$ with $|\s{U}| \leq |\sXL| + 2L$. The next proposition shows that the informational RDC function can be expressed as a minimisation over the achievable rate region $\RGW(\D)$.

\begin{lemma}\label{Lem:CachingGrayWyner}
\begin{equation*}
\R(\D,C) = \min_{(C,\RR) \in \RGW(\D)}\
\max_{\ell \in \L}\ R_\ell.
\end{equation*}
\end{lemma}

\begin{IEEEproof}
If $(C,\RR) \in \RGW(\D)$, then we can use the corresponding Gray-Wyner encoder and decoders to achieve a delivery phase-rate of $\max_\ell R_\ell$ in the caching problem; thus, $\R(\D,C)$ cannot be larger than the minimum in Lemma~\ref{Lem:CachingGrayWyner}. Now suppose $\R(\D,C)$ is strictly smaller than the above minimum: There would then exist an encoder and decoders in the Gray-Wyner problem that can operate outside of the rate region $\RGW(\D)$, which is a contradiction. 
\end{IEEEproof}

\section{A Strong Converse \mw{For} Separable Distortion Functions}\label{Sec:StrongConverse}

Consider the excess distortion problem formulation in Definition~\ref{Def:RDCFunc:ExcDist} for the special case of separable distortion functions $\bar{\bd}$ in~\eqref{Eqn:SeparableDistortionFunctionDefinition}. If the delivery-phase rate $R$ is strictly smaller than the informational RDC function $\R(\D,C)$, then the weak converse part of Theorem~\ref{Thm:RDC} shows that the excess-distortion probability of any sequence of $(n,\sMc,\sM)$ codes satisfying~\eqref{Eqn:Def:Ach:ExpDist:CacheCapacity} and \eqref{Eqn:Def:Ach:ExpDist:DeliveryRate} will be bounded away from zero; that is,
\begin{equation*}
\limsup_{n\to\infty}\ \Pr \left[\ \bigcup_{\ell \in \L} \Big\{ \dbl(\h{X}_\ell^n,X_\ell^n) \geq D_\ell \Big\} \right] > 0.
\end{equation*}
The next theorem strengthens this weak converse, and it will be used for $\f$-separable distortion functions in the next section. 

\begin{theorem}\label{Thm:Strong:Converse}
Fix any cache capacity $C$ and distortion tuple $\D$ such that $\R(\D,C) > 0$. Any sequence of $(n,\sMc,$ $\sM)$-codes satisfying
\begin{equation}\label{Eqn:StrongConverseConditionRate}
\limsup_{n\to\infty}\frac{1}{n} \log |\sM| < \R(\D,C) 
\end{equation}
and
\begin{equation}\label{Eqn:StrongConverseConditionCache}
\limsup_{n\to\infty} \frac{1}{n} \log |\sMc| \leq C 
\end{equation}
must also satisfy 
\begin{equation}\label{Eqn:StongConverse}
\limsup_{n\to\infty}\ \Pr \left[\ \bigcup_{\ell \in \L} \Big\{ \dbl(\h{X}_\ell^n,X_\ell^n) \geq D_\ell \Big\} \right] = 1.
\end{equation}
\end{theorem}

\begin{IEEEproof}
Theorem~\ref{Thm:Strong:Converse} is proved in Appendix~\ref{Thm:Strong:Converse:Proof}.
\end{IEEEproof} 

The strong converse in Theorem~\ref{Thm:Strong:Converse} applies to the \emph{union} of excess-distortion events in~\eqref{Eqn:StongConverse}. As an alternative to this union event, one could also consider the maximum excess distortion probability across source files. The following \mw{lemmas~\ref{Lem:Strong:Converse:Marginal} and \ref{Lem:Strong:Converse:Marginal:Cor}  present results on \emph{maximum} excess distortion probabilities. The next lemma follows by} modifying the strong converse for the usual point-to-point RD problem (see, for example, Kieffer~\cite{Kieffer-Mar-1991-A}). We omit its proof. 

\begin{lemma}\label{Lem:Strong:Converse:Marginal}
Fix any cache capacity $C$ and distortion tuple $\D$ such that $\R(\D,C) > 0$. Any sequence of $(n,\sMc,\sM)$-codes satisfying~\eqref{Eqn:StrongConverseConditionCache} and
\begin{equation*}
\limsup_{n\to\infty}\frac{1}{n} \log |\sM| + C
< 
\R_{X_\ell}(D_\ell)
\end{equation*}
for some $\ell \in \L$ must also satisfy 
\begin{equation}\label{Eqn:MarginalStrongConverseErrorProb}
\limsup_{n\to\infty}\ \Pr \left[ \dbl(\h{X}_\ell^n,X_\ell^n) \geq D_\ell \right] = 1.
\end{equation}
\end{lemma} 

Lemma~\ref{Lem:Strong:Converse:Marginal} and the definition of the critical cache capacity $\Kgenie(\D)$ in~\eqref{Eqn:Kgenie} immediately yield the following marginal strong converse \mw{for small cache sizes}. We omit its proof.

\begin{lemma}\label{Lem:Strong:Converse:Marginal:Cor}
Fix any cache capacity $C$ and distortion tuple~$\D$ such that $\R(\D,C) > 0$ and
$C \leq \Kgenie(\D)$. Any sequence of $(n,\sMc,\sM)$-codes satisfying~\eqref{Eqn:StrongConverseConditionCache} and
\begin{equation*}
\limsup_{n\to\infty}\frac{1}{n} \log |\sM| 
< 
\R(\D,C)
\end{equation*}
must also satisfy~\eqref{Eqn:MarginalStrongConverseErrorProb} for some $\ell \in \L$. 
\end{lemma}


\section{Examples (Separable Distortion Functions)}\label{Sec:Examples}


\subsection{Identical \mw{and Independent} Sources}

Suppose that $\X = (X_1,\ldots,X_L)$ is a string of $L$ mutually independent instances of a random variable $X$ on $\sX$. If $\d_1 = \cdots = \d_L = \d$ and $\D = (D,\ldots,D)$, then \mw{Theorem~\ref{Thm:RDC} specializes to:}
\begin{equation*}
\R(\D,C) = \Rsuper(\D,C) = \left[\R_{X}(D) - \frac{C}{L}\right]^+.
\end{equation*} 
The optimal caching strategy here is simple: Take an optimal RD code for $(X,\db,D)$; compress each source $X_\ell^n$ to the RD limit $\R_X(D)$; cache $C/L$ of the compressed bits; and transmit the remaining bits  during the delivery phase. 


\subsection{Multivariate Gaussian Sources}

The discussion so far has been restricted to sources defined on finite alphabets. However, it can be shown that Theorem~\ref{Thm:RDC} extends to multivariate Gaussian sources with squared-error distortions; for example, see the discussion in~\cite{Wyner-1978-A-II}. 

More formally, let $\X = (X_1,\ldots,X_L) \in \reals^L$ be a zero mean multivariate Gaussian with covariance matrix $\b{K}_{\X}$ and 
\begin{equation*}
\dl(\hx_\ell,x_\ell) = (\hx_\ell - x_\ell)^2,
\quad
\forall\
\ell \in \L.
\end{equation*} 
Let $\RexpG(\D,C)$ denote the corresponding operational RDC function w.r.t.~the expected distortion criteria: 
\begin{equation*}
\E \left[ \frac{1}{n} \sum_{i=1}^n (\hX_{\ell,i}\ssb{-}X_{\ell,i})^2\right] \leq D_i,
\quad 
\forall\
\ell \in \L.
\end{equation*}
Now let
\begin{equation}\label{Eqn:InformationRDCFunctionGaussian}
\RG(\D,C) = \inf_{(U,\h{\b{X}})}\ \max_{\ell \in \L} I(X_\ell;\hX_\ell|U),
\end{equation}
where the infinum is taken over all real tuples $(U,\h{\b{X}})$ jointly distributed with $\X$ such that 
\begin{subequations}\label{Eqn:GaussianConstraints}
\begin{equation}
I(\X;U) \leq C
\end{equation}
and 
\begin{equation}
\E\big[ (X_\ell - \hX_\ell)^2\big] \leq D_\ell,
\quad
\forall\ \ell \in \L.
\end{equation}
\end{subequations}
The next \mw{theorem} is the Gaussian counterpart of Theorem~\ref{Thm:RDC}, and we omit its proof.
\begin{theorem}\label{Prop:Gaussian}
\begin{equation*}
\RG(\D,C) = \RexpG(\D,C).
\end{equation*}
\end{theorem}

{Strictly speaking, the expression we give in \eqref{Eqn:InformationRDCFunctionGaussian} for the Gaussian RDC function is non-computable, because there is no bound on the cardinality on the auxiliary variable $U$ (and it is not clear if one can restrict the optimization domain to $U$'s that are jointly Gaussian with $\X$). 
As we will see in the next subsection, we can give explicit expressions for $\RexpG(\D,C)$ when there are only two sources and over wide ranges of parameters of the Gaussian sources.} 

\mw{For an arbitrary number of source files and for symmetric distortions, we have the following lower bound on $\RexpG(\D,C)$ in the next proposition. It is}
a Gaussian version of Lemma~\ref{Lem:SuperuserSLowerBound}.

For each subset $\S \subseteq \L$, let $\XS = (X_\ell;\ \ell \in \S)$ denote the tuple of random variables with indices in $\S$, and let $\KXS$ denote the covariance matrix of $\XS$. 

\begin{proposition}\label{Thm:Gaussian:Converse}
If $\D = (D,\ldots,D)$,  then
\begin{equation*}
\RG(\D,C) \geq 
\max_{\S \subseteq \L} 
\left[\frac{1}{2|\S|} \log \frac{\det \KXS}{D^{|\S|}} - \frac{C}{|\S|}\right]. 
\end{equation*}
\end{proposition}

\begin{IEEEproof}
Proposition~\ref{Thm:Gaussian:Converse} is proved in Appendix~\ref{Sec:Proof:Thm:Gaussian:Converse}.
\end{IEEEproof}


\subsection{Bivariate Gaussian Sources}

\begin{figure}
\begin{center}
\includegraphics[width=.6\columnwidth]{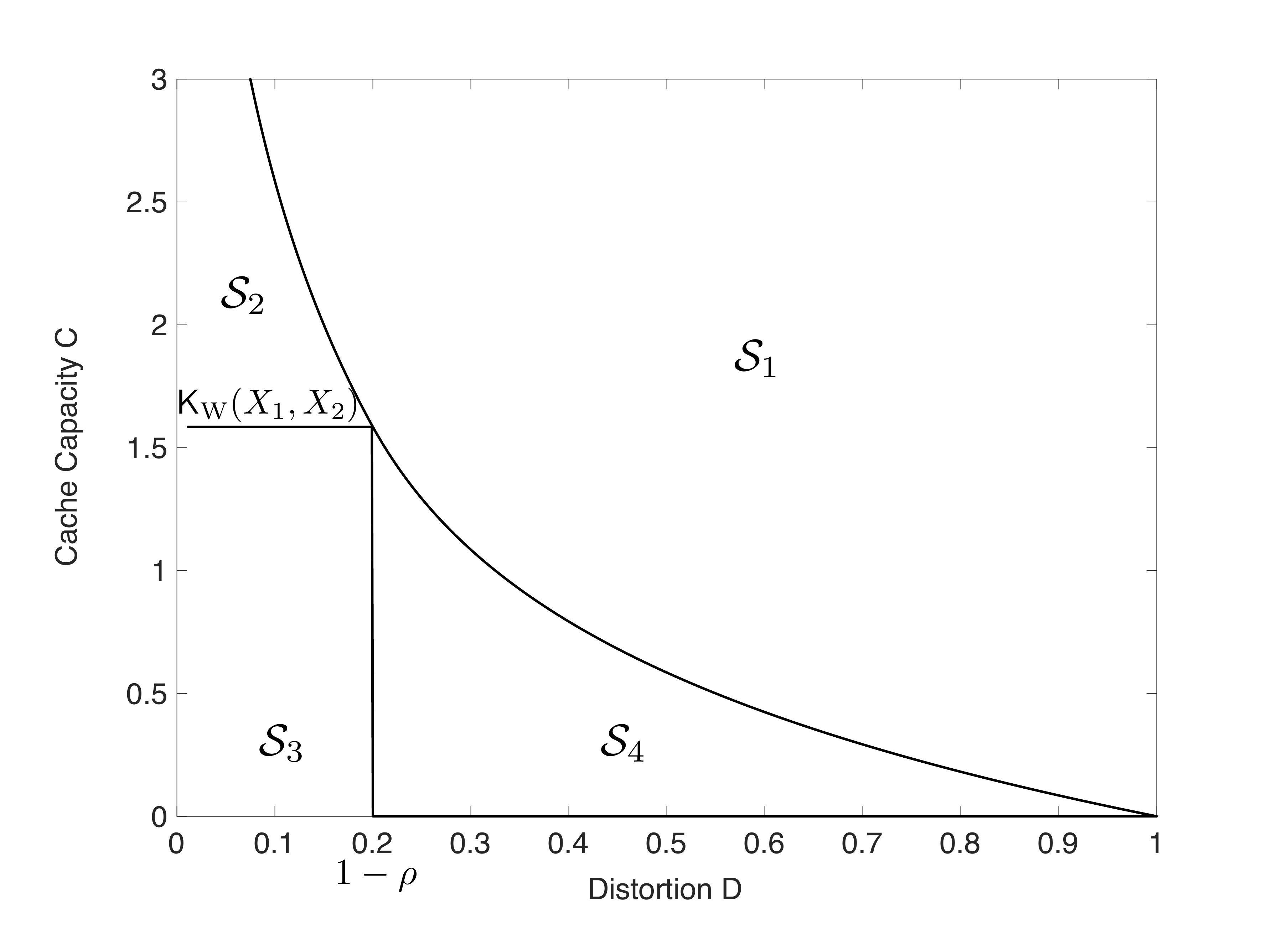}
\caption{Illustration of the distortion-cache regions $\s{S}_1$, $\s{S}_2$, $\s{S}_3$ and $\s{S}_4$ used in Theorem~\ref{Thm:BivariateGaussian} with $\rho = 0.8$.}
\label{Fig:GaussianRegions}
\end{center}
\end{figure}

Let us now fix $\rho \in (0,1)$ and consider a zero mean bivariate Gaussian source $\X = (X_1,X_2)$ with the covariance matrix
\begin{equation}\label{Eqn:BivariateGaussianSourceCovarianceTwo}
\Ktwo = 
\begin{bmatrix}
1 & \rho \\
\rho & 1
\end{bmatrix}. 
\end{equation}
We wish to evaluate the Gaussian RDC function in~\eqref{Eqn:InformationRDCFunctionGaussian} with symmetric distortions $D_1 = D_2 = D$. To do this, we will consider distortion-cache pairs $(D,C)$ separately for each one of the regions $\S_1,\S_2,\S_3$ and $\S_4$ that we define shortly. There are two key quantities defining these regions:   the Gaussian joint RD function $\RGjoint$ and the Wyner common information between $X_1$ and $X_2$. For symmetric\footnote{Here we only recall the joint RD function of $(X_1,X_2)$ for the case of symmetric distortions, $D_1 = D_2 = D$. A treatment of the RD function for arbitrary distortion pairs can be found in~\cite{Lapidoth-Jun-2010-A} and the references therein.} distortions $D_1 = D_2 = D$, the joint RD function $\RGjoint$ is given by~\cite[Thm.~III.1]{Lapidoth-Jun-2010-A}: 
\begin{enumerate}[(i)]
\item If $0 < D \leq 1 - \rho$, then
\begin{equation*}
\RGjoint(D,D) 
=
\dfrac{1}{2} \log \frac{1-\rho^2}{D^2}.
\end{equation*}

\item If $1 - \rho \leq D \leq 1$, then
\begin{equation*}
\RGjoint(D,D) 
=
\frac{1}{2} \log \frac{1 + \rho}{2D - (1 - \rho)}.
\end{equation*}

\item If $D > 1$, then 
\begin{equation*}
\RGjoint(D,D) = 0.
\end{equation*}
\end{enumerate} 
The Wyner common information of the Gaussian pair $X_1$ and $X_2$ is given by~\cite{Viswanatha-Jun-2014-A,Xu-Mar-2011-C}
\begin{equation}\label{Eqn:WynerCommonInformationGaussian}
\KW(X_1,X_2)
= \frac{1}{2} \log \frac{1 + \rho}{1 - \rho}.
\end{equation} 
Notice that for this symmetric Gaussian example, the original Wyner common information $\KW(X_1,X_2)$ equals the lossy Wyner common information $\KW(X_1,X_2; D,D)$ when $0<D<(1-\rho)$ \cite[Eq. (30)]{Viswanatha-Jun-2014-A}.

\mw{We can now define the four regions $\s{S}_1,\s{S}_2, \s{S}_3, \s{S}_4$:}
\begin{equation*}
\s{S}_1 := \Big\{ (D,C) : C \geq \RGjoint(D,D) \Big\},
\end{equation*}

\begin{equation*}
\s{S}_2: = 
\Big\{ 
(D,C) : 
\KW(X_1,X_2)
\leq 
C
\leq \RGjoint(D,D)
\Big\},
\end{equation*}

\begin{equation*}
\s{S}_3 := 
\Bigg\{ (D,C) :  D \leq 1 - \rho,\ 
C \leq \KW(X_1,X_2) \Bigg\},
\end{equation*}
and
\begin{equation*}
\s{S}_4 := \Bigg\{ (D,C) : 1 - \rho \leq D \leq 1,\ C \leq \RGjoint(D,D) \Bigg\}.
\end{equation*}
\mw{The four regions are illustrated in Figure~\ref{Fig:GaussianRegions}.} 


\begin{theorem}\label{Thm:BivariateGaussian}
For the zero mean bivariate Gaussian source $(X_1,X_2)$ with the covariance matrix $\Ktwo$ in~\eqref{Eqn:BivariateGaussianSourceCovarianceTwo} and squared error distortion constraints, we have
\begin{equation*}
\RG((D,D),C) 
=
\left\{
\begin{array}{ll}
0, 
& (C,D) \in \s{S}_1,\\[10pt]
\dfrac{1}{4} \log \dfrac{1 - \rho^2}{D^2} - \dfrac{C}{2},
& (C,D) \in \s{S}_2,
\end{array}
\right.
\end{equation*}
and
\begin{align*}
\RG((D,D),C) 
\leq
\dfrac{1}{2} \log\dfrac{1 - \frac{1}{2} (1 + \rho)(1-2^{-2C})}{D},\qquad
(C,D) \in \s{S}_3 \cup \s{S}_4.
\end{align*}
\end{theorem}

\begin{IEEEproof}
Theorem~\ref{Thm:BivariateGaussian} is proved in Appendix~\ref{Sec:Proof:Thm:BivariateGaussian}.
\end{IEEEproof}

\begin{figure}
\begin{center}
\includegraphics[width=.6\columnwidth]{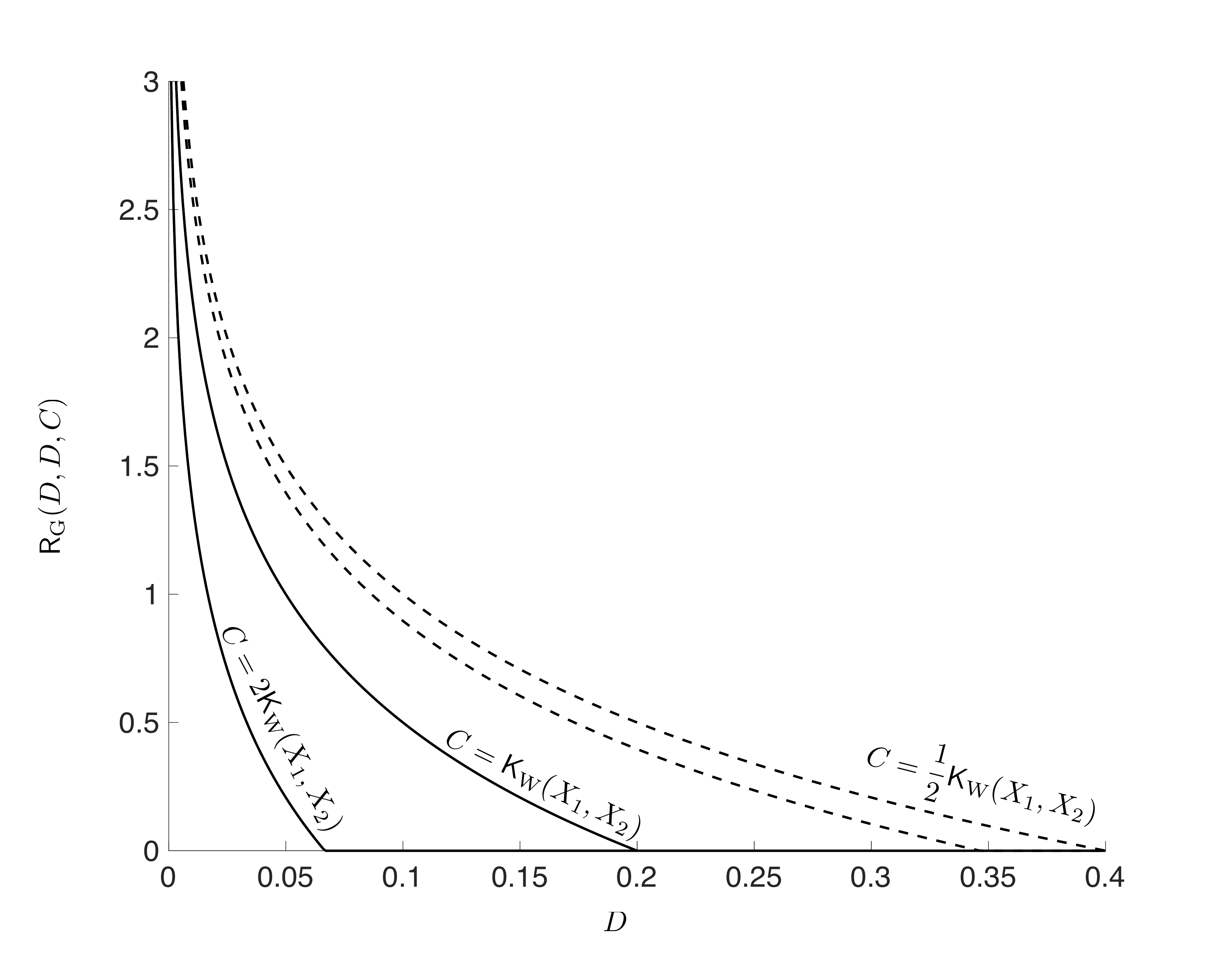}
\caption{Illustration of the RDC functions in Theorem~\ref{Thm:BivariateGaussian} for a zero mean bivariate Gaussian source $(X_1,X_2)$ with the covariance matrix $\Ktwo$ in~\eqref{Eqn:BivariateGaussianSourceCovarianceTwo} ($\rho = 0.8$), and symmetric distortion constraints $D_1 = D_2 = D$. The RDC function $\RG(D,D,C)$ is plotted as a function of the distortion $D$ for three difference cache capacities $C = 2 \KW(X_1,X_2)$, $\KW(X_1,X_2)$ and $(1/2) \KW(X_1,X_2))$, where $\KW(X_1,X_2)$ denotes the Wyner common information~\eqref{Eqn:WynerCommonInformationGaussian}. 
For $C = (1/2) \KW(X_1,X_2)$, Proposition~\ref{Thm:Gaussian:Converse} and Theorem~\ref{Thm:BivariateGaussian} only give lower and upper bounds, and these are shown with dashed lines. 
}
\label{Fig:GaussianCaches}
\end{center}
\end{figure}

Figure~\ref{Fig:GaussianCaches} illustrates the results in this Theorem~\ref{Thm:BivariateGaussian} and the previous Proposition~\ref{Thm:Gaussian:Converse} at hand of an example.

\subsection{Doubly Symmetric Binary Source}

\begin{figure}[t!]
\begin{center}
\includegraphics[width=.7\columnwidth]{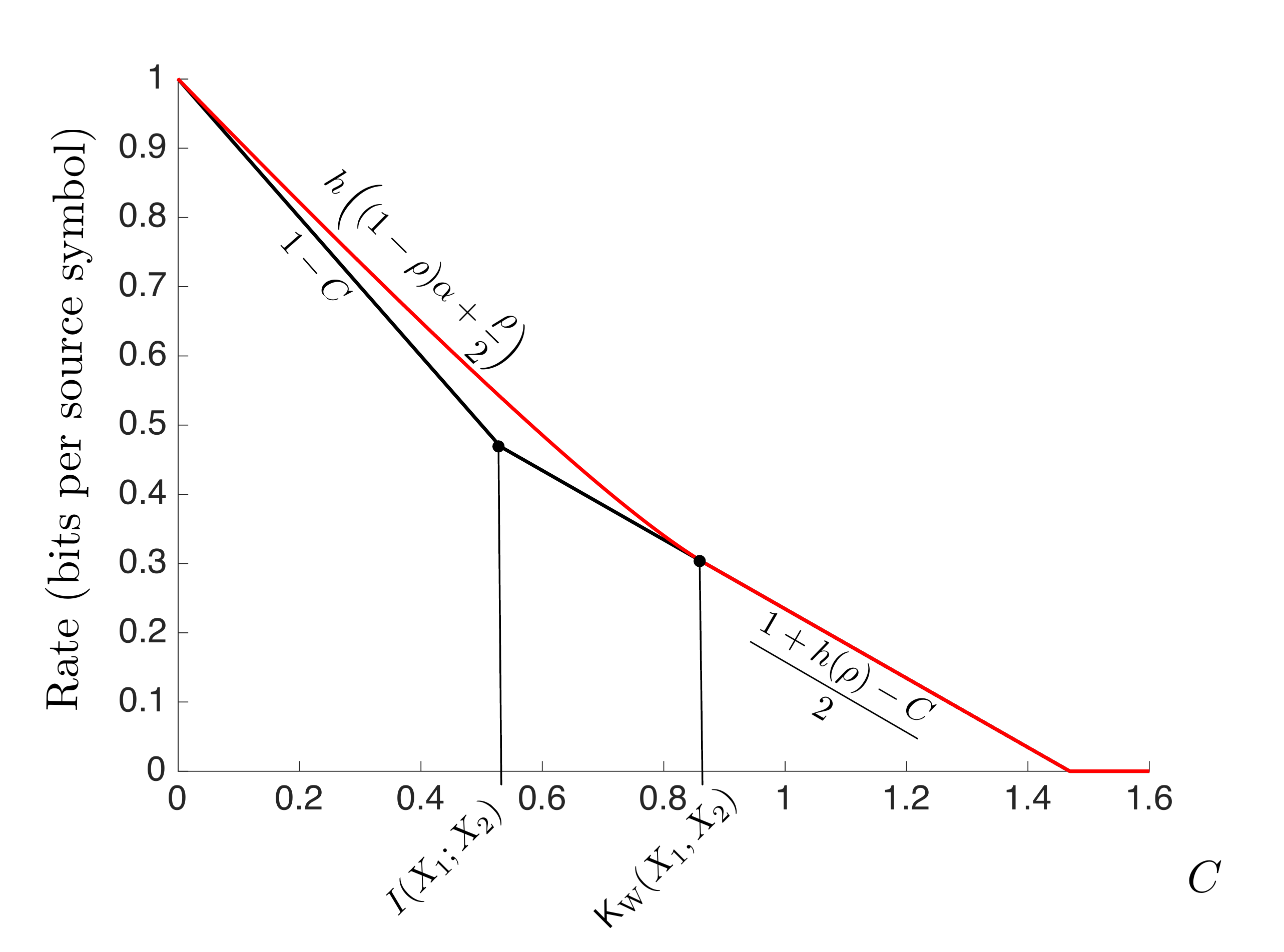}
\caption{Illustration of the upper (achievable) and lower (converse) bounds in Theorem~\ref{Thm:DSBS} for the DSBS RDC function $\R(\0,C)$ with $\rho = 0.1$.}
\label{Fig:DSBSRho01}
\end{center}
\end{figure}

We now evaluate the RDC function for a \emph{doubly symmetric binary source} (DSBS) under Hamming distortion functions. Fix $0 \leq \rho \leq 1/2$. Suppose that the library consists of two sources, $\s{X}_1  = \s{X}_2 = \hsX_1 = \hsX_2 = \{0,1\}$ and 
\begin{equation*}
p_{\X}(x_1, x_2) 
= 
\frac{1}{2}(1-\rho) \indicator{x_1 = x_2} 
+ \frac{1}{2} \rho \indicator{x_1 \neq x_2}.
\end{equation*}

Let 
\begin{equation*}
\rho^* = \frac{1}{2} - \frac{1}{2} \sqrt{1 - 2\rho},
\end{equation*}
and note that Wyner's common information reduces to
\begin{align*}
\mw{\KW(X_1,X_2)}
&=
\min_{U:\ X_1 \markov U \markov X_2} I(X_1,X_2;U)\\
&= 1 + h(\rho) - 2h(\rho^*).
\end{align*}

\begin{theorem}\label{Thm:DSBS}
The following is true for the DSBS:
\begin{enumerate}[(i)]

\item 
$\Kssuper(\b{0}) = \mw{\KW(X_1,X_2)}$ and
\begin{equation*}
\R(\0,C) = \frac{1 + h(\rho) - C}{2}
\end{equation*}
for all cache capacities $\mw{\KW(X_1,X_2)} \leq C \leq 1 + h(\rho)$.

\medskip

\item 
$\R(\0,0) = 1$, $\Kgenie(\b{0}) = 0$ and 
\begin{equation*}
\R(\0,C) > \big[1 - C\big]^+
\end{equation*}
for all cache capacities $0 < C < 1 + h(\rho)$.

\item 
\begin{equation*}
\frac{1 + h(\rho) - C}{2} \leq \R(\b{0},C) \leq h\left((1-\rho)\alpha + \frac{\rho}{2}\right),
\end{equation*}
for all cache capacities $0 < C  \leq \KW(X_1,X_2)$, where 
\begin{equation*}
\alpha = h^{-1}\left( \frac{1 - \rho - C}{1 - \rho} \right).
\end{equation*}
\end{enumerate}
The above bounds are illustrated in Figure~\ref{Fig:DSBSRho01}. 
\end{theorem}

\begin{IEEEproof}
The fact that $\R(\0,C)$ is strictly greater than $[1 - C]^+$ for all positive $0 < C < 1 + h(\rho)$ follows from Theorem~\ref{Thm:Kgenie}, Corollary~\ref{Thm:Kgenie:Cor}, and $\Ksgenie(\b{0}) = \KGK(\X;\b{0}) = 0$. The remaining assertions in Theorem~\ref{Thm:DSBS} can be proved using similar methods \mw{to evaluate the RDC function in Theorem~\ref{Prop:Gaussian}} as in~\cite[Sec.~1.5]{Gray-Nov-1974-A},\cite[Sec.~III.C]{Timo-Nov-2010-A} and~\cite[Ex.~1]{Wang-Apr-2015-A}, and we omit the details.
\end{IEEEproof}

It is worth pointing out that, in this special case, the RDC function $\R(\D,C)$ particularises to the same expression as in~\cite[Ex.~1]{Wang-Apr-2015-A} (see also~\cite[Sec.~1.5]{Gray-Nov-1974-A}). Of course, this does not mean that Theorem~\ref{Thm:DSBS} is a special case of the results in~\cite{Wang-Apr-2015-A}; rather, the equivalence appears to be a consequence of the symmetry of the problem.


\section{The RDC Function for $\bf$-Separable Distortion Functions}\label{Sec:fSeparableDistortions}

Consider the operational RDC functions in Definitions~\ref{Def:RDCFunc:ExpDist} and~\ref{Def:RDCFunc:ExcDist} for the general case of $\f$-separable distortion functions~\eqref{Eqn:fSeparableDistortionFunctions}, with $\bf$ denoting the $L$ continuous and increasing functions in~\eqref{Eqn:f-Functions} and $\bd$ denoting the $L$ single-symbol distortion functions in~\eqref{Eqn:SymbolDistortion}. 

For each request $\ell \in \L$ let 
\begin{equation*}
\d_\ell^* : \hsX_\ell \times \sX_\ell \to [0,\infty)
\end{equation*}
be the single-symbol distortion function obtained by setting 
\begin{equation}\label{Eqn:SingleSymbolDistortionFunctionDstar}
\d_\ell^*(\hx_\ell,x_\ell) = \fl\big(\dl(\hx_\ell,x_\ell)\big).
\end{equation}
Now let $\R_{\bd^*}(\bf(\D),C)$ denote the informational RDC function in~\eqref{Eqn:RDCFunctionSingleLetter} evaluated w.r.t.~the single-symbol distortion functions 
\begin{equation*}
\bd^* = (\d^*_1,\ldots,\d^*_L)
\end{equation*}
and distortion tuple
\begin{equation*}
\bf(\D) = \big(\f_1(D_1),\ldots,\f_L(D_L)\big).
\end{equation*}
Using Lemma~\ref{Lem:Strong:Converse:Marginal:Cor} and the ideas in~\cite{Shkel-Feb-2016-C}, it is not too difficult to obtain the following proposition. We omit the proof.

\begin{proposition}\label{Prop:RDCFSepMarginal}
For $\f$-separable distortion functions and all cache capacities $C \leq \Kgenie(\D)$, we have
\begin{equation*}
\Rexp(\D,C) = \Rexc(\D,C) = \R_{\bd^*}(\bf(\D),C).
\end{equation*}
\end{proposition}

Proposition~\ref{Prop:RDCFSepMarginal} is quite intuitive, and a natural question is whether or not it extends to cache capacities larger than $\Kgenie(\D)$. The next result considers such cases, but it requires a slightly more restricted version of the \emph{expected distortions} operational model. Specifically, let us consider the following definition in place of Definition~\ref{Def:RDCFunc:ExpDist}.

\begin{definition}\label{Def:RDCFunc:ExpMaxDist}
We say that a rate-distortion-cache tuple $(R,$ $\D,C)$ is \emph{achievable w.r.t.~the expected max-distortion criterium} if there exists a sequence of $(n,\sMc,\sM)$-codes such that~\eqref{Eqn:Def:Ach:ExpDist:CacheCapacity} and~\eqref{Eqn:Def:Ach:ExpDist:DeliveryRate} hold and
\begin{equation}\label{Def:RDCFunc:ExpMaxDist:Eqn}
\limsup_{n\to\infty}
\E \Big[ \max_{\ell \in \L} \big( \bfdl(\h{X}_\ell^n,X_\ell^n) - D_\ell \big) \Big] \leq 0.
\end{equation}
The \emph{RDC function w.r.t.~expected max-distortions criterion} is
\begin{align*}
\Rmaxexc(\D,C) := \inf\Big\{R \geq 0 : (R,\D,C) \text{ is achievable w.r.t.~expected max-distortions} \Big\}
\end{align*}
\end{definition}

Definition~\ref{Def:RDCFunc:ExpMaxDist} is more restrictive than Definition~\ref{Def:RDCFunc:ExpDist} in the following sense: Any tuple $(R,\D,C)$ that is achievable w.r.t.~the expected max-distortion criteria is also achievable w.r.t.~the expected distortion criteria. Therefore, 
\begin{equation*}
\Rexp(\D,C) \leq \Rmaxexc(\D,C).
\end{equation*}

\begin{theorem}\label{Thm:RDCFSep}
For $\f$-separable distortions we have
\begin{equation*}
\Rmaxexc(\D,C) = \Rexc(\D,C) = \R_{\bd^*}(\bf(\D),C).
\end{equation*}
\end{theorem}

\begin{IEEEproof}
Theorem~\ref{Thm:RDCFSep} is proved in Appendix~\ref{Thm:RDCFSep:Proof}.
\end{IEEEproof}


%
%
%
%


\section{Optimal Caching for Two Users (Separable Distortion Functions)}\label{Sec:TwoUsers}

\begin{figure}[t]
\begin{center}
\includegraphics[width=.6\columnwidth]{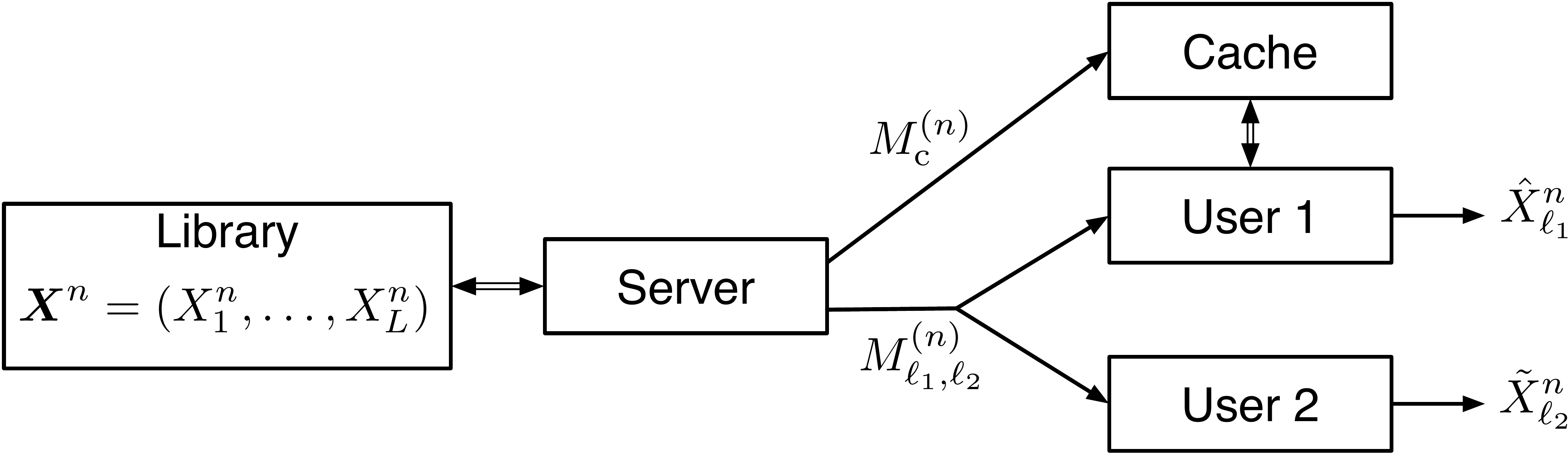}
\caption{Two-user RD cache problem.}
\label{two-user-one-cache}
\end{center}
\end{figure}

This section generalises some of the ideas in previous sections from one user to two users. In particular, we will assume that user~1 has a cache with capacity $C$ while user~2 does not have a cache as illustrated Figure~\ref{two-user-one-cache}. 


\subsection{Setup} 

Suppose that user~$1$ selects a source index $\ell_1$ from the set~$\L_1$ and user~$2$ selects an index $\ell_2$ from the set $ \L_2$, where $\L_1,\L_2 \subseteq \L$.  Let $L_1 = |\s{L}_1|$ and $L_2 = |\s{L}_2|$. We assume that the requests $\ell_1$ and $\ell_2$ are known by the server and both users during the delivery phase (this information can be shared, for example, with vanishing rate for finite $L$ and $n \to \infty$). A \emph{two-user RDC code} for a given blocklength $n$, which we call a two-user $(n,\sMc,\s{M}^{\ssb(n)})$-code, consists of the following mappings:
\begin{enumerate}[(i)]
\item A \emph{cache-phase encoder} at the server
\begin{equation*}
\encc \colon \b{\sX}^n \to \sMc.
\end{equation*} 
\item A \emph{delivery-phase encoder} at the server
\begin{equation*}
\phi\n_{(\ell_1 \ell_2)} \colon \b{\sX}^n \to \sM
\end{equation*} 
for each pair of user requests $(\ell_1,\ell_2) \in \L_1 \times \L_2$.
\item A  \emph{delivery-phase decoder} at user~$1$ 
\begin{equation*}
\varphi\n_{\ell_1\ell_2} : \sM  \times \sMc \to\hsX_{\ell_1}^n
\end{equation*}
for each pair of user requests $(\ell_1,\ell_2) \in \L_1 \times \L_2$.
\item A \emph{delivery-phase decoder} at user~$2$ 
\begin{equation*}
\psi\n_{\ell_1\ell_2} : \sM \to \tsX_{\ell_2}^{n}
\end{equation*}
for each pair of user requests $(\ell_1,\ell_2) \in \L_1 \times \L_2$.
\end{enumerate}

During the \emph{caching phase} the server places 
\begin{equation*}
\Mc = \encc(\X^n)
\end{equation*}
in the cache of user~1. After the request of both users $(\ell_1,\ell_2)\in\L_1 \times \L_2$ are revealed to the server and both users, the server sends
\begin{equation*}
M_{\ell_1\ell_2}\n = \phi\n_{\ell_1\ell_2}(\X^n)
\end{equation*}
over a common noiseless channel to both users. Users~1 and~2 respectively output
\begin{equation*}
\hX^{n}_{\ell_1} = \varphi\n_{\ell_1\ell_2}\big(M_{\ell_1\ell_2}\n,\Mc\big)
\end{equation*}
and
\begin{equation*}
\t{X}^{n}_{\ell_2} = \psi\n_{\ell_1\ell_2}\big(M_{\ell_1\ell_2}\n\big).
\end{equation*}

The users might have differing exigencies regarding the sources in the library. To account for this, we allow for different single-symbol distortion functions at each user:
\begin{equation*}
\d_{\ell_1} :\  \hsX_{\ell_1} \times \sX_{\ell_1} \to [0,\infty)
\quad
\text{(user 1)}
\end{equation*}
and
\begin{equation*}
\delta_{\ell_2}:\ \tsX_{\ell_2} \times \sX_{\ell_2} \to [0,\infty)
\quad
\text{(user 2)},
\end{equation*}  
where $\d_{\ell_1}$ and $\delta_{\ell_2}$ \mw{satisfy the two conditions preceding Definition~\ref{def:sep}. }

\begin{definition}\label{Def:CDAdmissible1}  
We say that a rate-distortion-cache tuple $(R,\D,\b{\Delta},C)$-tuple is $(\bd,\b{\delta})$-\emph{achievable w.r.t.~expected distortions} if there exists a sequence of two-user $(n,\sMc,\sM)$-codes such that 
\begin{subequations}
\begin{align}
\limsup_{n\to\infty} \frac{1}{n} \log |\sMc| &\leq C\\
\limsup_{n\to\infty} \frac{1}{n} \log |\sM| &\leq R\\
\limsup_{n\to\infty} \E \Big[ \db_{\ell_1} \big( \hX_{\ell_1}^n, X_{\ell_1}^n \big) \Big] &\leq D_{\ell_1},\quad \forall\ \ell_1 \in \L_1\\
\limsup_{n\to\infty} \E \Big[ \bar{\delta}_{\ell_2} \big( \tX_{\ell_2}^n, X_{\ell_2}^n \big) \Big] &\leq \Delta_{\ell_2},\quad \forall\ \ell_2 \in \L_2,
\end{align}
\end{subequations}
where $\db_{\ell_1}$ and $\bar{\delta}_{\ell_2}$ are the separable distortion functions (defined in the same way as~\eqref{Eqn:SeparableDistortionFunctionDefinition}) corresponding to $\d_{\ell_1}$ and $\delta_{\ell_2}$. The two-user \emph{operational RDC function w.r.t.~expected distortions} $\R^\dag(\D,\b{\Delta},C)$ is the infimum of all rates $R \geq 0$ such that the rate-distortion-cache tuple $(R,\D,\b{\Delta},C)$ is $(\bd,\b{\delta})$-achievable.
\end{definition}

Unfortunately, we have been unable to find a single-letter (or, computable) expression for the two-user operational RDC function. An achievable (upper) bound and a converse (lower) bound are presented in the next two subsections. 


\subsection{Achievable Bound}

As in the single-user problem, we would first like to jointly quantise the library $\X^n$ to a codeword $U^n$ that is placed in user~$1$'s cache. (The manner in which $U^n$ is placed in the cache is described next.) After the requests $\ell_1$ and $\ell_2$ are revealed to the server and both users, we then would like to communicate the approximations $\hX^{n}_{\ell_1}$ and $\tX^{n}_{\ell_2}$ to users $1$ and $2$ respectively.

Note that user~2 does not have a cache and, therefore, $\tX^{n}_{\ell_2}$ can always be recovered by user~$1$. Indeed, user~$1$ will be able to reconstruct $\tX^{n}_{\ell_2}$ before it attempts to recover $U^n$ from its cache. We may, therefore, view $\tX^{n}_{\ell_2}$ as a type of \emph{side information} at user $1$ that should be exploited by the cache encoder at the server to reduce the caching rate. 

For example, one could use a \emph{Wyner-Ziv-style code}~\cite{Wyner-Jan-1976-A} to compress $\X^n$ to a description~$U^n$ that is randomly binned at a rate matched to the bin size $I(U;\tX_{\ell_2})$. Unfortunately, the particular demand $\ell_2$ and, consequently, the bin size (or, binning rate) are not known to the server during the caching phase, and, for this reason, we need to use a slightly more sophisticated ``implicit'' binning technique to encode the cache. The technique is similar to that used by Tuncel in~\cite{Tuncel-Apr-2006-A}, and it leads to the following achievability result.

Let 
\begin{align}\label{eqn:worst-case}
\overline{\R}(\D, \b{\Delta},C)
= 
\min_{(U,\b{\hX},\b{\tX})}
\max_{(\ell_1, \ell_2)}
\max \Big\{ I(\X;U,\hX_{\ell_1},\tX_{\ell_2})
- C, 
I(U,\X;\tX_{\ell_2}) + I(\X;\hX_{\ell_1}|U,\tX_{\ell_2}),
\Big\} 
\end{align}
where the minimum is taken over all tuples $(U,\b{\hX},\b{\tX})$ jointly distributed with $\X$ such that 
\begin{subequations}
\begin{align}
\label{Thm:TwoUserAchievability:D1}
\forall\ \ell_1 \in \L_1:\quad \E[ \d_{\ell_1}(\hX_{\ell_1},X_{\ell_1})] &\leq D_{\ell_1}\\
\label{Thm:TwoUserAchievability:D2}
\forall\ \ell_2 \in \L_2:\quad \E[ \delta_{\ell_2}(\hX_{\ell_2},X_{\ell_2})] &\leq \Delta_{\ell_2}.
\end{align}
\end{subequations}

\begin{theorem}\label{Thm:TwoUsersAchievability} 
\begin{equation*}
\overline{\R}(\D, \b{\Delta},C) \geq \R^\dag(\D, \b{\Delta},C).
\end{equation*}
with equality when $\L_1= \{\ell\}$ or $\L_2=\{\ell\}$ for some $\ell \in \L$.
\end{theorem}

\begin{IEEEproof}
An outline of the proof of Theorem~\ref{Thm:TwoUsersAchievability} is given in Appendix~\ref{Thm:TwoUsersAchievability:Sec}.
\end{IEEEproof}


\subsection{Genie-Aided Lower Bound}

If both users' demands were revealed by a genie to the server even before the caching phase, then the caching problem would coincide with a ``worst-demands'' of the classic successive-refinement of information problem~\cite[Sec.~13.5]{El-Gamal-2011-B}. The RD function corresponding to this worst-demands problem thus forms a lower bound on  $\R^\dag(\D, \b{\Delta},C)$. More specifically, let
\begin{align*}
\underline{\R}(\D, \b{\Delta},C) 
= 
\min_{(\h{\X},\t{\X})}
\max_{(\ell_1, \ell_2)} 
\max\Big\{ 
I(\X;\tX_{\ell_2}),
I(\X;\hX_{\ell_1},\tX_{\ell_2}) - C 
\Big\}, 
\end{align*}
where the minimum is taken over all $(\h{\X},\t{\X})$ jointly distributed with $\X$ such that~\eqref{Thm:TwoUserAchievability:D1} and~\eqref{Thm:TwoUserAchievability:D2} hold.

\begin{theorem}\label{Thm:TwoUsersConverse}
\begin{equation*}
\underline{\R}(\D, \b{\Delta},C) \leq  \R^\dag(\D,\b{\Delta},C).
\end{equation*}
\end{theorem}


\subsection{(Almost) Lossless Reconstructions at both users}

We now specialise Theorems~\ref{Thm:TwoUsersAchievability} and~\ref{Thm:TwoUsersConverse} to the (almost) lossless reconstructions setup.

\begin{corollary}
For Hamming distortions
\label{fix2}
\begin{align}\label{eq:bound1}
\R^\dag(\b{0}, \b{0},C) 
\geq &
\underline{\R}(\b{0}, \b{0},C) \\
= &
\max_{(\ell_1, \ell_2)} \; \max\Big\{ H({X}_{\ell_2}),\ H({X}_{\ell_1},{X}_{\ell_2})-C \Big\} 
\end{align}
and 
\begin{align}\label{eq:bound2}
\R^\dag(\b{0}, \b{0},C) 
\leq&
\overline{\R}(\b{0}, \b{0},C)\\
=& \min_U \max_{(\ell_1, \ell_2)} \max \Big\{H(X_{\ell_2}) + H(X_{\ell_1} | U, X_{\ell_2}),
H(U,{X}_{\ell_1},{X}_{\ell_2})-C \Big\}
\end{align}
where the minimisation is over all auxiliaries $U$. Moreover, \eqref{eq:bound1} holds with equality when $\L_1=\{\ell\}$ for some $\ell\in\L$ or $\L_1= \L_2 = \{\ell,\ell'\}$ for $\ell, \ell' \in \L$.
\end{corollary}

Interestingly, in these cases there is no  penalty in the rate-distortions function even though the server does not know the users' demands during the caching phase.


\subsection{(Almost) Lossless Reconstruction at user $2$}

We finally consider the setup where reconstruction at user~$1$ is lossy with a prescribed distortion, and reconstruction at user $2$ is lossless ($\b\Delta=\0$). We  specialise Theorems~\ref{Thm:TwoUsersAchievability}, \ref{Thm:TwoUsersConverse}, and  compare the obtained lower and upper bounds for a doubly symmetric binary source (DSBS). Then, we prove a stronger lower bound and show that it matches the upper bound of Theorem \ref{Thm:TwoUsersAchievability} for the studied DSBS.
\begin{corollary}
\label{cor:D0}
Theorems~\ref{Thm:TwoUsersAchievability} and \ref{Thm:TwoUsersConverse} provide the following upper and lower bound on the rate-distortion-memory function when $\Delta=0$ under the Hamming distortion function.
\begin{align}\label{eq:conv}
\R^\dag(\b{D}, \b{0},C) 
\geq &
\underline{\R}(\b{D}, \b{0},C) \\
= &
\min_{\b{\hX}}\max_{(\ell_1, \ell_2)} \; H({X}_{\ell_2}) + \max\Big\{ 0,\ I(\X;\hat{X}_{\ell_1}|X_{\ell_2})-C \Big\} 
\end{align}
and
\begin{align}
R^\dag(\b{D}, \b{0},C) \label{eq:cor-D0}
\leq&
\overline{\R}(\b{D}, \b{0},C)\\
=& \min_{(U,\b{\hX})} \max_{(\ell_1, \ell_2)} \max \Big\{I(\X;U,\hat{X}_{\ell_1},{X}_{\ell_2})-C,
H(X_{\ell_2})+ I(\X;\hat{X}_{\ell_1} | U, X_{\ell_2}), \Big\}
\end{align}
where the minimum is taken over all $(U,\b{\hX})$ jointly distributed with $\X$ such that~\eqref{Thm:TwoUserAchievability:D1} holds.
\end{corollary}

For the case where the library consists of two files, which are the outcomes of a DSBS,  we use Corollary \ref{cor:D0} to  find lower and upper bounds on the rate-distortion-cache function. The bounds meet for all distortions below a certain threshold.

\begin{corollary}
\label{cor:DBSB2user}
For the DSBS $(X_\ell,X_\ell')$ with parameter $\rho$, $0\leq\rho\leq \frac{1}{2}$, with symmetric distortion criteria $D_\ell=D_\ell'=D$ under the Hamming distortion function, we have
\begin{align}
\R^\dag(\b{D}, \b{0},C) 
\geq  1 + \left(\ h(\rho)-h(D)-C \right)^+ 
\end{align}
and
\begin{align}
&R^\dag(\b{D}, \b{0},C)\nonumber \\
&\leq\left\{\!\!\begin{array}{l}1 + \left(\ h(\rho)-h(D)-C \right)^+ \\\hspace{5cm} \text{If }D\leq D^\star \\1+\left(h(D)\!-\!\rho\!-\!(1\!-\!\rho)h\left(\frac{2D-\rho}{2(1-\rho)}\right)\!-\!C\right)^+\\\hspace{5cm}\text{If }D^\star<D\leq \frac{1}{2}\end{array}\right.
\end{align}
where \begin{align}\label{Dstar}D^\star=\frac{1}{2}(1-\sqrt{1-2\rho)}.\end{align}
\end{corollary}
\begin{IEEEproof}
The proof is outlined in Appendix \ref{RoyPaper}, and uses connections to the problem of broadcasting to two users with complementary side information \cite{Timo-Nov-2010-A}.
\end{IEEEproof}

Finally, we find a lower bound on the delivery rate by considering an average-case scenario, rather than a worst-case scenario, for user $2$. This type of lower bounds has also appeared in \cite{Wang2016}, \cite{Saeedi2016}, and \cite{Saeedi2016B}.
\begin{theorem}
\label{thm:Loweravg}
For any distribution $p_I(.)$ on $\{1,\ldots, L_2\}$, we have 
\begin{align}
R^\dag(\b{D}, \b{0},C) 
\geq
\min_{(U,\b{\hX})} \max_{\ell_1} \max \Big\{I(\X;U,\hat{X}_{\ell_1},{X}_{I})-C,
H(X_{I}|I)+I(\X;\hat{X}_{\ell_1} | U, X_{I}), \Big\}\label{eqn:Loweravg-case}
\end{align}
where $X_{I}$ is indexed by the random variable $I$ and the minimisation is over all $p_{U\b{\hX}|\X I}$ such that 
\begin{align}p_{U\b{\hX}|\X I}=p_{\b{\hX}|\X}\times p_{U|\b{\hX}\X I}\label{probdis}\end{align} and~\eqref{Thm:TwoUserAchievability:D1} holds. To compute the above optimization, the cardinality of  $\mathcal{U}$ may be restricted to $\mathcal{U}\leq |\mathcal{X}|+2L_1$ without any loss of generality.
\end{theorem}
\begin{IEEEproof}
Theorem \ref{thm:Loweravg} is proved in Appendix \ref{app:conv-2user-avg}.
\end{IEEEproof}
\begin{remark}
For symmetric sources where $p_{X_1}(.)=\ldots=p_{X_L}(.)$, \eqref{eqn:Loweravg-case} reduces to
\begin{align}
R^\dag(\b{D}, \b{0},C) 
\geq
\min_{(U,\b{\hX})} \max_{\ell_1} \max \Big\{I(\X;U,\hat{X}_{\ell_1},{X}_{I})-C,
H(X_{I})+I(\X;\hat{X}_{\ell_1} | U, X_{I}) \Big\}.\label{eqn:avg-case}
\end{align}
Note that \eqref{eqn:avg-case}  looks similar to \eqref{eq:cor-D0} but  $X_{I}$ implies an average demand criterion at user $2$ and may in general be weaker than~\eqref{eqn:worst-case}.
\end{remark}
\begin{remark}
The lower bound in Theorem \ref{thm:Loweravg} is larger than or equal to the genie-aided lower bound in Theorem \ref{Thm:TwoUsersConverse}.
\end{remark}

\appendices

\section{Proof of Theorem~\ref{Thm:RDC}}\label{Thm:RDC:Sec}


\subsection{Achievability (Expected and Excess Distortions)}

Pick any test channel $q_{U|\X}$ from $\sXL$ to $\s{U}$ such that $(\X,U)$ $\sim p_\X\ q_{U|\X}$ satisfies $I(\X;U) < C$. Build a codebook for the cache by randomly generating $2^{nR'}$ codewords 
\begin{equation*}
\s{C}_\text{cache} := \Big\{U^n(j) = \big(U_1(j),U_2(j),\ldots,U_n(j)\big)\Big\}_{j=1}^{2^{nR'}},
\end{equation*}
each of length $n$, by selecting symbols iid $U \sim q_U$ from $\s{U}$. Give $\s{C}_\text{cache}$ to the server and user. 

The server observes the entire library $\X^n$, and it looks for a unique codeword index $j$ such that $(\X^n,U^n(j))$ is jointly typical (for example, we may use letter typicality~\cite{Kramer-2008-A} or strong typicality~\cite{Cover-2006-B}). If successful, the transmitter places $j$ in the cache; if unsuccessful, the server places $j = 1$ in the cache. If $R' > I(\X;U)$, then the probability of a \emph{cache encoding error} (i.e., such a $j$ cannot be found) vanishes exponentially fast in the blocklength $n$.  

The user reads the index $j$ from its cache, and it recovers the codeword $U^n(j)$ from the cache codebook $\s{C}_\text{cache}$. If the server's cache-encoding step was successful, then $U^n(j)$ will be jointly typical with $X^n_\ell$. The server-user compression problem then reduces to the standard conditional rate-distortion problem~\cite{Gray-Oct-1972-A} with $U^n(j)$ as side information.  \qed


\subsection{Weak Converse (Expected Distortions)}

Fix the distortion tuple $\D$ and cache capacity $C$ arbitrarily, and suppose that $R$ is $(\D,C)$-admissible. Suppose that we have a sequence of $(n,\s{M},\sMc)$-codes satisfying~\eqref{Eqn:Def:Ach:ExpDist}. For any $\ve > 0$ there exists a sufficiently large $n^*$ depending on $\epsilon$ such that for all $n \geq n^*$ we have
\begin{align}
C + \ve
\notag
&\geq 
\frac{1}{n} \log |\sMc|\\[5pt]
\notag
&\geq 
\frac{1}{n} H(\Mc)\\[5pt]
\notag
&\geq 
\frac{1}{n}
I(\X^n; \Mc)\\[5pt]
\notag
&= 
\frac{1}{n}
\sum_{i=1}^n I(\X_i; \Mc|\X_1^{i-1})\\
\notag
&\step{a}{=} 
\frac{1}{n}
\sum_{i=1}^n I(\X_i; \Mc,\X_1^{i-1})\\
\notag
&
\step{b}{=} 
\frac{1}{n}
\sum_{i=1}^n I(\X_i; U_i),
\end{align}
where (a) follows because $\X^n$ is iid, and (b) follows by setting 
\begin{equation*}
U_i = (\Mc,\X_1^{i-1})
\quad
\text{on}
\quad
\s{U}_i = \sMc \times \sXL^{i-1}.
\end{equation*}

Similarly, for each index $\ell \in \L$ and $n \geq n^*$ we have
\begin{align}
R + \ve  
\notag
&\geq
\frac{1}{n} \log |\sM|\\[5pt]
\notag
&\geq 
\frac{1}{n} H(\Ml)\\[5pt]
\notag
&\geq 
\frac{1}{n} I(\X^n;\Ml|\Mc)\\[5pt]
\notag
&\step{a}{\geq} 
\frac{1}{n} I(\X^n;\h{X}^n_\ell|\Mc)\\
\notag
&\geq
\frac{1}{n} 
\sum_{i=1}^n
I(\X_i;\h{X}^n_\ell|\Mc,\X_1^{i-1})\\
\notag
&\step{b}{=}
\frac{1}{n} 
\sum_{i=1}^n 
I(\X_i;\h{X}^n_\ell | U_i)\\
\notag
&\geq
\frac{1}{n}
\sum_{i=1}^n 
I(X_{\ell i};\h{X}_{\ell i} | U_i),
\end{align}
where (a) follows because $\hX^n_\ell$ is a function of the delivery-phase message $\Ml$ and the cached message $\Mc$ and (b) substitutes $U_i$.   

Consider each index $i$ in the above sum and, in particular, the tuple $(\X_i,U_i,\h{\X}_i)$. We can write the joint pmf of this tuple as 
\begin{align*}
p_{\X_i U_i \h{\X}_i}(\x_i,u_i,\h{\x}_i)
= p_{\X}(\x_i)\ q_{U_i|\X_i}(u_i|\x_i)
q_{\hXL_i |\X_i,U_i}(\h{\x}_i|\x_i,u_i).
\end{align*}
Let 
\begin{equation*}
\s{U}^* = \bigcup_{i = 1}^n \s{U}_i,
\end{equation*}
and construct a new joint pmf on $\tilde{p}_{\X U \h{\X}}$ on $\sXL \times \s{U} \times \hsXL$ by setting
\begin{align*}
p^*_{\X U \h{\X}}(\x,u,\h{\x}) =
p_{\X}(\x)
\frac{1}{n}
\sum_{i = 1}^n 
\indicator{u \in \s{U}_i} q_{U_i|\X_i}(u_i|\x_i) 
q_{\h{\X}_i|\X_i}(\h{\x}_i|u,\x).
\end{align*}
Then, 
\begin{align*}
\frac{1}{n} \sum_{n=1}^n I(\X_i;U_i) &= I_{p^*_{\X U \h{\X}}}(\X;U)\\
\frac{1}{n} \sum_{n=1}^n I(X_{\ell,i};\h{X}_{\ell,i}|U_i) &= I_{p^*_{\X U \h{\X}}}(X_\ell;\h{X}_\ell|U)
\end{align*}
and
\begin{equation*}
D_\ell + \epsilon \geq \E\left[ \frac{1}{n} \sum_{i=1}^n \dl(\h{X}_{\ell,i},X_{\ell,i}) \right]
= \E_{p^*_{\X U \h{\X}}}\Big[\dl(\h{X}_\ell,X_\ell)\Big]
\end{equation*}
for all $\ell \in \L$. To complete the converse, we need only show that the alphabet $\s{U}^*$ and the joint pmf $p^*_{\X U \h{\X}}$ on $\sXL \times \s{U}^* \times \hsXL$ can be replaced by an alphabet $\s{U}$ of cardinality $|\s{U}| \leq |\sXL| + 2L$ and a joint pmf $q_{\X U \h{\X}}$ on $\sXL \times \s{U} \times \hsXL$ such that
\begin{enumerate}[(i)]
\item $I_{p^*_{\X U \h{\X}}}(\X;U) = I_{q_{\X U \h{\X}}}(\X;U)$
\item $\forall\ \ell \in \L :\ $ $I_{p^*_{\X U \h{\X}}}(X_\ell;\h{X}_\ell|U) = I_{q_{\X U \h{\X}}}(X_\ell;\h{X}_\ell|U)$
\item $\forall\ \ell \in \L :\ $  $ \E_{p^*_{\X U \h{\X}}} \big[\dl(\h{X}_\ell,X_\ell)\big] =  \E_{q_{\X U \h{\X}}} \big[\dl(\h{X}_\ell,X_\ell)\big]$.
\end{enumerate}
Let $q \in \s{P}^*$ and $\x^* \in \sXL$ be arbitrary, and define the following functions.
\begin{enumerate}

\item 
For each $\x \in \sXL$, let
\begin{equation*}
g^1_\x(q) 
:= 
\sum_{\h{\x} \in \hsXL}\ q(\x,\h{\x})
\end{equation*}

\item 
\begin{equation*}
g^2(q)
 := 
- \sum_{\x \in \sXL} 
\left(\sum_{\h{\x} \in \hsXL} q(\x,\h{\x}) \right)
\log 
\left(\sum\limits_{\h{\x} \in \hsXL} q(\x,\h{\x}) \right).
\end{equation*}

\item 
For each $\ell \in \L$, let
\begin{align*}
g^3_\ell(q)
:=
\sum_{x_\ell \in \sX_\ell}
\sum_{\h{x}_\ell \in \hsX_\ell}
q(x_\ell,\h{x}_\ell) 
\cdot\ \log \frac{q(x_\ell,\h{x}_\ell)}{
\Big(\sum\limits_{\h{x}_\ell \in \hsX_\ell} q(x_\ell,\h{x}_\ell)\Big)
\Big(\sum\limits_{x_\ell \in \sX_\ell} q(x_\ell,\h{x}_\ell)\Big)
}
\end{align*}

\item 
For each $\ell \in \L$, let
\begin{equation*}
g^4_\ell(q)
:=
\sum_{x_\ell \in \sX_\ell}
\sum_{\h{x}_\ell \in \hsX_\ell}
q(x_\ell,\h{x}_\ell)\
\dl(\h{x}_\ell,x_\ell).
\end{equation*}
\end{enumerate}

Define a Borel measure $\mu$ on $\s{P}^*$ by 
\begin{equation*}
\mu(p^*_{|U=u}) := p^*_{U}(u),
\quad u \in \s{U}^*.
\end{equation*}
Then, 
\begin{align*}
\int\ g^1_\x\ d\mu &= p^*_\X(\x),\quad \forall\ \x \in \sXL.\\
\int\ g^2\ d\mu &= H_{p^*_{\X,U}}(\X|U)\\
\int\ g^3_\ell\ d\mu &= I_{p^*_{\X,U,\h{\X}}}(X_\ell;\h{X}_\ell|U),\quad \forall\ \ell \in \L.\\
\int\ g^4_\ell\ d\mu &= \E_{p^*_{X_\ell,U,\h{X}_\ell}}\ \dl(\h{X}_\ell,X_\ell),\quad \forall\ \ell \in \L.
\end{align*}

By Carath\'{e}odory's theorem, it is possible to assign probabilities to at most $|\sXL| + 2L$ points in $\s{U}^*$ and preserve the area of the above functions. \qed


\subsection{Weak Converse (Excess Distortions)}

The weak converse for excess distortions follows immediately from Assertion IV of Lemma~\ref{Lem:BasicProperties} and the weak converse for expected distortions given above. \qed


\section{Proof of Theorem~\ref{Thm:Kgenie}}


\subsection{Proof of Theorem~\ref{Thm:Kgenie}}\label{Thm:Kgenie:Sec}

Choose the cache capacity to be $C = \Kgenie(\D)$ \mw{and assume that $\R(\D, \Kgenie(\D))>0$. By the definition of $\Kgenie(\D)$:
\begin{equation}\label{eq:recall1}
\R(\D,C)=\max_{\ell\in\s{L}}  \R_{X_{\ell}}(D_{\ell}) -C. 
\end{equation}}

\mw{Let $U$ be an optimal auxiliary random variable for the informational RDC function $\R(\D,C)$, i.e., $U$ is so that
\begin{equation}\label{eq:recall2}
\R(\D,C)= \max_{\ell\in\s{L}} \R_{X_{\ell}|U}(D_{\ell})
\end{equation}
and 
\begin{equation}\label{eq:recall3}
I(\X;U)\leq C.
\end{equation}}

\mw{Let  $\ell^*\in\L^*$, i.e., $\ell^*$ attains the maximum in \eqref{eq:recall1}. We have the following: }
\begin{align*}
\R(\D,C)
&\step{a}{=}
\max_{\ell \in \L} 
\R_{X_{\ell}|U}(D_{\ell})\\[5pt]
&\geq
\R_{X_{\ell^*}|U}(D_{\ell^*})\\[5pt]
&=
\mw{\min_{q_{\hat{X}_{\ell^*}|X_{\ell^*},U} \colon \E[d(\hat{X}_{\ell^*},{X}_{\ell^*})]\leq D_\ell} I(X_{\ell^*};\hat{X}_{\ell^*}|U) }
\\[5pt]
&\step{b}{\geq}\mw{\min_{q_{\hat{X}_{\ell^*}|X_{\ell^*},U} \colon \E[d(\hat{X}_{\ell^*},{X}_{\ell^*}]\leq D_\ell} I(X_{\ell^*};U, \hat{X}_{\ell^*})}
 - I(\X;U)
\\[5pt]
&\step{c}{\geq}\mw{
\R_{X_{\ell^*}}(D_{\ell^*}) - I(\X;U)}
\\[5pt]
&\step{d}{\geq}
\R_{X_{\ell^*}}(D_{\ell^*}) - C\\[5pt]
&\step{e}{\mw{=}}
\R(\D,C),
\end{align*}
where (a) is identical to \eqref{eq:recall2}; (b) follows by adding the negative term $I(X_{\ell^*};U)-I(\X;U)$; (c) holds because $I(X_{\ell^*};U, \hat{X}_{\ell^*}) \geq I(X_{\ell^*}; \hat{X}_{\ell^*})$; (d) holds by \eqref{eq:recall3}; and (e) holds by \eqref{eq:recall1} and because $\ell^*\in\L^*$. 

The above inequalities must all hold with equality and so the chosen $U$ must satisfy $I(\X;U) = C\mw{=\Kgenie(\D)}$, \eqref{Eqn:GenieCondition1} and~\eqref{Eqn:GenieCondition2}. Therefore, 
\begin{equation}\label{Eqn:GenieCacheCapacityProof2}
\Kgenie(\D) 
\leq 
\Ksgenie(\D).
\end{equation}

Choose now the cache capacity $C=\Ksgenie(\D)$, and let $U$ be an optimal auxiliary random variable for $\Ksgenie(\D)$. \mw{That means, $U$ satisfies~\eqref{Eqn:GenieCondition1} and \eqref{Eqn:GenieCondition2} and 
\begin{equation}\label{eq:recall4}
I(\X;U)=\Ksgenie(\D)=C.
\end{equation}}
 The following holds for all $\ell^* \in \L^*$:
\begin{align*}
\R(\D,C) 
&\step{a}{\leq} 
\max_{\ell \in \L} 
\R_{X_\ell|U}(D_\ell)\\
&\step{b}{=}
\R_{X_{\ell^*}|U}(D_{\ell^*})\\
&\step{c}{=}
\R_{X_{\ell^*}}(D_{\ell^*}) - \mw{I(\X;U)}\\
&\step{d}{=}
\mw{\R_{X_{\ell^*}}(D_{\ell^*}) -C},
\end{align*}
where (a) follows because $U$ need not be optimal for $\R(\D,C)$, (b) follows from~\eqref{Eqn:GenieCondition2}, (c)  follows from~\eqref{Eqn:GenieCondition1}, and (d) from~\eqref{eq:recall4}. 

Therefore, at the cache capacity $C = I(\X;U) = \Ksgenie(\D)$ we have $
\R(D,C) = \R_{X_{\ell^*}}(D_{\ell^*}) - C$ and consequently  
\begin{equation}\label{Eqn:GenieCacheCapacityProof3}
\Kgenie(\D) \geq \Ksgenie(\D). 
\end{equation}
The theorem follows from~\eqref{Eqn:GenieCacheCapacityProof2} and~\eqref{Eqn:GenieCacheCapacityProof3}. \qed


\subsection{Proof of Corollary~\ref{Thm:Kgenie:Cor}}\label{Thm:Kgenie:Cor:Sec}

The conditional RD function particularises to the conditional entropy function: $\R_{X_{\ell}|U}(0) = H(X_{\ell} | U)$. Similarly, the constraint~\eqref{Eqn:GenieCondition1} particularises to 
\begin{equation*}
I(\X;U) = H(X_{\ell^*}) - H(X_{\ell^*} | U) = I(X_{\ell^*};U), 
\end{equation*}
which is equivalent to $U \markov X_{\ell^*} \markov X_{\L \backslash \ell^*}$. \qed


\section{Proof of Theorem~\ref{Thm:KgenieKGK}}\label{Thm:KgenieKGK:Sec}
\mw{Let $\ell\in\L$.}
For any $(\X,U) \sim p_\X\ p_{U|\X}$ on $\sXL \times \sU$, the following inequalities hold:
\begin{align}
\R_{\X_\ell | U}(D_\ell) 
\notag
&=
\min_{q_{\hX_\ell' | U,X_\ell} :\ \E[ \dl(\hX_\ell',X_\ell)] \leq D_\ell}\ I(X_\ell;\hX'_\ell|U) \\
\notag
&\geq
\min_{q_{\hX_\ell' | X_\ell} :\ \E[ \dl(\hX_\ell',X_\ell)] \leq D_\ell} I(X_\ell;\hX'_\ell) - I(\X;U)\\
\label{Thm:KgenieKGK:Eqn1}
&= 
\R_{X_\ell}(D_\ell) - I(\X;U).
\end{align}

Now suppose that we have $(U,\hXL) \sim p_{\hXL,U|\X}$ on $\sU \times \hsXL$ satisfying conditions (i), (ii), (iii), \mw{and (iv)} in Definition~\ref{Def:GacsKorner:Lossy}.  Then, 
\begin{align}
\R_{X_\ell | U}(D_\ell) 
\notag
&\step{a}{\leq}
I(X_\ell;\hX_\ell | U) \\
\notag
&\step{b}{=}
I(X_\ell ; \hX_\ell) - I(\X ; U) \\
\label{Thm:KgenieKGK:Eqn2}
&\step{c}{=}
\R_{X_\ell}(D_\ell) - I(\X;U),
\end{align}
where (a)  follows from property (iii) of Definition~\ref{Def:GacsKorner:Lossy}; (b) follows by properties (i) and (ii) of Definition~\ref{Def:GacsKorner:Lossy}; and \mw{(c) follows from property (iv) of Definition~\ref{Def:GacsKorner:Lossy}}. 

\mw{Inequalities~\eqref{Thm:KgenieKGK:Eqn1} and~\eqref{Thm:KgenieKGK:Eqn2} combine to
\begin{equation}
\R_{X_\ell | U}(D_\ell)  = \R_{X_\ell}(D_\ell) - I(\X;U), \quad \forall \ell\in\L.
\end{equation}
Thus, 
 the pair $(U,\hXL)$ satisfies~\eqref{Eqn:GenieCondition1}. Moreover,  since the mutual information $I(\X;U)$ does not depend on $\ell \in \L$, the conditional rate-distortion function $\R_{X_\ell | U}(D_\ell) $ is largest for the same indices $\ell$ as the standard rate-distortion function $\R_{X_\ell}(D_\ell)$.  Since $\R_{X_\ell}(D_\ell)$ is maximum for indices $\ell^*\in\L^*$, this proves that the pair  $(U,\hXL)$ also satisfies~\eqref{Eqn:GenieCondition2}.} To conclude: If $(U,\hXL)$ satisfies (i), (ii), (iii), and (iv) in Definition~\ref{Def:GacsKorner:Lossy}, then $U$ is a valid tuple for $\Ksgenie(\D)$ and $\KGK(\D) \leq \Ksgenie(\D)$. 

Now suppose that 
\begin{equation*}
\R_{X_1}(D_1) = \R_{X_2}(D_2) = \cdots = \R_{X_L}(D_L)
\end{equation*}
and, therefore, $\L^* = \L$. Let $U \sim p_{U|\X}$ on $\sU$ be any auxiliary random variable satisfying~\eqref{Eqn:GenieCondition1} for every $\ell \in \L$. (Condition~\eqref{Eqn:GenieCondition2} automatically follows because $\L^* = \L$.) For each $\ell \in \L$, let $p_{\hX_\ell | U X_\ell}$ be any test channel that is optimal for the informational conditional RD function 
\begin{equation*}
\R_{X_\ell | U}(D_\ell) = \min_{q_{\hX'_\ell | U X_\ell} :\ \E[\dl(\hX'_\ell,X_\ell)] \leq D_\ell} I(X_\ell ; \hX'_\ell | U).
\end{equation*}
Now consider the tuple
\begin{equation*}
(\X,U,\hXL) \sim p_\X\ p_{U|\X}\ \prod_{\ell \in \L} p_{\hX_\ell | U X_\ell}.
\end{equation*}
For all $\ell \in \L$ we have
\begin{align*}
\R_{X_\ell | U}(D_\ell) 
&\step{a}{=} 
\R_{X_\ell}(D_\ell) - I(\X;U)\\\
&\step{b}{\leq} 
I(X_\ell;\hX_\ell) -  I(\X;U)\\
&\leq 
I(X_\ell;\hX_\ell|U)\\
&\step{c}{=} 
\R_{X_\ell | U}(D_\ell),
\end{align*} 
where (a) follow because $U$ was originally chosen to satisfy~\eqref{Eqn:GenieCondition1}; (b) follows because $(\X,U,\hXL)$ need not be optimal for the informational RD functions $\R_{X_\ell}(D_\ell)$; and (c) follows because  $p_{\hX_\ell | U X_\ell}$ achieves $\R_{X_\ell | U}(D_\ell)$. The above inequalities must be equalities and, therefore, $(\X,U,\hXL)$ satisfies the following four conditions:
\begin{itemize}
\item $\forall\ \ell \in \L:\ $ $U \markov X_\ell \markov X_\ellc$ 
\item $\forall\ \ell \in \L:\ $ $U \markov \hX_\ell \markov X_\ell$ 
\item $\forall\ \ell \in \L:\ $ $I(X_\ell;\hX_\ell) = \R_{X_\ell}(D_\ell)$
\item $\forall\ \ell \in \L:\ $ $\E[\dl(\hX_\ell,X_\ell)] \leq D_\ell$. 
\end{itemize}

To conclude: Given any $(\X,U) \sim p_\X\ p_{U|\X}$ satisfying~\eqref{Eqn:GenieCondition1} for all $\ell \in \L$ we can always find a test channel $p_{\hXL | U \X}$ such that $(\X,U,\hXL) \sim p_\X\ p_{U|\X}\ p_{\hXL | U \X}$ satisfies the conditions of Definition~\ref{Def:GacsKorner:Lossy}. 
\qed  


\section{Proof of Lemma~\ref{Lem:GacsKornerMarkovDefinition}}\label{Lem:GacsKornerMarkovDefinition:Sec}
\mw{For Hamming distortion functions, it is not too hard to see that Definition~\ref{Def:GacsKorner:Lossy} particularises to 
\begin{equation*}
\KGK(\b{0}) 
= \max_{U:\ U \markov X_\ell \markov X_{\L\backslash \ell},\ \forall \ell \in \L}\ I(\X;U).
\end{equation*}}

Clearly, we have
\begin{align*}
\max_{U:\ H(U|X_\ell) = 0,\ \forall \ell \in \L}\ H(U)
\leq
\max_{U:\ U \markov X_\ell \markov X_{\L\backslash \ell},\ \forall \ell \in \L}\ I(\X;U) 
\end{align*}
since any $U$ satisfying $H(U|X_\ell) = 0$ for all $\ell \in \L$ must also satisfy $U \markov X_\ell \markov X_{\L \backslash \ell}$ for all $\ell \in \L$. The reverse inequality (and therefore Lemma~\ref{Lem:GacsKornerMarkovDefinition}) follows by the next lemma, which is a multivariate extension of~\cite[Lem.~A.1]{Prabhakaran-Jun-2014-A}. \qed

\begin{lemma}\label{Lem:Bernhard}
If $U$ is jointly distributed with $\X$ such that $U \markov X_\ell \markov X_\ellc$ for all $\ell \in \L$, then there exists $U'$ jointly distributed with $(U,\X)$ such that $
U \markov U' \markov \X$ and $H(U'|X_\ell) = 0$ for all $\ell \in \L$.
\end{lemma}

\begin{IEEEproof}
Let $p_{U|\X}$ denote the conditional distribution of $U$ given $\X$, and suppose that 
\begin{equation}\label{Eqn:Lem:Bernhard:MC}
U \markov X_\ell \markov X_\ellc,
\quad 
\forall\ \ell \in \L.
\end{equation}
We first generate an $L$-partite graph 
\begin{equation*}
\s{G} = (\s{V},\s{E}),
\end{equation*}
with vertices  
\begin{equation*}
\s{V} = \bigcup_{\ell \in \L} \sX_\ell.
\end{equation*}
The edge set $\s{E}$ contains an edge
\begin{equation*}
\big\{x, x'\big\},\quad
x \in \sX_i,\ x' \in \sX_j,
\quad
i,j \in \L \text{ with } i \neq j,
\end{equation*}
if and only if there exists an $\tilde{\x} \in \sXL$ with $\tilde{x}_i = x$ and $\tilde{x}_j = x'$ and $p_\X(\tilde{\x}) > 0$. 

Let $\s{C}_1,\s{C}_2,\ldots,\s{C}_{\Ncc}$ denote the connected components of $\s{G}$, and let $c(x)$ denote the index of the connected component that contains vertex $x$. 

Let us now construct a new auxiliary random variable $U'$ on $\{1,$ $\ldots,\Ncc\}$ that is jointly distributed with $\X$ by setting 
\begin{equation*}
U' = c(X_1).
\end{equation*}
Now, for any $\x \in \sXL$ with $p_\X(\x) > 0$, the corresponding set of vertices $\{x_1,\ldots,x_L\}$ forms a clique and, therefore, is a subgraph of some connected component. Therefore,
\begin{equation*}
U' = c(X_\ell)\quad \text{a.s.},\ \forall\ \ell \in \{2,\ldots,L\}.
\end{equation*}
This, of course, implies $H(U'|X_\ell) = 0$ for all $\ell$. 

To complete the proof, we need only to show that $U$ can be generated by some conditional distribution $q_{U|U'}: \{1,\ldots,$ $\Ncc\} \to \s{U}$. We first notice that the Markov chain~\eqref{Eqn:Lem:Bernhard:MC} is equivalent to the following condition: For all $\x \in \sXL$ with $p_\X(\x) > 0$, we have
\begin{equation*}
p_{U|\X}(u|\x) = p_{U|X_1}(u|x_1) = \cdots = p_{U|X_L}(u|x_L),
\ \ \forall\ u \in \s{U}.
\end{equation*}
Now consider any connected component $\s{C}_i$ and any $u \in \s{U}$. By the above method of constructing $\s{G}$, we may conclude that
\begin{equation*}
p_{U|X_\ell}(u|x_\ell) = \text{constant},
\quad
\forall\ 
\ell \in \L \text{ and } x_\ell \in \s{C}_i \cap \sX_\ell.
\end{equation*}
That is, $p_{U|X_\ell}(u|x_\ell)$ depends only on the connected component $c(x_\ell)$ and the particular $u \in \s{U}$, and we can write the above constant as $q_{c(x_\ell)}(u)$. Choose $p_{U|U'}(u|u') := q_{u'}(u)$ to complete the proof. 
\end{IEEEproof}


\section{Proof of Theorem~\ref{Thm:Ksuper}}\label{Thm:Ksuper:Sec}

Choose the cache capacity $C = \Ksuper(\D)$. Let $U$ be an optimal auxiliary random variable for the informational RDC function; that is, 
\begin{equation*}
\R(\D,C) = \max_{\ell \in \L} \R_{X_\ell | U}(D_\ell).
\end{equation*} 
Now, for each $\ell \in \L$, let $p_{\hX_\ell | U}$ be an optimal test channel for the informational conditional RD function $\R_{X_\ell | U}(D_\ell)$. Define 
\begin{equation*}
(\X,U,\hXL) \sim p_\X\ p_{U  | \X}\ \prod_{\ell \in \L} p_{\hX_\ell | U X_\ell},
\end{equation*}
and note that 
\begin{equation}\label{Eqn:Superuser:MarkovChain}
\hX_\ell \markov (U,X_\ell) \markov (X_\ellc, \hX_\ellc),
\quad 
\forall\ \ell \in L.
\end{equation}
Then, 
\begin{align*}
\R(\D,C) 
&=
\max_{\ell \in \L} 
I(X_\ell ; \hX_\ell | U)\\
&\geq
\frac{1}{L}
\sum_{\ell = 1}^L
I(X_\ell ; \hX_\ell | U)\\
&\step{a}{\geq} 
\frac{1}{L}
\sum_{\ell = 1}^L
I(\X ; \hX_\ell | U,\hX_1^{\ell - 1})\\
&= 
\frac{1}{L}
I(\X ; \hXL | U)\\
&\step{b}{\geq}
\frac{1}{L}\Big( I(\X ; \hXL) - \Ksuper(\D) \Big)\\
&\step{c}{\geq}
\frac{1}{L}\Big(\R_\X(\D) - \Ksuper(\D) \Big)\\
&\step{d}{=}
\R(\D,C),
\end{align*}
where (a) follows from~\eqref{Eqn:Superuser:MarkovChain}; (b) follows because $I(\X;U) \leq \Ksuper(\D)$; (c) follows because $\E[\dl(X_\ell,\hX_\ell)] \leq D_\ell$; and $(d)$ follows from the definition of $\Ksuper(\D)$. 

The above inequalities are equalities and consequently $I(X_1;\hX_1 | U) = \cdots = I(X_L;\hX_L | U)$, $\hX_\ell \markov U \markov \hX_{\ell-1}$ (and therefore $\hX_\ell \markov U \markov \hX_\ellc$ since the chain rule expansion order is arbitrary), $\X \markov \hXL \markov U$ and $C = I(\X;U)$. We can thus conclude that the tuple $(\X,U,\hXL)$ satisfies conditions (i)--(v) in the definition of $\Kssuper(\D)$  and $\Kssuper(\D) \leq \Ksuper(\D)$. 

Now suppose that $(\X,U,\hXL)$ satisfies conditions (i)--(v) in the definition of $\Kssuper(\D)$ and $I(\X;U) = \Kssuper(\D)$. Then
\begin{align*}
\R(\D,C)
&\leq 
\max_{\ell \in \L} I(X_\ell ; \hX_\ell | U)\\
&\step{a}{=}
\frac{1}{L} \sum_{\ell = 1}^L I(X_\ell ; \hX_\ell | U)\\
&\step{b}{\leq}
\frac{1}{L} \sum_{\ell = 1}^L I(\X;\hX_\ell | U, \hX_1^{\ell-1}) \\
&=
\frac{1}{L} \Big( I(\X;\hXL,U) - I(\X;U) \Big)\\
&\step{c}{\leq}
\frac{1}{L} \Big( \R_\X(\D) - \Kssuper(\D) \Big),
\end{align*}
where (a) follows because from condition (ii); (b) follows from condition (iii); (c) follows from conditions (i) and (v). Thus, we can achieve the superuser bound at $C = \Kssuper(\D)$ and $\Kssuper(\D) \leq \Ksuper(\D)$. \qed


\section{Proof of Lemma~\ref{Lem:SuperuserSLowerBound}}\label{Lem:SuperuserSLowerBound:Sec}

First suppose that $\s{S} = \{1,2,\ldots,S\}$ for some $1 \leq S \leq L$. Then 
\begin{equation*}
\DS = (D_1,\ldots,D_S)
\quad
\text{and}
\quad
\XS = (X_1\ldots,X_S).
\end{equation*}
Let 
\begin{equation*}
\R_\XS(\DS) = \min_{p_{\hX_\S | \XS} : \ \E[ \dl(\hX_\ell,X_\ell)] \leq \dl,\ \forall \ell \in \L} I(\XS ; \hX_\S)
\end{equation*}
denote the joint RD function of the source $\XS$ w.r.t.~the $S$ distortion functions $\d_1,\d_2,\ldots,\d_S$. A proof of the next lemma is essentially given in~\cite[Thm.~3.1]{Gray-Jul-1973-A} and is omitted.  

\begin{lemma}\label{Lem:JointAndConditionalRDFunction}
The joint RD function is upper bounded by
\begin{equation*}
\R_\XS(\DS) \leq \sum_{\ell = 1}^S \R_{X_\ell | X_1^{\ell-1}}(D_\ell),
\end{equation*}
where 
\begin{equation*}
\R_{X_\ell|X_1^{\ell-1}}(D_\ell) 
= 
\min_{p_{\hX_\ell|X_1^\ell} :\ \E \dl (\hX_\ell , X_\ell) \leq D_\ell} I(X_\ell;\hX_\ell|X_1^{\ell-1})
\end{equation*}
denotes the conditional RD function for compressing a source $X_\ell$ with side information $X_1^{\ell - 1}$.
\end{lemma}

Let us now return to Lemma~\ref{Lem:SuperuserSLowerBound}. We first notice that the information RDC function can be written as
\begin{equation}\label{Prop:SuperuserSLowerBound:RDCFunction}
\R(\D,C) = \min_{p_{\hXL U|\X} \in \s{P}_{C}(\D)} \max_{\ell \in \L} I(X_\ell ; \hX_\ell | U),
\end{equation}
where $\s{P}_{C}(\D)$ denotes the set of all test channels $p_{\hXL U|\X}$ from $\sXL$ to $\hsXL \times \sU$ satisfying the cache-capacity constraint 
\begin{equation}\label{Prop:SuperuserSLowerBound:RDCFunctionCache}
I(\X;U) \leq C,
\end{equation} 
the expected-distortion constraints  
\begin{equation}\label{Prop:SuperuserSLowerBound:RDCFunctionDist}
\E\big[ \dl(\hX_\ell,X_\ell) \big] \leq \dl,
\quad
\forall\ \ell \in \L,
\end{equation}
and the Markov chains
\begin{equation}\label{Prop:SuperuserSLowerBound:RDCFunctionMC}
\hX_\ell \markov (X_\ell,U) \markov (X_{\L \backslash \ell}, \hX_{\L \backslash \ell} ),
\quad \forall\ \ell \in \L.
\end{equation}
Intuitively, the Markov chains in~\eqref{Prop:SuperuserSLowerBound:RDCFunctionMC} can be imposed without changing the minimisation because each conditional mutual information $I(X_\ell;\hX_\ell | U)$ depends only on the marginal distribution of $(X_\ell,U,\hX_\ell)$. Moreover, the minimum in~\eqref{Prop:SuperuserSLowerBound:RDCFunction} exists because the random variables are all defined on finite alphabets and conditional mutual information is continuous in $p_{\hXL U|\X}$ and bounded from below. Pick any test channel that achieves this minimum and, with a slight abuse of notation, let $(\X,U,\hXL)$ denote the resulting tuple of random variables. We then have
\begin{align*}
\notag
&\R(\D,C)
\geq
\max_{\ell \in \S} 
I(X_\ell;\hX_\ell|U)\\
&\geq 
\frac{1}{S} 
\sum_{\ell = 1}^S
I(X_\ell;\hX_\ell|U)\\
&\step{a}{\geq}
\frac{1}{S} 
\sum_{\ell = 1}^S 
I(X_\ell;\hX_\ell|U,X_1^{\ell-1}) \\
&\geq
\frac{1}{S}
\sum_{\ell = 1}^S
I(X_\ell;\hX_\ell|X_1^{\ell-1})
- 
\frac{1}{S}
\sum_{\ell = 1}^S
I(X_\ell; U|X_1^{\ell-1})\\
&\step{b}{\geq}
\frac{1}{S}
\sum_{\ell = 1}^S\
I(X_\ell;\hX_\ell|X_1^{\ell-1}) 
- 
\frac{1}{S} C\\
&\step{c}{\geq}
\frac{1}{S}
\sum_{\ell = 1}^S
\R_{X_\ell|X_1^{\ell-1}}(D_\ell) 
- 
\frac{1}{S}C\\
&\step{d}{\geq}
\frac{1}{S}
\big( \R_{X_\S}(D_\S) - C \big), 
\end{align*}
where (a) follows from the Markov chains in~\eqref{Prop:SuperuserSLowerBound:RDCFunctionMC}, (b) follows from~\eqref{Prop:SuperuserSLowerBound:RDCFunctionCache}, (c) follows from the distortion constraints~\eqref{Prop:SuperuserSLowerBound:RDCFunctionDist} and the definition of the informational conditional RD function $\R_{X_\ell | X_1^{\ell - 1}}(D_\ell)$; and (d) follows from Lemma~\ref{Lem:JointAndConditionalRDFunction}. 

The lower bound for the remaining subsets $\S \subseteq \L$ can be proved by repeating the above arguments with an appropriate relabelling of the variables. \qed


\section{Proof of Theorem~\ref{Thm:Strong:Converse}}\label{Thm:Strong:Converse:Proof}

\mw{We need the following lemma.}
\begin{lemma}\label{Lem:TypeStrongConverse}
Take any sequence of $(n,\sMc,\sM)$-codes and any positive real sequence $\{\alpha_n\} \downarrow 0$. If \mw{for every sufficiently large blocklength $n$ we have}
\begin{equation*}
\Pr \left[\ \bigcap_{\ell \in \L} \Big\{ \dbl(\h{X}_\ell^n,X_\ell^n) < D_\ell \Big\} \right] \geq 2^{-n \alpha_n},
\end{equation*}
then there exists real sequence $\{\zeta_n\} \to 0$ such that
\begin{equation*}
\frac{1}{n} \log |\sM| \geq \R\left(\D + \zeta_n,\frac{1}{n} \log |\sMc| + \zeta_n\right) - \zeta_n.
\end{equation*}
\end{lemma}
\begin{IEEEproof}
Lemma~\ref{Lem:TypeStrongConverse} is proved in Appendix~\ref{Lem:TypeStrongConverse:Proof}.
\end{IEEEproof}

Now consider Theorem~\ref{Thm:Strong:Converse} and any sequence of $(n,\sMc,$ $\sM)$-codes satisfying~\eqref{Eqn:StrongConverseConditionRate} and~\eqref{Eqn:StrongConverseConditionCache}. 
Pick a positive real sequence $\{\alpha_n\} \downarrow 0$ satisfying 
\begin{equation*}
\lim_{n \to \infty} 2^{-n \alpha_n} = 0.
\end{equation*}
Suppose that \mw{there exists a large blocklength $n^*$ so that for all $n> n^*$:}
\begin{equation}\label{Eqn:StrongConverseAssumptionForContradiction}
\Pr\left[\ \bigcap_{\ell \in \L} \Big\{ \dbl(\hX^n_\ell,X^n_\ell) < D_\ell \Big\} \right] \geq 2^{-n \alpha_n}.
\end{equation}

\mw{Pick $\gamma>0$ arbitrarily. By~assumptions \eqref{Eqn:StrongConverseConditionRate} and \eqref{Eqn:StrongConverseConditionCache}, and by Lemma~\ref{Lem:TypeStrongConverse}, we can pick $n^*$ sufficiently large so that $\forall n \geq n^*$ the following chain of inequalities holds:}
\begin{align}
\notag
\R(\D, C) + \gamma
\notag
&\step{a}{>}
\frac{1}{n} \log |\sM|\\[5pt]
\notag
&\step{b}{\geq}
\R\left(\D + \gamma,\frac{1}{n} \log |\sMc| + \gamma \right) - \gamma\\
\label{Eqn:StrongConverseContradiction}
&\step{c}{\geq}
\R\left(\D + \gamma, C + 2\gamma \right) - \gamma,
\end{align}
where step (a) follows by assumption~\eqref{Eqn:StrongConverseConditionRate}; step (b) follows from Lemma~\ref{Lem:TypeStrongConverse}; and step (c) follows by assumption~\eqref{Eqn:StrongConverseConditionCache} and the fact that the informational RDC function is non-increasing in the cache capacity. 

Since the RDC function $\R(\D,C)$ is a continuous function of $\D \in [0,\infty)^L$ and $C \in [0,\infty)$ and by choosing $\gamma$ sufficiently close to $0$, for any desired $\epsilon>0$ we can obtain from \eqref{Eqn:StrongConverseContradiction} that 
\begin{IEEEeqnarray}{rCl}\label{Eqn:StrongConverseLargeBlocklengths}
\lefteqn{\R(\D,C) - \frac{1}{n} \log |\sM| } \nonumber \quad \\
&\leq & \R(\D,C) - \R\left(\D + \gamma, C + 2\gamma \right) + \gamma \nonumber \\
 & < &  \epsilon. 
\end{IEEEeqnarray}
This contradicts assumption~\eqref{Eqn:StrongConverseConditionRate}.
We therefore conclude that assumption~\eqref{Eqn:StrongConverseAssumptionForContradiction} was wrong and holds with a strict inequality in the reverse direction for some $n \geq n^*$ and 
consequently 
\begin{equation}
\limsup_{n\to\infty} \Pr\left[\ \bigcup_{\ell \in \L} \Big\{ \dbl(\hX^n_\ell,X^n_\ell) \geq D_\ell \Big\} \right] = 1.
\end{equation}
\qed


\section{Proof of Lemma~\ref{Lem:TypeStrongConverse}}\label{Lem:TypeStrongConverse:Proof}


\subsection{Proof Setup and Outline}
Assume that we have a sequence of $(n,\sMc,\sM)$-codes for the RDC problem. For each blocklength $n$ and RDC code $(\encc,\enc,\dec)$, let
\begin{align*}
\s{G}\n := \Big\{ \x^n \in \sXL^n : \dbl\Big(\dec\big( \f(\x^n),\encc(\x^n)\big),x_\ell^n)\Big) < D_\ell,
\forall\ \ell \in \L \Big\} 
\end{align*}
denote the set of all ``good'' sequences that the code will reconstruct with acceptable distortions. Let $\{\alpha_n\} \downarrow 0$ be a sequence of positive real numbers, and suppose that the above mentioned sequence of RDC codes satisfies 
\begin{equation}\label{Eqn:EpsAssumption}
\Pr\big[ \X^n \in \s{G}\n \big] \geq 2^{-n \alpha_n}
\end{equation}
for every blocklength $n$. For example, we are free to choose $\{\alpha_n\}$ such that $\{2^{-n\alpha_n}\} \to 0$ or $\{2^{-n\alpha_n}\} \to 1$. 

The basic idea of the following proof is to show that~\eqref{Eqn:EpsAssumption} implies that the delivery-phase rate of the sequence of RDC codes satisfies 
\begin{equation}
\frac{1}{n} \log |\sM| \geq \R\left(\D + \zeta_n,\frac{1}{n}\log|\sMc| + \zeta_n\right) \mw{-\zeta_n}
\end{equation}
for some sequence $\{\zeta_n \} \to 0$. The key idea in proving this inequality will be to use the RDC code on a hypothetical ``perturbed'' source that is constructed from the good set $\s{G}\n$ and the DMS of pmf $p_\X$.


\subsection{Construction of the Perturbed Source}

The following construction is similar to that used by Watanabe~\cite{Watanabe-Aug-2015-A} and Gu and Effros~\cite{Gu-Jun-2009-C}.
Let us call the DMS 
\begin{equation*}
\X^n \sim p^n_{\X}(\x^n) = \prod_{i=1}^n p_\X(\x_i),
\quad
\x^n \in \sXL^n
\end{equation*}
the \emph{real source}. The \emph{perturbed source} 
\begin{equation*}
\Y^n \sim q_{\Y^n}(\y^n) = \Pr[\Y^n=\y^n] 
\quad
\y^n \in \sXL^n
\end{equation*}
is defined as follows: If $\y \in \s{G}_n$, then
\begin{subequations}\label{Eqn:PerturbedSourceDistributionDef}
\begin{equation}
q_{\Y^n}(\y^n) = \dfrac{2^{n (\alpha_n + \frac{1}{\sqrt{n}}) } p^n_\X(\y^n)}{2^{n (\alpha_n + \frac{1}{\sqrt{n}}) } \Pr[\X^n \in \s{G}_n] + \Pr[\X^n \notin \s{G}_n]}.
\end{equation}
Otherwise if $\y^n \notin \s{G}_n$, then
\begin{equation}
q_{\Y^n}(\y^n) = \dfrac{ p^n_\X(\y^n)}{2^{ n (\alpha_n + \frac{1}{\sqrt{n}}) } \Pr[\X^n \in \s{G}_n] + \Pr[\X^n \notin \s{G}_n]}
\end{equation}
\end{subequations}
It is worth noting that $q_{\Y^n}$ need not be a product distribution on $\sXL^n$. It is, however, not too difficult to see that $q_{\Y^n}$ is ``close'' to the product distribution $p^n_\X$ of the real DMS in the following sense. 
For every sequence $\y^n \in \sXL^n$:
\begin{align}\label{Eqn:PerturbedSourceProbabilityBounds}
2^{-n (\alpha_n + \frac{1}{\sqrt{n}})}\ p^n_\X(\y^n) \leq q_{\Y^n}(\y^n) 
\leq 2^{n (\alpha_n + \frac{1}{\sqrt{n}})}\ p^n_\X(\y^n).
\end{align}


\subsection{Caching the Perturbed Source --- Distortion Bounds}

We now take the $(n,\sMc,\sM)$-code $(\encc,\enc,\dec)$ from the above mentioned sequence, and use it to cache the perturbed source $\Y^n \sim q_{\Y^n}$. For each $\ell \in \L$, let 
\begin{equation*}
\h{Y}^n_\ell = \dec\big(\encc(\Y^n),\enc(\Y^n)\big)
\end{equation*} 
denote the corresponding output at the decoder. A lower bound on the probability of the decoding success for this RDC code on $\Y^n$ can be obtained as follows:
\begin{align}
\notag
& \Pr\big[\Y^n \in \s{G}\n\big] \\[5pt]
\notag
&= 
\sum_{\y^n \in \s{G}\n} q_{\Y^n}(\y^n)\\
\notag
&\step{a}{=}
\sum_{\y^n \in \s{G}\n} 
\dfrac{2^{n(\alpha_n + \frac{1}{\sqrt{n}})} p_\X(\y^n)}
{2^{n(\alpha_n + \frac{1}{\sqrt{n}})}\ \Pr[\X^n \in \s{G}\n]  + 1 - \Pr[\X^n \in \s{G}\n]}\\
\notag
&=
\dfrac{2^{n(\alpha_n + \frac{1}{\sqrt{n}})} \Pr[\X^n \in \s{G}\n]}
{2^{n(\alpha_n + \frac{1}{\sqrt{n}})}\ \Pr[\X^n \in \s{G}\n]  + 1 - \Pr[\X^n \in \s{G}\n]}\\
\notag
&=
\dfrac{2^{n(\alpha_n + \frac{1}{\sqrt{n}})}}
{2^{n(\alpha_n + \frac{1}{\sqrt{n}})}  + \frac{1}{\Pr[\X^n \in \s{G}\n]} - 1}\\
\notag
&\step{b}{\geq}
\dfrac{2^{n(\alpha_n + \frac{1}{\sqrt{n}})}}
{2^{n(\alpha_n + \frac{1}{\sqrt{n}})}  + 2^{n\alpha_n} - 1}\\
\notag
&=
\dfrac{2^{\sqrt{n}}}
{2^{\sqrt{n}}  + 1 - 2^{-n\alpha_n}}\\
\notag
&\geq
\dfrac{2^{\sqrt{n}}}
{2^{\sqrt{n}}  + 1},
\end{align}
where (a) substitutes the definition of $q_{\Y^n}(\y^n)$ from~\eqref{Eqn:PerturbedSourceDistributionDef} and (b) invokes the assumption~\eqref{Eqn:EpsAssumption}. Therefore,
\begin{equation*}
\lim_{n \to \infty} \Pr\big[\Y^n \in \s{G}\n\big] = 1.
\end{equation*} 

The expected distortion performance of the RDC code on $\Y^n \sim q_{\Y^n}$ can be upper bounded by
\begin{align}
\E\Big[ \dbl\big(\h{Y}_\ell^n,Y_\ell^n\big) \Big]
\notag
&= 
\E\Big[ \dbl\big(\h{Y}_\ell^n,Y_\ell^n\big) \Big| \Y^n \in \s{G}\n \Big] \Pr\Big[ \Y^n \in \s{G}\n \Big]\\
\notag
&\ \
+
\E\Big[ \dbl\big(\h{Y}_\ell^n,Y_\ell^n\big) \Big| \Y^n \notin \s{G}\n \Big] \Pr\Big[ \Y^n \notin \s{G}\n \Big]\\
\label{Eqn:PerturbedSourceExpectedDistortion}
&\leq 
D_\ell + \Dmax \left( 1 - \dfrac{2^{\sqrt{n}}}
{2^{\sqrt{n}}  + 1}\right).
\end{align}
Therefore, 
\begin{equation*}
\limsup_{n\to\infty} \E\big[ \dbl(\h{Y}_\ell^n,Y_\ell^n)\big] \leq D_\ell,
\quad 
\forall\ \ell \in \L.
\end{equation*}


\subsection{Caching the Perturbed Source --- A Lower Bound on the Caching Rate} 

We now use a slight modification of the converse proof in Appendix~\ref{Thm:RDC:Sec} to give a single-letter lower bound on the caching rate for the perturbed source. Let $\Mc = \encc(\Y^n)$ in $\sMc$ denote the corresponding cache message. We have
\begin{align}
\notag
& 
\frac{1}{n} \log |\sMc|
\geq 
\frac{1}{n} H(\Mc)
\notag
\geq 
\frac{1}{n} I(\Y^n ; \Mc)\\
\notag
&= 
\frac{1}{n} 
\sum_{i=1}^n I(\Y_i ; \Mc | \Y_1^{i-1}) \\
\notag
&\step{a}{=} 
\frac{1}{n} 
\sum_{i=1}^n I(\Y_i ; \Mc , \Y_1^{i-1}) - I(\Y_i ; \Y_1^{i-1})\\
\notag
&\step{b}{=} 
\frac{1}{n} 
\sum_{i=1}^n I(\Y_i ; U_i) - \frac{1}{n} \sum_{i=1}^n H(\Y_i) + \frac{1}{n} \sum_{i=1}^n H(\Y_i | \Y_1^{i-1})\\
\label{Eqn:PerturbedSourceCacheRateBound}
&= 
\frac{1}{n} 
\sum_{i=1}^n I(\Y_i ; U_i) - \frac{1}{n} \sum_{i=1}^n H(\Y_i) + \frac{1}{n} H(\Y^n),
\end{align}
where in (a) we note that $q_{\Y^n}$ need not be a product measure and (b) substitutes 
\begin{equation*}
U_i = (\Mc,\Y_1^{i-1}) 
\quad 
\text{on}
\quad
\s{U}_i = \sMc \times \sXL^{i-1}
\end{equation*}
in the same way as the weak converse in Appendix~\ref{Thm:RDC:Sec}. 


\subsection{Caching the Perturbed Source --- A Lower Bound on the Delivery Rate} 

Now consider an arbitrary request $\ell \in \L$, and let $M\n_\ell = \phi_\ell^{(n)}(\Y^n)$ in $\sM$ denote the corresponding delivery phase message. The delivery-phase rate can be lower bound as follows:
\begin{align}
\frac{1}{n} \log |\sM| 
\notag
&\geq 
\frac{1}{n} H(M\n_\ell | \Mc) \\
\notag
&\geq 
\frac{1}{n} I(\Y^n; M\n_\ell | \Mc) \\
\notag
&\step{a}{\geq} 
\frac{1}{n} I(\Y^n; \h{Y}^n_\ell | \Mc) \\
\notag
&= 
\frac{1}{n} \sum_{i=1}^n I(\Y_i;  \h{Y}^n_\ell | \Mc,\Y_1^{i-1}) \\
\label{Eqn:PerturbedSourceDeliveryRateBound}
&\geq
\frac{1}{n} \sum_{i=1}^n I(Y_{\ell,i};  \h{Y}_{\ell,i} | U_i),
\end{align}
where (a) follows because $\h{Y}^n_\ell \leftrightarrow (M\n_\ell,\Mc) \leftrightarrow \Y^n$ forms a Markov chain; and (b) substitutes $U_i$ as in the proof of the weak converse in Appendix~\ref{Thm:RDC:Sec}. 


\subsection{Caching the Perturbed Source --- Timesharing and Cardinality Reduction} 

Consider the tuple of random variables $(\Y^n,U^n,\hY{}^n)$ constructed in the above sections. Let $J \in \{1,2,\ldots,n\}$ be a uniform random variable that is independent of $(\Y^n,U^n,\hY{}^n)$, and let 
\begin{equation*}
\bar{\s{U}}\n = \left( \bigcup_{i=1}^n\ \s{U}_i \right) \times \{1,2,\ldots,n\}.
\end{equation*}
Let $\big(\bar{\Y}, \bar{U}, \hat{\bar{\Y}}\big) \in \sXL \times \bar{\s{U}} \times \hsXL$, 
denote the random tuples generated by setting 
\begin{equation*}
\bar{\Y} = \Y_J,\ \
\bar{U} = (U_J,J) \text{ and}\ \
\hat{\bar{\Y}} = \hat{\Y}_J.
\end{equation*}
With this choice, it then follows from~\eqref{Eqn:PerturbedSourceCacheRateBound} that 
\begin{align}
\frac{1}{n} \log |\sMc| 
\notag
&\geq
I(\Y_J ; U_J | J) - H(\Y_J) + \frac{1}{n} H(\Y^n) \\
\label{Eqn:PerturbedSourceCacheRateBound2}
&=
I(\bar{\Y} ; \bar{U}) - H(\bar{\Y}) + \frac{1}{n} H(\Y^n)
\end{align} 
and from~\eqref{Eqn:PerturbedSourceDeliveryRateBound} that 
\begin{align}
\frac{1}{n} \log |\sM| 
\notag
&\geq 
I(Y_{\ell,J} ; \hat{Y}_{\ell,J} | U_J,J)\\
\label{Eqn:PerturbedSourceDeliveryRateBound2}
&= 
I(\bar{Y}_{\ell} ; \hat{\bar{Y}}_{\ell} | \bar{U}).
\end{align}
Finally, from~\eqref{Eqn:PerturbedSourceExpectedDistortion} the expected distortion for satisfies
\begin{align}
\E\big[ \dl(\h{\bar{Y}}_\ell,\bar{Y}_\ell) \big] 
\notag
&= 
\E\Big[ \db(\hat{Y}^n_\ell,Y^n_\ell)\Big]\\
\label{Eqn:PerturbedSourceExpectedDistortion2}
&\leq 
D_\ell +
\Dmax \left( 1 - \dfrac{2^{\sqrt{n}}} {2^{\sqrt{n}}  + 1}\right).
\end{align}

Let $q_{\bar{\Y} \bar{U} \hat{\bar{\Y}}}$ denote the joint distribution of the variables $(\bar{\Y},\bar{U},\hat{\bar{\Y}})$. The cardinality of $\bar{\s{U}}\n$ grows without bound in $n$, and the next lemma uses the convex cover method~\cite[Appendix~C]{El-Gamal-2011-B} (see also the arguments in~\ref{Thm:RDC:Sec}) to bound this cardinality by a finite number.

\begin{lemma}\label{Lem:SupportLemmaStrongCoverse}
There exists a random tuple $(\bbar{\Y},\bbar{U},\hat{\bbar{\Y}}) \sim q_{\bbar{\Y} \bbar{U} \hat{\bbar{\Y}}}$ defined on $\sXL \times \bbar{\s{U}} \times \hsXL$ for which the following is true:
\begin{itemize}
\setlength{\itemsep}{5pt}
\item $|\bbar{\s{U}} | \leq |\sXL| + 2L$,
\item  $q_{\bbar{\Y}} = q_{\bar{\Y}}$,
\item $I(\bbar{\Y} ; \bbar{U}) = I(\bar{\Y} ; \bar{U})$, 
\item $I(\bbar{Y}_{\ell} ; \hat{\bbar{Y}}_{\ell} | \bbar{U}) = I(\bar{Y}_{\ell} ; \hat{\bar{Y}}_{\ell} | \bar{U})$ for all $\ell \in \L$, and
\item $\E[ \dl(\hat{\bbar{Y}}_\ell,\bbar{Y}_\ell)] = \E[ \dl(\h{\bar{Y}}_\ell,\bar{Y}_\ell)] $ for all $\ell \in \L$.
\end{itemize}
\end{lemma}

Combining Lemma~\ref{Lem:SupportLemmaStrongCoverse} with~\eqref{Eqn:PerturbedSourceCacheRateBound2}, \eqref{Eqn:PerturbedSourceDeliveryRateBound2} and~\eqref{Eqn:PerturbedSourceExpectedDistortion2} yields the following: There exists some tuple 
\begin{equation*}
(\bbar{\Y},\bbar{U},\hat{\bbar{\Y}}) \sim q_{\bbar{\Y} \bbar{U} \hat{\bbar{\Y}}} 
\quad
\text{on}
\quad \sXL \times \bbar{\s{U}} \times \hsXL
\end{equation*}
such that cache rate is lower bounded by
\begin{equation}\label{Eqn:PerturbedSourceCacheRateBound3}
\frac{1}{n} \log |\sMc| \geq I(\bbar{\Y} ; \bbar{U}) - H(\bbar{\Y}) + \frac{1}{n} H(\Y^n);
\end{equation}
the expected distortion is upper bounded by
\begin{equation}\label{Eqn:PerturbedSourceExpectedDistortion3}
\E\Big[ \dl(\hat{\bbar{Y}}_\ell,\bbar{Y}_\ell)\Big] 
\leq 
D_\ell + \Dmax \left( 1 - \dfrac{2^{\sqrt{n}}}
{2^{\sqrt{n}}  + 1}\right);
\end{equation}
and the delivery phase rate is lower bounded by
\begin{align}
\frac{1}{n} \log |\sM|
\notag
&\geq 
I(\bbar{Y}_{\ell} ; \h{\bbar{Y}}_{\ell} | \bbar{U})\\
\label{Eqn:PerturbedSourceDeliveryRateBound3}
&\geq 
\R_{\bbar{Y}_\ell | \bbar{U}}\left(D_\ell + \Dmax \left( 1 - \dfrac{2^{\sqrt{n}}}
{2^{\sqrt{n}}  + 1}\right)\right),
\end{align}
where the second inequality follows from the definition of the conditional RD function.


\subsection{Convergence of $H(\bbar{\Y})$ to $H(\X)$}

Fix $\gamma > 0$ arbitrarily small. The set of \emph{$\gamma$-letter typical sequences}~\cite{Kramer-2008-A} with respect to the DMS $p^n_\X$ will be useful in the following arguments. This set is given by
\begin{align*}
\s{A}\n_\gamma(p^n_\X)
= 
\Bigg\{
\x^n \in \sXL^n : 
\left| \frac{1}{n} \N(\a|\x^n) - p_\X(\a) \right| \leq \gamma\ p_\X(\a),
\
\forall\ \a \in \sXL
\Bigg\}.
\end{align*}

\begin{lemma}\label{Lem:LetterTypicalityLemmaKramer}
The probability that the real DMS $\X^n \sim p_\X$ does not emit an $\gamma$-letter typical sequence satisfies~\cite[Thm.~1.1]{Kramer-2008-A}
\begin{equation*}
\Pr\Big[ \X^n \notin \s{A}\n_\gamma(p_\X)\Big] 
\leq
2|\sXL|2^{-n\gamma^2 \mu(p_\X)},
\end{equation*} 
where $\mu(p_\X)$ is the smallest value of $p_\X$ on its support set $\supp{p_\X}$.
\end{lemma} 

Let us now return to the perturbed source $\Y^n \sim q_{\Y^n}$. For each $\a \in \sXL$ we have
\begin{align}
\notag
&
q_{\bbar{\Y}}(\a)\\
\notag
&\step{a}{=} q_{\bar{\Y}}(\a)\\ 
\notag
&\step{b}{=} 
\sum_{\y^n \in \sXL^n} 
q_{\Y^n}(\y^n)  \Pr\big[ \bar{\Y} = \a \big| \Y^n = \y^n \big]\\
\notag
&\step{c}{=} 
\sum_{\y^n \in \sXL^n} 
q_{\Y^n}(\y^n) 
\dfrac{\N(\a|\y^n)}{n}\\
\notag
&= 
\sum_{\y^n \in \s{A}\n_\gamma} 
q_{\Y^n}(\y^n) 
\dfrac{N(\a|\y^n)}{n}
+
\sum_{\y^n \notin \s{A}\n_\gamma} 
q_{\Y^n}(\y^n) 
\dfrac{N(\a|\y^n)}{n} \\
\notag
&\step{d}{\leq} 
p_\X(\a) (1 + \gamma) \Pr\big[ \X^n \in \s{A}\n_\gamma \big] 
+
 \Pr\big[ \X^n \notin \s{A}\n_\gamma \big] \\
\label{Eqn:PerturbedSourceSingleLetterDistributionBound1}
&\step{e}{\leq} 
p_\X(\a) (1 + \gamma) 
+
2|\sXL| 2^{-n\gamma^2 \mu(p_\X)}
\end{align}
where (a) applies Lemma~\ref{Lem:SupportLemmaStrongCoverse}; (b) and (c) use the fact that $\bar{\Y}$ is generated by uniformly at random selecting symbols from $\Y^n$ (the timesharing argument above);  (d) uses the definition of $\gamma$-letter typical sequences; and (e) invokes Lemma~\ref{Lem:LetterTypicalityLemmaKramer}. Using similar arguments, we obtain
\begin{equation}\label{Eqn:PerturbedSourceSingleLetterDistributionBound2}
q_{\bbar{\Y}}(\a) 
\geq 
p_\X(\a) (1 - \gamma) \big(1-2^{-n\gamma^2 \mu(p_\X)}\big).
\end{equation}

From~\eqref{Eqn:PerturbedSourceSingleLetterDistributionBound1} and~\eqref{Eqn:PerturbedSourceSingleLetterDistributionBound2}, we have
\begin{align}\label{Eqn:SequenceOfSingleLetterQbbar1}
(1-\gamma) p_\X(\a) 
\leq 
\liminf_{n \to \infty} q_{\bbar{\Y}}(\a) 
\leq 
\limsup_{n \to \infty} q_{\bbar{\Y}}(\a) 
\leq 
(1+\gamma) p_\X(\a).
\end{align}
Since~\eqref{Eqn:SequenceOfSingleLetterQbbar1} holds for every $\gamma > 0$, and the sequence $\{q_{\bbar{\Y}}\}$ does not dependent on $\gamma$, we have
\begin{equation}\label{Eqn:StrongConverseConvergence1a}
\lim_{n \to \infty} q_{\bbar{\Y}}(\a) = p_\X(\a),\quad \forall\ \a \in \sXL.
\end{equation}
Therefore, by the continuity of entropy~\cite[Chap.~2.3]{Yeung-2008-B} we have
\begin{equation}\label{Eqn:StrongConverseConvergence1}
\lim_{n \to \infty} H(\bbar{\Y}) = H(\X).
\end{equation}


\subsection{Convergence of $(1/n)H(\Y^n)$ to $H(\X)$}

It follows from~\eqref{Eqn:PerturbedSourceProbabilityBounds} that for all $\a^n \in \sXL^n$ we have
\begin{align}\label{Eqn:PerturbedSourceProbabilityBoundsLogVersion}
-\alpha_n - \frac{1}{\sqrt{n}} \leq \frac{1}{n}\log p^n_\X(\a^n)  - \frac{1}{n}\log q_{\Y^n}(\a^n) \\ \leq \alpha_n + \frac{1}{\sqrt{n}}.
\end{align}
Moreover, for every $\a^n \in \s{A}\n_\gamma(p_\X)$ we have
\begin{align}
\frac{1}{n} \log \frac{1}{p^n_\X(\a^n)} 
\notag
&\step{a}{=} 
\frac{1}{n}
\log \left(\prod_{i=1}^n \frac{1}{p_\X(\a_i)}\right) \\
\notag
&= 
\frac{1}{n}
\sum_{i = 1}^n \log \frac{1}{p_\X(\a_i)} \\
\notag
&= 
\frac{1}{n}
\sum_{\a' \in \sXL} \N(\a'|\a^n)  \log \frac{1}{p_\X(\a')}\\\
\notag
&\step{b}{\leq}
\big( 1 + \gamma\big) 
\sum_{\a' \in \sXL} p_\X(\a') \log \frac{1}{p_\X(\a')}\\
\label{Eqn:WeakTypicality1}
&=
\big( 1 + \gamma\big) 
H(\X),
\end{align}
where (a) follows because $p^n_\X$ is a product measure and (b) follows because $\a^n \in \s{A}\n_\gamma(p_\X)$. Similarly, we have
\begin{equation}\label{Eqn:WeakTypicality2}
\frac{1}{n} \log \frac{1}{p^n_\X(\a^n)} \geq (1 - \gamma) H(\X)
\end{equation}
for all $\a^n \in \sXL^n$.

\begin{floatEq}
\begin{align}
\frac{1}{n} H(\Y^n) 
\notag
&= 
 \frac{1}{n}
\sum_{\a^n \in \supp{q_{\Y^n}}} 
q_{\Y^n}(\a^n) \log \frac{1}{q_\Y(\a^n)}\\
\notag
&\step{a}{\leq}
\sum_{\a^n \in \supp{q_{\Y^n}}} 
q_{\Y^n}(\a^n) \left( \frac{1}{n} \log \frac{1}{p^n_\X(\a^n)} + \alpha_n + \frac{1}{\sqrt{n}}\right)\\
\notag
&=
\sum_{\a^n \in \s{A}\n_\gamma(p_\X)\ \cap\ \supp{q_{\Y^n}} }
q_{\Y^n}(\a^n) \left( \frac{1}{n} \log \frac{1}{p^n_\X(\a^n)} + \alpha_n + \frac{1}{\sqrt{n}}\right)\\
\notag
&\hspace{40mm}
+
\sum_{\a^n \notin \s{A}\n_\gamma(p_\X)\  \cap\ \supp{q_{\Y^n}}  }
q_{\Y^n}(\a^n) \left( \frac{1}{n} \log \frac{1}{p^n_\X(\a^n)} + \alpha_n + \frac{1}{\sqrt{n}}\right)\\
\notag
&\step{b}{\leq}
\sum_{\a^n \in \s{A}\n_\gamma(p_\X)\ \cap\ \supp{q_{\Y^n}} }
q_{\Y^n}(\a^n) \left( (1+\gamma)H(\X) + \alpha_n + \frac{1}{\sqrt{n}}\right)\\
\notag
&\hspace{45mm}
+
\sum_{\a^n \notin \s{A}\n_\gamma(p_\X)\  \cap\ \supp{q_{\Y^n}}  }
q_{\Y^n}(\a^n) 
\left( \log \frac{1}{\mu(p_\X)} + \alpha_n + \frac{1}{\sqrt{n}}\right)\\
\label{Eqn:HYnConvergenceToHX1}
&\step{c}{\leq} 
(1+\gamma)H(\X) + \alpha_n + \frac{1}{\sqrt{n}}
+
2|\X|2^{-n\gamma\mu(p_\X)} \left( \log \frac{1}{\mu(p_\X)} + \alpha_n + \frac{1}{\sqrt{n}}\right)
\end{align}
\end{floatEq}

Now consider the joint entropy $H(\Y^n)$. With a few manipulations, we obtain the upper bound in~\eqref{Eqn:HYnConvergenceToHX1}. Here step (a) uses~\eqref{Eqn:PerturbedSourceProbabilityBoundsLogVersion}. Step (b) uses the upper bound in~\eqref{Eqn:WeakTypicality1} on the first logarithmic term, and 
\begin{align}
\notag
\frac{1}{n} \log \frac{1}{p^n_\X(\a^n)} 
\notag
&= 
\frac{1}{n} \sum_{i=1}^n \log \frac{1}{p_\X(\a_i)}\\
\notag
&\leq 
\frac{1}{n} \sum_{i=1}^n \log \frac{1}{\mu(p_\X)}\\
\notag
&= 
\log \frac{1}{\mu(p_\X)}
\end{align}
on the second term\footnote{If $p^n_\X(\a^n) = 0$, then by definition $q_{\Y^n}(\a^n) = 0$ and $\a^n \notin \supp{q_{\Y^n}}$.}. Finally, step (c) applies Lemma~\ref{Lem:LetterTypicalityLemmaKramer}. Using similar arguments, we also have
\begin{align}
\notag
&\frac{1}{n} H(\Y^n)\\
\notag
&=
\frac{1}{n} \sum_{\a^n \in \supp{q_{\Y^n}}} 
q_\Y^n(\a^n) \log \frac{1}{q_\Y(\a^n)}\\
\notag
&\step{a}{\geq}
\sum_{\a^n \in\s{A}\n_\gamma(p_\X)} 
q_\Y^n(\a^n) \left( \frac{1}{n} \log \frac{1}{p^n_\X(\a^n)} - \alpha_n - \frac{1}{\sqrt{n}} \right)\\
\notag
&\step{b}{\geq} 
\sum_{\a^n \in \s{A}\n_\gamma(p_\X)} 
q_\Y^n(\a^n) \left((1-\gamma) H(\X)- \alpha_n - \frac{1}{\sqrt{n}} \right)\\
\label{Eqn:HYnConvergenceToHX2}
&\step{c}{\geq} 
\left((1-\gamma) H(\X)- \alpha_n - \frac{1}{\sqrt{n}} \right)
\Big(1 - 2|\X|2^{-n\gamma \mu(p_\X)}\Big).
\end{align}
Step (a) follows from~\eqref{Eqn:PerturbedSourceProbabilityBoundsLogVersion}; step (b) follows from~\eqref{Eqn:WeakTypicality2}; and step (c) applies Lemma~\ref{Lem:LetterTypicalityLemmaKramer}. From~\eqref{Eqn:HYnConvergenceToHX1} and~\eqref{Eqn:HYnConvergenceToHX2} we have for every fixed $\gamma > 0$ 
\begin{align*}
(1-\gamma) H(\X)
\leq 
\liminf_{n\to\infty} \frac{1}{n} H(\Y^n) 
\leq
\limsup_{n\to\infty} \frac{1}{n} H(\Y^n) 
\leq 
(1+\gamma) H(\X),
\end{align*}
which, in turn, implies 
\begin{equation}\label{Eqn:StrongConverseConvergence2}
\lim_{n \to \infty} \frac{1}{n} H(\Y^n) = H(\X).
\end{equation}


\subsection{Completing the Proof}

The above arguments show that there exists a sequence of random variables\footnote{Here, for clarity, we have added the subscript $n$ on the random variables to identify the corresponding blocklength $n$.} 
\begin{equation*}
\Big\{ (\bbar{\Y}_n,\bbar{U}_n) \sim q_{\bbar{\Y}_n}(\cdot)\ q_{\bbar{U}_n | \bbar{\Y}_n} (\cdot|\cdot)\Big\},
\end{equation*}
with each $(\bbar{\Y}_n,\bbar{U}_n)$ defined on $\sXL \times \sU$, such that 
\begin{equation*}
\lim_{n\to\infty} q_{\bbar{\Y}_n}(\a) = p_\X(\a),
\quad
\forall\ \a \in \sXL
\end{equation*}
and
\begin{align*}
\frac{1}{n} \log |\sMc| &\geq I(\bbar{\Y};\bbar{U}) - \eps_{1,n}\\
\frac{1}{n} \log |\sMc| &\geq \R_{\bbar{Y}_{\ell,n}|\bbar{U}_n}(D_\ell + \eps_{2,n}), \quad \forall\ \ell \in \L,
\end{align*}
where 
\begin{align}
\eps_{1,n} &= \Big| \frac{1}{n} H(\Y^n) - H(\bbar{\Y}) \Big|\\
\eps_{2,n} &= \Dmax \left(\mw{1-\frac{2^{\sqrt{n}}}{2^{\sqrt{n}} - 1}}\right).
\end{align}
Let $(\X,\bbar{U}_n) \sim p_\X(\cdot)\ q_{\bbar{U}_n | \bbar{\Y}_n}(\cdot|\cdot)$, and define
\begin{equation*}
\eps_{3,n} = \Big| \R_{\bbar{\Y}_n|\bbar{U}_n}(D_\ell + \eps_{2,n}) - \R_{\X|\bbar{U}_n}(D_\ell + \eps_{2,n}) \Big|.
\end{equation*}
Finally, choose $\zeta_n = \max\{\eps_{1,n},\eps_{2,n},\eps_{2,n}\}$ so that the lemma follows from~\eqref{Eqn:StrongConverseConvergence1a}, \eqref{Eqn:StrongConverseConvergence1} and \eqref{Eqn:StrongConverseConvergence2} and the continuity of the informational conditional RD function.
\qed


\section{Proof of Theorem~\ref{Thm:RDCFSep}}\label{Thm:RDCFSep:Proof}

The proof of Theorem~\ref{Thm:RDCFSep} will bootstrap the achievability part of Theorem~\ref{Thm:RDC} and the strong converse in Theorem~\ref{Thm:Strong:Converse}. 

Take the single-symbol distortion functions $\bd^*$ from~\eqref{Eqn:SingleSymbolDistortionFunctionDstar}, and consider $\R_{\bd^*}^\dag(\D,C)$ and $\R_{\bd^*}^\ddag(\D,C)$ --- the respective operational RDC functions  in the expected and excess distortion settings w.r.t.~the separable distortion functions 
\begin{equation*}
\bar{\bd}^* = (\db_1^*,\ldots,\db_L^*),
\end{equation*}
where
\begin{equation*}
\db^*_\ell(\hx_\ell^n,x_\ell^n) 
= \frac{1}{n} \sum_{i=1}^n \d^*_\ell(\hx_{\ell,i},x_{\ell,i})
= \frac{1}{n} \sum_{i=1}^n \fl\big(\d_\ell(\hx_{\ell,i},x_{\ell,i})\big).
\end{equation*}

\begin{lemma}\label{Lem:ThmFSepLem1}
\begin{equation*}
\R_{\bd^*}^\dag(\D,C) = \R_{\bd^*}^\ddag(\D,C) = \R_{\bd^*}(\D,C).
\end{equation*}
\end{lemma}

\begin{IEEEproof}
Apply Theorem~\ref{Thm:RDC} with $\bar{\bd}^*$.
\end{IEEEproof}

\begin{lemma}\label{Lem:ThmFSepLem2}
\begin{equation*}
\Rexc(\D,C) = \R_{\bd^*}^\ddag\big(\bf(\D),C\big).
\end{equation*}
\end{lemma}

\begin{IEEEproof}
For every $(n,\sMc,\sM)$-code we have
\begin{align}
\notag
&\Pr\left[\ \bigcup_{\ell \in \L}
\left\{
\bfdl(\hX^n_\ell,X^n_\ell) \geq D_\ell
\right\}
\right]\\
\notag
&\step{a}{=}
\Pr\left[\ \bigcup_{\ell \in \L} 
\left\{ 
\fl^{-1}\left( 
\frac{1}{n} \sum_{i=1}^n \fl \Big( \dl(\hX_\ell^n,X_\ell^n)\Big) 
\right) 
\geq D_\ell \right\}\right]\\
\notag
&=
\Pr\left[\ \bigcup_{\ell \in \L} 
\left\{ 
\frac{1}{n} \sum_{i=1}^n \d^*_\ell(\hX_\ell^n,X_\ell^n) 
\geq \fl\big(D_\ell\big) \right\}\right]\\
\notag
&\step{b}{=}
\Pr\left[\ \bigcup_{\ell \in \L} 
\left\{ 
\bar{\d}^*_\ell(\hX^n_\ell,X^n_\ell) 
\geq \fl\big(D_\ell\big) \right\}\right].
\end{align}
The left hand side of (a) corresponds to the excess-distortion event for $\Rexc(\D,C)$, and the right hand side of (b) corresponds to the excess-distortion event for $\R_{\bd^*}^\ddag(\bf(\D),C)$. Therefore, a sequence of $(n,\sMc,\sM)$-codes can achieve vanishing error probabilities w.r.t. the~$\f$-separable distortion functions $\bar{\bd}_\bf$ if and only if it achieves vanishing error probabilities w.r.t.~the separable distortion functions $\bar{\bd}^*$. 
\end{IEEEproof}

\begin{lemma}\label{Lem:ThmFSepLem3}
\begin{equation*}
\Rmaxexc(\D,C) \leq \Rexc(\D,C).
\end{equation*}
\end{lemma}

\begin{IEEEproof}
Recall Definition~\ref{Def:RDCFunc:ExcDist} and fix the distortion tuple $\D$ and cache capacity $C$. If $R > \Rexc(\D,C)$ then there exists a sequence of $(n,\sMc,\sM)$-codes satisfying~\eqref{Eqn:Def:Ach:ExpDist:CacheCapacity}, \eqref{Eqn:Def:Ach:ExpDist:DeliveryRate} and \eqref{Eqn:Def:Ach:ExcDist:Distortions}. For this sequence of codes, let 
\begin{equation*}
\s{G}_n = \bigcap_{\ell \in \L} \Big\{ \bar{\d}_\fl(\hX^n_\ell,X^n_\ell) < D_\ell \Big\},
\end{equation*}
and let $\s{G}^c_n$ denote the complement of $\s{G}_n$. Then  
\begin{align}
\notag
&\E \left[ \max_{\ell \in \L} \big(\bfdl(\hX^n_\ell,X^n_\ell) - D_\ell\big) \right] \\
\notag
&= 
\E \Big[ \max_{\ell \in \L} \big(\bfdl(\hX^n_\ell,X^n_\ell) - D_\ell\big) \Big| \s{G}_n \Big] \Pr\Big[ \s{G}_n \Big]\\
\notag
& \qquad
+ \E \Big[ \max_{\ell \in \L} \big(\bfdl(\hX^n_\ell,X^n_\ell) - D_\ell\big)  \Big| \s{G}^c_n \Big] \Pr\Big[ \s{G}^c_n \Big]\\
&\leq 
\Dmax\ \Pr[\s{G}^c_n].
\end{align}
Since $\Dmax$ is finite and $\Pr[\s{G}^c_n] \to 0$ by~\eqref{Eqn:Def:Ach:ExcDist:Distortions}, we have
\begin{equation*}
\limsup_{n \to \infty} \E \left[ \max_{\ell \in \L} \big( \bfdl(\hX^n_\ell,X^n_\ell) - D_\ell \Big) \right] \leq 0 
\end{equation*}
and $R \geq \Rmaxexc(\D,C)$ by Definition~\ref{Def:RDCFunc:ExpMaxDist}. 
\end{IEEEproof}

\begin{lemma}\label{Lem:ThmFSepLem4}
\begin{equation*}
\Rmaxexc(\D,C) \geq \R_{\bd^*}(\bf(\D),C).
\end{equation*}
\end{lemma}

\begin{IEEEproof}
If $\R_{\bd^*}(\D^*,C) = 0$, then the lemma immediately follows because we always have $\Rmaxexc(\D,C) \geq 0$. We henceforth restrict attention to the nontrivial case $\R_{\bd^*}(\bf(\D),C) > 0$. 

Suppose, to the contrary of Lemma~\ref{Lem:ThmFSepLem4}, that $\Rmaxexc(\D,C)$ is strictly smaller than $\R_{\bd^*}(\bf(\D),C)$ and, therefore, there exists some $\gamma > 0$ such that 
\begin{equation}\label{Lem:ThmFSepLem4:Eqn:Dgamma}
\Rmaxexc(\D,C) \leq \R_{\bd^*}(\bf(\D),C) - \gamma.
\end{equation}
By the continuity and monotonicity of $\R_{\bd^*}(\bf(\D),C)$ and each $\fl$, there exists some distortion tuple $\D'$ such that
\begin{equation}\label{Lem:ThmFSepLem4:Eqn:Ddash}
\R_{\bd^*}(\bf(\D'),C) = \R_{\bd^*}(\bf(\D),C) - \frac{\gamma}{2}
\end{equation}
where $D'_\ell > D_\ell$ for all $\ell \in \L$.

Now recall Definition~\ref{Def:RDCFunc:ExpMaxDist} and the operational meaning of $\Rmaxexc(\D,C)$. There exists a sequence of $(n,\sMc,\sM)$-codes satisfying~\eqref{Eqn:Def:Ach:ExpDist:CacheCapacity}, \eqref{Eqn:Def:Ach:ExpDist:DeliveryRate} and~\eqref{Def:RDCFunc:ExpMaxDist:Eqn}. On combining~\eqref{Eqn:Def:Ach:ExpDist:DeliveryRate}, \eqref{Lem:ThmFSepLem4:Eqn:Dgamma} and~\eqref{Lem:ThmFSepLem4:Eqn:Ddash}, we see that the delivery-phase rates of this sequence of codes satisfy 
\begin{equation}\label{Eqn:AsymptoticDeliveryPhaseRateFSep}
\limsup_{n\to\infty} \frac{1}{n} \log |\sM| \leq \R_{\bd^*}\big(\bf(\D'),C\big) - \frac{\gamma}{2}.
\end{equation}

Now consider the excess-distortion performance of the sequence of $(n,\sMc,\sM)$-codes w.r.t.~the separable distortion functions $\bd^*$. Let
\begin{equation*}
\s{B}_n = 
\bigcup_{\ell \in \L} \Big\{ \bar{\d}_\ell^*(\hX^n_\ell,X^n_\ell) \geq \f_\ell(D_\ell') \Big\},
\end{equation*}
and let $\s{B}^c_n$ denote the complement of $\s{B}_n$. Notice that we have
\begin{equation*}
\s{B}_n = \bigcup_{\ell \in \L} \Big\{ \bfdl (\hX^n_\ell,X^n_\ell) \geq D_\ell' \Big\}.
\end{equation*}
Since the asymptotic delivery-phase rate is strictly smaller than the informational RDC function~\eqref{Eqn:AsymptoticDeliveryPhaseRateFSep}, the strong converse in Theorem~\ref{Thm:Strong:Converse} yields 
\begin{equation*}
\limsup_{n\to\infty}
\Pr\big[\s{B}_n\big] = 1.
\end{equation*}

Let 
\begin{equation*}
\zeta = \min_{\ell \in \L} \big(D'_\ell - D_\ell\big).
\end{equation*}
We now have
\begin{align}
\notag
&
\E \Big[ \max_{\ell \in \L} \big(\bfdl(\hX^n_\ell,X^n_\ell) - D_\ell \big) \Big] \\ 
\notag
&=
\E \Big[ \max_{\ell \in \L} \big( \bfdl(\hX^n_\ell,X^n_\ell) - D_\ell \big) \Big| \s{B}_n\Big]\ \Pr\Big[\s{B}_n\Big]\\
\notag
& \quad
+ \E \Big[ \max_{\ell \in \L} \big( \bfdl(\hX^n_\ell,X^n_\ell) - D_\ell \big) \Big| \s{B}^c_n\Big]\ \Pr\Big[\s{B}^c_n\Big] \\
\notag
&\step{a}{\geq} 
\E \Big[ \max_{\ell \in \L} \big( \bfdl(\hX^n_\ell,X^n_\ell) - D_\ell \big) \Big| \s{B}_n\Big]\ \Pr\Big[\s{B}_n\Big]\\
\notag
&
- \big(\min_{\ell \in \ell} D_\ell\big) \Pr\Big[\s{B}^c_n\Big]\\
\label{Eqn:INeedMoreCoffee}
&\step{b}{\geq} 
\zeta\ \Pr\Big[\s{B}_n\Big] 
- \big(\min_{\ell \in \ell} D_\ell\big) \Pr\Big[\s{B}^c_n\Big],
\end{align}
where (a) follows because $\bfdl(\hX^n_\ell,X^n_\ell)$ is nonnegative; and (b) follows because, conditioned on $\s{B}_n$, there must exist at least one $\ell' \in \L$ such that 
\begin{equation*}
\bfdlp(\hX^n_{\ell'},X^n_{\ell'}) \geq D'_{\ell'} > D_{\ell'}
\end{equation*} 
and thus 
\begin{equation*}
\max_{\ell \in \L} \big(\bfdl(\hX^n_\ell,X^n_\ell) - D_\ell\big) \geq {\bfdlp}(\hX^n_{\ell'},X^n_{\ell'}) - D_{\ell'} > \zeta.
\end{equation*}
Finally, we have
\begin{align*}
0 &\step{a}{=}
\limsup_{n\to\infty} \E \Big[ \max_{\ell \in \L} \big( \bfdl(\hX^n_\ell,X^n_\ell) - D_\ell \big) \Big] \\
&\step{b}{\geq} 
\limsup_{n\to\infty} \Big[\zeta\ \Pr\big[ \s{B}_n \big] - (\min_{\ell \in \L} D_\ell) \Pr\big[ \s{B}^c_n \big] \Big]\\
&\step{c}{>} 
0,
\end{align*}
where (a) follows from~\eqref{Def:RDCFunc:ExpMaxDist:Eqn}, (b) follows from~\eqref{Eqn:INeedMoreCoffee}, and (c) follows because $\Pr[\s{B}_n] \to 1$ by the strong converse Theorem~\ref{Thm:Strong:Converse} and $\zeta > 0$. The above contradiction implies that $\Rmaxexc(\D,C)$ \emph{cannot} be strictly smaller than $\R_{\bd^*}(\bf(\D),C)$. 
\end{IEEEproof}

To complete the proof of Theorem~\ref{Thm:RDCFSep:Proof} we need only combine the above lemmas:
\begin{align*}
\Rmaxexc(\D,C) 
&\step{a}{\leq}
\Rexc(\D,C)\\
&\step{b}{=}
\R_{\bd^*}^\ddag\big(\bf(\D),C\big) \\
&\step{c}{=}
\R_{\bd^*}(\bf(\D),C)\\
&\step{d}{\leq}
\Rmaxexc(\D,C),
\end{align*}
where (a) uses Lemma~\ref{Lem:ThmFSepLem3}, (b) uses Lemma~\ref{Lem:ThmFSepLem2}, (c) uses Lemma~\ref{Lem:ThmFSepLem1}, and (d) uses Lemma~\ref{Lem:ThmFSepLem4}. \qed


\section{Proof of Theorem~\ref{Thm:Gaussian:Converse}}\label{Sec:Proof:Thm:Gaussian:Converse}

Fix $\D = (D,D,\ldots,D)$ for some $D \geq 0$, and consider any tuple $(U,\hXL)$ satsifying~\eqref{Eqn:GaussianConstraints}. \mw{Fix $\s{S}\subseteq \L$ and let $S:=|\s{S}|$}. Then
\begin{align}
\notag
\max_{\ell \in \L} &\
I(X_\ell ; \hX_\ell | U)\\
\notag
&\geq 
\max_{\ell \in \S}
\Big[
h(X_\ell | U) - h(X_\ell | \hX_\ell) 
\Big]
\\
\notag
&\step{a}{\geq} 
\frac{1}{S} 
h(\XS | U) - \frac{1}{2} \log(2\pi e D)
\\
\notag
&\step{b}{\geq} 
\frac{1}{S} \left( \frac{1}{2} \log \big((2\pi e)^S \det \KXS\big) - C\right) 
- \frac{1}{2} \log(2\pi e D)\\
\notag
&=
\frac{1}{2S} \log \frac{\det \KXS}{D^S} - \frac{C}{S}.
\end{align}
Step (a) follows because 
\begin{align*}
h(X_\ell|\hX_\ell) 
&\step{a.1}{=} 
h(X_\ell - \hX_\ell | \hX_\ell) \\
&\step{a.2}{\leq}
h\big(\s{N}(0,\E(\hX_\ell - X_\ell)^2) \big) \\
&\step{a.3}{\leq}
h\big(\s{N}(0,D) \big) \\
&\step{a.4}{\leq}
\frac{1}{2} \log (2\pi e D),
\end{align*}
where (a.1) follows by the \emph{translation property of differential entropy}~\cite[Thm.~10.18]{Yeung-2008-B}; (a.2) uses the fact that the normal distribution maximises differential entropy for a given second moment~\cite[Thm.~10.43]{Yeung-2008-B}, and (a.3) invokes the distortion constraint in~\eqref{Eqn:GaussianConstraints}. Moreover, for the first term, we have
\begin{equation*}
\max_{\ell \in \S} h(X_\ell|U)
\step{a.5}{\geq}
\frac{1}{S}\ \sum_{\ell \in \S} h(X_\ell|U)
\step{a.6}{\geq}
\frac{1}{S}\ h(\XS|U),
\end{equation*}
where (a.5) follows because the maximum cannot be smaller than the average, and (a.6) follows by the \emph{independence bound for differential entropy}~\cite[Thm.~10.34]{Yeung-2008-B} 

Step (b) follows from the cache capacity constraint in~\eqref{Eqn:GaussianConstraints}
\begin{align*}
C 
&\geq 
I(\X;U)
\geq 
I(\XS;U)\\
&= 
h(\XS) - h(\XS|U)\\
&= 
\frac{1}{2} \log \big( (2\pi e)^S \det\KXS \big) - h(\XS|U)
\end{align*}
\qed


\section{Proof of Theorem~\ref{Thm:BivariateGaussian}}\label{Sec:Proof:Thm:BivariateGaussian}


\subsection{Case 1: $(D,C) \in \s{S}_1$}

If $(D,C) \in \s{S}_1$, then it trivially follows from the definition of $\RGjoint(D,D)$ that $\RG(D,D,C) = 0$. 


\subsection{Case 2: $(D,C) \in \s{S}_2$}

Since $\RGjoint(D,D)$ is strictly decreasing in $D$, it follows that for a given $C \leq \RGjoint(D,D)$ the distortion $D$ must satisfy
\begin{equation*}
0 < D \leq 2^{-C} \sqrt{1-\rho^2}.
\end{equation*}
Define
\begin{equation}\label{Eqn:GaussianAlphaCoding}
\alpha = 1 - \rho - 2^{-C} \sqrt{1 - \rho^2}
\end{equation}
and note that $0 \leq \alpha < 1 - \rho$ for all finite
\begin{equation*}
C > \frac{1}{2} \log \frac{1 + \rho}{1 - \rho}.
\end{equation*}

Now let $W,N_1,N_2,\tN_1,\tN_2,Z_1$ and $Z_2$ be mutually independent standard Gaussians $\s{N}(0,1)$, and notice that our bivariate Gaussian source $(X_1,X_2)$ can be written as 
\begin{align*}
X_i = \sqrt{\rho}\ W 
+ \sqrt{\alpha}\ N_i 
+ \sqrt{1 - \rho - D - \alpha}\ \tN_i 
+ \sqrt{D}\ Z_i,
\quad
i = 1,2
\end{align*}
 
Choose $U = (U_1,U_2)$, where
\begin{equation*}
U_i = \sqrt{\rho}\ W + \sqrt{\alpha}\ N_i,
\quad 
i = 1,2.
\end{equation*}
Define the reconstructions $\hX_1$ and $\hX_2$ to be
\begin{equation*}
\hX_i := U_i + \sqrt{1 - \rho - \alpha - D}\ \tN_i,
\quad i = 1,2.
\end{equation*}
We notice that
\begin{equation}\label{Eqn:GaussainMCCoding1}
X_1 \markov \hX_1 \markov U_1 \markov U \markov U_2 \markov \hX_2 \markov X_2
\end{equation}
forms a Markov chain. Additionally, 
\begin{align*}
\notag
& I(X_1,X_2;U)\\ 
&= 
h(X_1,X_2) - h(X_1,X_2|U)\\
&\step{a}{=}
h(X_1,X_2) - h(X_1|U) - h(X_2|U)\\
&\step{b}{=}
h(X_1,X_2) - h(X_1|U_1) - h(X_2|U_2)\\
&=
h(X_1,X_2) - 2h(X_1|U_1)\\
&\step{c}{=}
h(X_1,X_2) - 2 h(X_1-U_1|U_1)\\
&=
\frac{1}{2} \log\big((2\pi \e)^2 (1-\rho^2)\big) 
- \log\big( 2\pi\e(1-\rho-\alpha)\big)\\
&=
\frac{1}{2} \log \frac{1-\rho^2}{(1-\rho-\alpha)^2}\\
&\step{d}{=} 
C,
\end{align*}
where (a) and (b) follow from~\eqref{Eqn:GaussainMCCoding1}, (c) follows by symmetry, and (d) substitutes~\eqref{Eqn:GaussianAlphaCoding}. Similarly, 
\begin{align*}
I(X_1;\h{X}_1|U)
&= 
h(X_1|U) - h(X_1|\hX_1,U)\\
&\step{a}{=} 
h(X_1|U_1) - h(X_1|\hX_1)\\
&= 
h(X_1-U_1|U_1) - h(X_1-\hX_1|\hX_1)\\
&=
\frac{1}{2} \log\Big( 2\pi\e (1-\rho-\alpha) \Big) 
- \frac{1}{2} \log\Big( 2\pi\e D\Big)\\
&=
\frac{1}{2} \log\left( \frac{1-\rho-\alpha}{D} \right)\\
&= 
\frac{1}{4} \log \left(\frac{1-\rho^2}{D^2}\right)   - \frac{C}{2},
\end{align*}
where (a) uses the Markov chain~\eqref{Eqn:GaussainMCCoding1} and (b) substitutes~\eqref{Eqn:GaussianAlphaCoding}.
Finally, we notice that the above achievable rate is equal to the superuser lower bound from Theorem~\ref{Thm:Gaussian:Converse}.


\subsection{Case 3: $(D,C) \in \s{S}_3$}

Let
\begin{equation*}
\alpha = \frac{1}{2} (1 + \rho) (1 - 2^{-2C}),
\end{equation*}
and note that $0 \leq \alpha \leq \rho$. Now let $W,\tilde{W},Z_1,Z_2,N_1$ and $N_2$ be mutually independent standard Gaussians $\s{N}(0,1)$. Choose
\begin{equation*}
U = \sqrt{\alpha}\ W + \sqrt{\rho - \alpha}\ \tilde{W}
\end{equation*}
and
\begin{equation*}
\hX_i = \sqrt{\rho}\ W + \sqrt{1 - \rho - D}\ Z_i,
\quad
i = 1,2.
\end{equation*}
We may now write our bivariate Gaussian source $(X_1,X_2)$ as 
\begin{equation*}
X_i = \hX_i + \sqrt{D}\ N_i,
\quad 
i = 1,2.
\end{equation*}
The pair $(X_1,U)$ and the pair $(X_2,U)$ are both zero mean bivariate Gaussians with identical covariance matrices
\begin{equation*}
\b{K}_{X_1,U} 
= \b{K}_{X_2,U}
=
\begin{bmatrix}
1 & \sqrt{\alpha \rho}\\
\sqrt{\alpha \rho} & \rho
\end{bmatrix}.
\end{equation*}
Similarly, $(X_1,X_2,U)$ is a zero mean multivariate normal with the covariance matrix 
\begin{equation*}
\b{K}_{X_1 X_2 U} = 
\begin{bmatrix}
1 & \rho & \sqrt{\alpha \rho}\\
\rho & 1 & \sqrt{\alpha \rho}\\
\sqrt{\alpha \rho} & \sqrt{\alpha \rho} & \rho
\end{bmatrix}.
\end{equation*}
Thus, 
\begin{align*}
I(X_1,X_2;U) 
&= 
h(X_1,X_2) + h(U) - h(X_1,X_2,U)\\
&= 
\frac{1}{2} \log \big( (2\pi e)^2\det \b{K}_{X_1X_2} \big)
+ \frac{1}{2} \log \big( 2\pi e \rho \big)\\
&\hspace{25mm}
- \frac{1}{2} \log \big( (2\pi e)^3\det \b{K}_{X_1X_2U} \big)\\
&= 
\frac{1}{2} \log \frac{1 + \rho}{1 + \rho - 2 \alpha}\\
&= 
C,
\end{align*}
and
\begin{align*}
I(X_1;\hX_1|U) 
&= 
h(X_1|U) - h(X_1|\hX_1,U)\\
&= 
h(X_1,U) - h(U) - h(X_1|\hX_1)\\
&= 
\frac{1}{2} \log \big((2\pi e)^2 \det \b{K}_{X_1U} \big)
- \frac{1}{2} \log(2\pi e \rho) \\
&\hspace{45mm}
- \frac{1}{2} \log(2\pi e D)
\\
&= 
\frac{1}{2} \log \frac{1 - \alpha}{D}.
\end{align*}


\subsection{Case 4: $(D,C) \in \s{S}_4$}

Suppose that $(D,C) \in \s{S}_4$. Since $(D,C)$ lies below the Gaussian joint RD function $\RGjoint(D,D)$, it follows that for any given distortion $D \in [1-\rho,1]$ the cache capacity $C$ must lie within 
\begin{equation*}
0 \leq C \leq \frac{1}{2} \log \frac{1 + \rho}{2D- 1 + \rho}.
\end{equation*}
Define
\begin{equation*}
\alpha = \frac{1}{2} (1 + \rho) (1 - 2^{-2C})
\end{equation*}
and
\begin{equation*}
\beta = 1 - \alpha - D,
\end{equation*}
where we notice that 
\begin{equation*}
0 \leq \alpha,\beta \leq 1  - D
\quad 
\text{and}
\quad 
\alpha + \beta = 1 - D \leq \rho.
\end{equation*}
In this case, we may write
\begin{equation*}
X_i = \sqrt{\alpha}\ A + \sqrt{\beta}\ B + \sqrt{\rho - (\alpha + \beta)}\ W + \sqrt{1 - \rho}\ N_i,
\quad i = 1,2,
\end{equation*}
where $A,B,W,N_1$ and $N_2$ are mutually independent standard Gaussians $\s{N}(0,1)$. Now let 
\begin{equation*}
U = \sqrt{\alpha}\ A,
\end{equation*}
and 
\begin{equation*}
\hX_1 = \hX_2 = \hX := U + \sqrt{\beta}\ B.
\end{equation*}

Here $(X_1,X_2,U)$ is a zero mean multivariate Gaussian with covariance matrix
\begin{equation*}
\b{K}_{X_1,X_2,U} 
= 
\begin{bmatrix}
1 & \rho & \alpha \\
\rho & 1 & \alpha \\
\alpha & \alpha & \alpha.
\end{bmatrix}
\end{equation*} 
Then, 
\begin{align*}
I(X_1,X_2;U) 
&=
h(X_1,X_2) - h(X_1,X_2,U) - h(U)\\
&= 
\frac{1}{2}\log\big((2\pi\e)^2 \det \b{K}_{X_1,X_2} \big)\\
&\hspace{10mm}
- \frac{1}{2}\log\big((2\pi\e)^3 \det \b{K}_{X_1,X_2,U} \big)\\
&\hspace{40mm}
+\frac{1}{2} \log\big(2\pi\e \rho \big)\\
&=
\frac{1}{2} \log \frac{1 + \rho}{1 + \rho - 2 \alpha}\\
&= 
C.
\end{align*}
Moreover, 
\begin{align*}
I(X_1;\hX|U) 
&= 
h(X_1|U) - h(X_1|U,\hX)\\
&= 
h(X_1|U) - h(X_1|\hX)\\
&=
\frac{1}{2} \log \big( 2\pi\e (1 - \alpha) \big) 
- \frac{1}{2} \log \big( 2\pi\e (1 - \alpha - \beta) \big)\\
&=
\frac{1}{2} \log\frac{1 - \alpha }{D}.
\end{align*}
\qed


\section{Proof of Theorem~\ref{Thm:TwoUsersAchievability} (Outline)}\label{Thm:TwoUsersAchievability:Sec} 

Choose $(U,\b{\hX}, \b{\tX})$ such that~\eqref{Thm:TwoUserAchievability:D1} and \eqref{Thm:TwoUserAchievability:D2} hold. We need only find a scheme that has an arbitrarily small average error probability whenever 
\begin{align*}
C & \geq 
I(U;\X) -I(U;\tX_{\ell_2})\\
R &\geq 
I(U,\X;\tX_{\ell_2})
+I(\X;\hX_{\ell_1}|U,\tX_{\ell_2})
\end{align*}
holds for all $(\ell_1,\ell_2) \in \mathcal{L}_1\times\mathcal{L}_2$. 


\subsection{Code Construction} 

Fix $\epsilon>0$ arbitrarily small. Generate a $U$-codebook
\begin{equation*}
\big\{ U^n(m_u) 
= 
\big(U_1(m_u),U_2(m_u),\ldots,U_n(m_u)\big)
\big\}
\end{equation*}
indexed by
\begin{equation*}
m_u =1,2,\ldots,2^{n(I(\X;U)+\eps)}
\end{equation*}
by randomly selecting symbols from $\sU$ in an iid manner using $U \sim p_U$. For each index $\ell_2$, generate an $\tX_{\ell_2}$-codebook
\begin{equation*}
\big\{ 
\tX^n_{\ell_2}(m_{\ell_2}) = 
\big( 
\tX_{\ell_2,1}(m_{\ell_2}),
\ldots,
\tX_{\ell_2,n}(m_{\ell_2})\big)
\big\}
\end{equation*}
indexed by
\begin{equation*}
m_{\ell_2} =1,2,\ldots,2^{n(I(U,\X;\tX_{\ell_2})+\epsilon)}
\end{equation*}
by selecting symbols from~$\tsX_{\ell_2}$ iid $\tX_{\ell_2} \sim p_{\tX_{\ell_2}}$. Finally, for each $(\ell_2,m_u,m_{\ell_2})$, generate an $\hX_{\ell_1}$-codebook
\begin{align*}
\Big\{\hX^n_{\ell_1}(m_u,m_{\ell_2},\mw{\ell_2},m_{\ell_1}) = \big(
\hX_{\ell_1,1}(m_u,m_{\ell_2},\mw{\ell_2},m_{\ell_1}),\ldots,
\hX_{\ell_1,n}(m_u,m_{\ell_2},\mw{\ell_2},m_{\ell_1}) \big)
\Big\}
\end{align*}
indexed by
\begin{equation*}
m_{\ell_1}  = 1,2,\ldots, 2^{n(I(\X;\hX_{\ell_1}|U,\tX_2)+\epsilon)}
\end{equation*}
by selecting symbols from~$\hsX_{\ell_1}$  independently (in a memoryless manner) according to 
\begin{equation*}
\prod_{i=1}^n p_{\hX_{\ell_1}|U,\tX_{\ell_2}}\big(\cdot \big| u_i(m_u),\tx_{\ell_2i}(m_{\ell_2})\big).
\end{equation*}


\subsection{Cache Encoder at Server}

Find a tuple of indices $m_u$ and $(m_{\ell_2}; \ \ell_2 \in \L_2)$ such that 
\begin{equation*}
(U^n(m_u),\X^n,\tX^n_{\ell_2}(m_{\ell_2}))
\end{equation*}
is jointly typical for every $\ell_2\in\mathcal{L}_2$. If there are one or more index tuples pick one uniformly at random; otherwise, declare an error. Represent $m_u$ as a string of binary bits, and let $m_c$ denote the first $nC$ bits of $m_u$.


\subsection{Delivery-Phase Encoder at Server}
Given $(m_u,m_{\ell_2})$,  find $m_{\ell_1}$ such that
\begin{equation*}
(U^n(m_u),\X^n,\tX^n_{\ell_2}(m_{\ell_2}),\hX^{n}_{\ell_1}(m_u,m_{\ell_2},\mw{\ell_2},m_{\ell_1}))
\end{equation*}
is jointly typical. If there are one or more indices, pick one uniformly at random and set $m=(m_{\ell_1},m_{\ell_2})$. If there is no such index tuple declare an error.


\subsection{Delivery-Phase Decoders}

\subsubsection{User 1}

Find an index $\hat{m}_u$ with the same first $nC$ bits as in $m_\text{c}$ such that 
\begin{equation*}
(U^n(\h{m}_u),\tX^{n}_{\ell_2}(m_{\ell_2}))
\end{equation*}
is jointly typical. If there is no such index declare error. If there are multiple such indices choose one uniformly at random. Output $\hX^{n}_{\ell_1}(\hat{m}_u,m_{\ell_2},\mw{\ell_2},m_{\ell_1})$.

\subsubsection{User $2$} 

Output $\tX^{n}_{\ell_2}(m_{\ell_2})$.

\section{Proof of Corollary~\ref{cor:DBSB2user}}\label{RoyPaper}
Starting from \eqref{eq:conv}, one can write
\begin{align}
&\R^\dag(\b{D}, \b{0},C) \nonumber\\
&\geq 
\underline{\R}(\b{D}, \b{0},C) \\
&\geq \label{a}
1+\max_{(\ell_1, \ell_2)} \;\max\Big\{ 0,\ R_{X_{\ell_1}|X_{\ell_2}}(D_{\ell_1})-C \Big\} \\
&\geq 
1+ \;\max\Big\{ 0,\ R_{X_{\ell}|X_{\ell^\prime}}(D)-C,R_{X_{\ell^\prime}|X_{\ell}}(D)-C \Big\} \\
&\geq 
1+ \;\max\Big\{ 0,h(\rho)-h(D)-C \Big\} \label{b}
\end{align}
where \eqref{a} is because $\min-\max$ is larger than or equal to $\max-\min$ and \eqref{b} is because the conditional rate-distortion function of a doubly symmetric source is given by $h(\rho)-h(D)$.

Also, we specialise \eqref{eq:cor-D0} by choosing $U=(\hat{X}_{\ell},\hat{X}_{\ell^\prime})$ to obtain
\begin{align}
&R^\dag(\b{D}, \b{0},C)\nonumber\\
&\leq
\overline{\R}(\b{D}, \b{0},C)\\
&\leq  \min_{\b{\hX}}\max_{ \ell_2} \max \Big\{I(\X;\hat{X}_{\ell},\hat{X}_{\ell^\prime},{X}_{\ell_2})-C,
H(X_{\ell_2})\Big\}\\
&= 1+\min_{\b{\hX}}\max \left\{0,I(\X;\hat{X}_{\ell},\hat{X}_{\ell^\prime}|{X}_{\ell^\prime})-C,\right.\nonumber\nonumber\\&\qquad\qquad\qquad\qquad\qquad \left.I(\X;\hat{X}_{\ell},\hat{X}_{\ell^\prime}|{X}_{\ell})-C \right\}\\
&\leq \left\{\!\!\begin{array}{l}1 + \left(\ h(\rho)-h(D)-C \right)^+ \\\hspace{5cm} \text{If }D\leq D^\star \\1+\left(h(D)\!-\!\rho\!-\!(1\!-\!\rho)h\left(\frac{2D-\rho}{2(1-\rho)}\right)\!-\!C\right)^+\\\hspace{5cm}\text{If }D^\star<D\leq \frac{1}{2}\end{array}\right.\label{c}
\end{align}
where $D^\star$ is defined in \eqref{Dstar} and  \eqref{c} follows by the choice of $\b\hX$ that is made in the proof of \cite[Theorem 3]{Timo-Nov-2010-A}.

\section{Proof of Theorem~\ref{thm:Loweravg}}
\label{app:conv-2user-avg}
For any $\epsilon>0$, let the distortion vector $\b{\epsilon}=(\epsilon,\ldots,\epsilon)$. We have
\begin{align}
&n(R(\D,\b{\epsilon},C)+C)\nonumber\\&\geq H(M_{\ell_1,\ell_2}^{(n)},M^{(n)}_c)\\
&\geq I(M_{\ell_1,\ell_2}^{(n)},M^{(n)}_c;\X^n)\\
&=\sum_{i=1}^n I(M_{\ell_1,\ell_2}^{(n)},M^{(n)}_c,\X^{i-1};\X_i)\\
&=\sum_{i=1}^n I(\hat{X}^n_{\ell_1},\tilde{X}^n_{\ell_2},M^{(n)}_c,\X^{i-1};\X_i)\label{d1}\\
&\geq\sum_{i=1}^n I(\hat{X}^n_{\ell_1},{X}^n_{\ell_2},M^{(n)}_c,\X^{i-1};\X_i)-n\delta_1(\epsilon)\label{e1}\\
&=\sum_{i=1}^n I(\hat{X}^n_{\ell_1},{X}_{\ell_2,i},{X}^n_{\ell_2,i+1},M^{(n)}_c,\X^{i-1};\X_i)-n\delta_1(\epsilon)\\
&\geq \sum_{i=1}^n I(\hat{X}_{\ell_1,i},{X}_{\ell_2,i},U_{\ell_2,i};\X_i)-n\delta_1(\epsilon)\\
&=nI(\hat{X}_{\ell_1,Q},{X}_{\ell_2,Q},U_{\ell_2,Q};\X_Q|Q)-n\delta_1(\epsilon)\\
&=nI(\hat{X}_{\ell_1,Q},{X}_{\ell_2,Q},U_{\ell_2,Q},Q;\X_Q)-n\delta_1(\epsilon)\label{avg1}
\end{align}
where we have defined $U_{\ell_2,i}=(M^{(n)}_c,{X}^n_{\ell_2,i+1},\X^{i-1})$, and $Q$ is a random variable that is independent of everything else and takes values in $\{1,\ldots,n\}$ uniformly at random. In the above chain of inequalities, \eqref{d1} is because $\tilde{X}_{\ell_2}^n$ and $\hat{X}_{\ell_1}^n$ are functions of $M_{\ell_1,\ell_2}^{(n)}$ and $(M_c^{(n)},M_{\ell_1,\ell_2}^{(n)})$, respectively, and \eqref{e1} is because $\bar{d}_{\ell_2}(X_{\ell_2}^n,\tilde{X}_{\ell_2}^n)\leq \epsilon$. Here, $\delta_1(\epsilon)\to 0$ as $\epsilon\to 0$.

Similarly, we have
\begin{align}
&nR(\D,\b{\epsilon},C)\nonumber\\
&\geq I(M_{\ell_1,\ell_2}^{(n)};M^{(n)}_c,\X^n)\\
&= I(M_{\ell_1,\ell_2}^{(n)},\tilde{X}_{\ell_2}^n;M^{(n)}_c,\X^n)\\
&\geq I(M_{\ell_1,\ell_2}^{(n)},{X}_{\ell_2}^n;M^{(n)}_c,\X^n)-n\delta_2(\epsilon)\\
&\geq I({X}_{\ell_2}^n;M^{(n)}_c,\X^n)+I(M_{\ell_1,\ell_2}^{(n)};M^{(n)}_c,\X^n|{X}_{\ell_2}^n)-n\delta_2(\epsilon)\\
&\geq I({X}_{\ell_2}^n;M^{(n)}_c,\X^n)+I(M_{\ell_1,\ell_2}^{(n)};\X^n|M^{(n)}_c,{X}_{\ell_2}^n)-n\delta_2(\epsilon)\\
&\geq I({X}_{\ell_2}^n;M^{(n)}_c,\X^n)+I(\hat{X}^n_{\ell_1};\X^n|M^{(n)}_c,{X}_{\ell_2}^n)-n\delta_2(\epsilon)\\
&= \sum_{i=1}^nI({X}_{\ell_2,i};M^{(n)}_c,\X^n|{X}_{\ell_2,i+1}^{n})\nonumber\\&\quad+\sum_{i=1}^nI(\hat{X}^n_{\ell_1};\X_i|M_c^{(n)},{X}^n_{\ell_2},\X^{i-1})-n\epsilon\\
&= \sum_{i=1}^nI({X}_{\ell_2,i};M^{(n)}_c,\X^n,{X}_{\ell_2,i+1}^{n})\nonumber\\&\quad+\sum_{i=1}^nI(\hat{X}^n_{\ell_1};\X_i|M_c^{(n)},{X}^n_{\ell_2},\X^{i-1})-n\epsilon\\
&\geq \sum_{i=1}^nI({X}_{\ell_2,i};M^{(n)}_c\X^{i-1}{X}_{\ell_2,i+1}^{n})\nonumber\\&\quad+\sum_{i=1}^nI(\hat{X}_{\ell_1,i};\X_i|M_c^{(n)},\X^{i-1},{X}^n_{\ell_2,i+1},{X}_{\ell_2,i})-n\delta_2(\epsilon)\\
&= \sum_{i=1}^nI({X}_{\ell_2,i};U_{\ell_2,i})\nonumber\\&\quad+\sum_{i=1}^nI(\hat{X}_{\ell_1,i};\X_i|U_{\ell_2,i},{X}_{\ell_2,i})-n\delta_2(\epsilon)\\
&=nI({X}_{\ell_2,Q};U_{\ell_2,Q}|Q)\nonumber\\&\quad+nI(\hat{X}_{\ell_1,Q};\X_Q|U_{\ell_2,Q},{X}_{\ell_2,Q},Q)-n\delta_2(\epsilon)\\
&=nI({X}_{\ell_2,Q};U_{\ell_2,Q}Q)\nonumber\\&\quad+nI(\hat{X}_{\ell_1,Q};\X_Q|U_{\ell_2,Q},Q,{X}_{\ell_2,Q})-n\delta_2(\epsilon)\label{avg2}
\end{align}

Let us now average \eqref{avg1} and \eqref{avg2} over $\ell_2\in\{1,\ldots,L_2\}$ weighted by $p_I(\ell_2)$. We thus have
\begin{align}
&n(R(\D,\b{\epsilon},C)+C)\nonumber\\&\geq n\sum_{\ell_2}P_I(\ell_2)I(\hat{X}_{\ell_1,Q},{X}_{\ell_2,Q},U_{\ell_2,Q},Q;\X_Q)-n\delta_1(\epsilon)\\
&= nI(\hat{X}_{\ell_1,Q},{X}_{I,Q},U_{I,Q},Q;\X_Q|I)-n\delta_1(\epsilon)
\end{align}
and similarly, we have
\begin{align}
&nR(\D,\b{\epsilon},C)\nonumber\\
&\geq nI({X}_{I,Q};U_{I,Q},Q|I)\nonumber\\&\quad+nI(\hat{X}_{\ell_1,Q};\X_Q|U_{I,Q},Q,{X}_{I,Q},I)-n\delta_2(\epsilon)
\end{align}

Finally, we verify that \eqref{Thm:TwoUserAchievability:D1} holds. Since the rate-distortion-memory $(R,\D,C)$ is admissible, for any $\epsilon^\prime>0$ we have
\begin{align}
(D+\epsilon^\prime)&\geq\mathbb{E}\left[\bar{d}_{\ell_1}(X_{\ell_1}^n,\hat{X}^n_{\ell_1})\right]\\
&=\frac{1}{n}\sum_{i=1}^n\mathbb{E}\left[{d}_{\ell_1}(X_{\ell_1,i},\hat{X}_{\ell_1,i})\right]\\
&=\mathbb{E}\left[{d}_{\ell_1}(X_{\ell_1,Q},\hat{X}_{\ell_1,Q})\right]
\end{align}

We now define $U=(U_{I,Q},Q)$, and rename $\hat{X}_{\ell_1,Q}$, $X_{I,Q}$, and $\X_{Q}$ to $\hat{X}_{\ell_1}$, $X_{I}$, and $\X$, respectively. It is not difficult to see that \eqref{probdis} holds.



\end{document}